\documentclass[useAMS,usenatbib,usegraphicx]{mn2e}

\def\gtrsim{\mathrel{\hbox{\rlap{\hbox{\lower3pt\hbox{$\sim$}}}\hbox{\raise2pt\hbox{$>$}}}}}

\title[Stirring in massive, young debris discs]{Stirring in massive, young debris discs from spatially resolved 
Herschel images\thanks{This work is based on observations made with the Herschel Space
Observatory. Herschel is an ESA
space observatory with science instruments provided by European-led
Principal Investigator consortia and with important participation from
NASA.}}
\author[A. Mo\'or et al.]{A. Mo\'or$^{1}$\thanks{E-mail:moor@konkoly.hu}, 
\'A. K\'osp\'al$^{1,2}$, P. \'Abrah\'am$^{1}$, D. Apai$^{3}$, Z. Balog$^{4}$, C. Grady$^{5}$, Th. Henning$^{4}$,
\newauthor A. Juh\'asz$^{6}$, Cs. Kiss$^{1}$, A.~V. Krivov$^{7}$, N. Pawellek$^{7}$ and Gy.~M. Szab\'o$^{1,8}$ \\  
$^{1}$Konkoly Observatory, Research Centre for Astronomy and Earth Sciences, \\ 
Hungarian Academy of Sciences, PO Box 67, 
H-1525 Budapest, Hungary\\
$^{2}$European Space Agency (ESA-ESTEC, SRE-S),
P.O. Box 299, 2200 AG, Noordwijk, The Netherlands; ESA fellow.\\
$^{3}$Department of Astronomy and Department of Planetary Sciences, The University of Arizona, Tucson, AZ 85721 USA\\
$^{4}$Max-Planck-Institut f\"ur Astronomie, K\"onigstuhl 17, 69117 Heidelberg, Germany\\
$^{5}$Exoplanets and Stellar Astrophysics Laboratory, Code 667, Goddard Space Flight Center, Greenbelt, MD 20771 USA\\
$^{6}$Leiden Observatory, Leiden University, Niels Bohrweg 2, NL-2333 CA Leiden, The Netherlands \\
$^{7}$Astrophysikalisches Institut und Universit\"atssternwarte, 
Friedrich-Schiller-Universit\"at Jena, Schillerg\"a\ss chen 2--3, 07745 Jena, Germany \\
$^{8}$ ELTE, Gothard Astrophysical Observatory, H-9704 Szombathely, Szent Imre herceg \'ut 112, Hungary \\
}

\begin{document}

\date{Accepted ... Received ...; in original form ...}

\pagerange{\pageref{firstpage}--\pageref{lastpage}} \pubyear{2013}

\maketitle

\label{firstpage}

\begin{abstract}
A significant fraction of main-sequence stars are encircled by
dusty debris discs, where the short-lived dust particles are
replenished through collisions between planetesimals.
Most destructive collisions occur when the orbits of smaller bodies
are dynamically stirred up, either by the gravitational effect of locally formed  
Pluto-sized planetesimals (self-stirring scenario), or via secular perturbation caused by an inner 
giant planet (planetary stirring). The relative importance of these
scenarios in debris systems is unknown. Here we present
new {\sl Herschel Space Observatory} imagery of 11 discs selected 
from the most massive and extended known debris systems.
All discs were found to be extended at far-infrared wavelengths, 
five of them being resolved for the first time.
We evaluated the feasibility of the self-stirring scenario by comparing
the measured disc sizes with the predictions of the model calculated for 
the ages of our targets. 
We concluded that the self-stirring explanation works for seven discs. However, in four
cases, the predicted pace of outward propagation of the stirring front, assuming reasonable
initial disc masses, was far too low to explain the radial extent of the cold dust.
Therefore, for HD\,9672, HD\,16743, HD\,21997, and HD\,95086, another explanation is needed. We
performed a similar analysis for $\beta$\,Pic and HR\,8799, reaching the same conclusion. We argue
that planetary stirring is a promising possibility to explain the disk properties in these
systems.
In HR\,8799 and HD\,95086 we may already know the potential 
perturber, since their known outer giant planets could be responsible for the stirring process.
Interestingly, the discs around HD\,9672, HD\,21997, and $\beta$\,Pic are also unique in harbouring
detectable amount of molecular CO gas.
Our study demonstrates that among the largest and
most massive debris discs self-stirring may not be the only
active scenario, and potentially planetary stirring is responsible for destructive collisions and debris dust production
in a number of systems.

\end{abstract}


\section{Introduction} \label{intro}

Many main-sequence stars host circumstellar dust discs
whose particles re-emit the absorbed stellar light and produce thermal 
emission at infrared (IR) and millimeter wavelengths. 
Grain-grain collisions and dynamical interactions with the stellar radiation lead to the removal of the
emitting dust particles on a timescale significantly shorter than the age of the star. 
It is believed that the particles are continuously replenished by collisional 
erosion of previously formed larger bodies. Thus, these
{\sl debris} discs are composed of second generation material, and their presence implies 
the existence of a
significant planesimal population.
 Because of the close link between the dust 
and large bodies the detailed investigation of the debris discs
can also provide information on the characteristics and evolution of the 
parent planetesimal belt(s) and even on the formation and evolution 
of the underlying planetary system \citep[e.g.][]{wyatt2008}. 

For collisions to produce a copious amount of dust, 
the relative velocities between the colliding larger bodies must exceed 
a critical value. This requires a dynamical stirring of the planetesimals' 
motion. In the {\sl self-stirring} scenario proposed 
by \citet{kb2004}, the gradual build-up of large planetesimals via collisional 
coagulation of smaller bodies eventually leads to an enhanced dust production. 
According to this model, the emergent largest planetesimals ("oligarchs") perturb the 
orbits of neighbouring smaller bodies, increasing their inclination 
and eccentricities. 
This results in destructive collisions and initiates a collisional 
cascade. 
In a specific disc region the peak dust production roughly coincides with the 
formation of $\sim$1000~km radius  ($\sim$Pluto-sized) planetesimals.   
Alternatively, in the {\sl planetary or binary stirring} scenario 
\citep{wyatt2005b,mustill2009}, a giant planet or a stellar companion triggers the
collisions. In this case, the companion can dynamically excite  the motion of
planetesimals via secular perturbations even if the companion is located far from the
planetesimals.
Self-stirring is an inside-out process, i.e. the collisional cascade is
ignited in the inner disc first and then the active dust production
propagates outward. The same is true for planetary stirring if the
perturber is located closer to the star than the planetesimal belt.
{\sl Stellar flybys} can also 
initiate more energetic collisions between planetesimals \citep{kb2002}.  
The scarcity of stellar encounters among old field stars, however, suggests 
that this mechanism is likely limited to debris systems 
located in dense young clusters \citep{matthews2014}. 

Very little is known about the relative contributions of self-stirring and planetary stirring in 
observed debris systems. \citet{kennedy2010} concluded 
that the observational statistics provided by the {\sl Spitzer Space Telescope} (hereafter {\sl Spitzer}) 
for debris discs around A-type stars can be reproduced by simulating the evolution of 
a sample of discs assuming either self-stirring or planetary stirring.
Although the importance of the different mechanisms cannot be established 
by studying the whole debris population, there are some individual systems 
where self-stirring can be excluded with high probability. 
In self-stirring models, the pace of the outward propagation of the formation of
1\,000~km-sized planetesimals depends on the disc mass: the more massive the 
disc initially, the faster the outwards spread.  Therefore, the existence of dust-producing 
planetesimals at large stellocentric radii around a relatively young star would require 
an unrealistically massive initial protoplanetary disc. 
In these cases, planetary stirring by an inner planet is a natural candidate because its timescale 
in the outer disc could be faster than that of the growth of large planetesimals 
needed for self-stirring. 
Indeed, based on this consideration, \citet{mustill2009} identified debris discs around 
two young moving group members where self-stirring is unlikely. 

The location of the dust grains in a debris disc is generally estimated from their 
temperature and
the star's luminosity, assuming thermal equilibrium. However, the 
temperature of a dust grain depends not only on its 
distance from the star, but also on its size and optical properties. 
Because of this well known degeneracy \citep[e.g.][]{krivov2010}, modeling of the spectral 
energy distribution (SED) cannot provide an 
unambiguous picture on the spatial distribution of dust 
in debris systems. To break this degeneracy and to estimate 
the location of the emitting grains reliably and precisely, spatially resolved 
images are needed. Due to the unprecedented spatial resolution 
(6--11\arcsec) and sensitivity of the {\sl Herschel Space
Observatory} \citep[hereafter {\sl Herschel,}][]{pilbratt2010} at far-infrared (far-IR) 
wavelengths, the number of resolved debris discs increased significantly in the
last few years \citep{matthews2014,pawellek2014}. 
These studies showed that disc sizes derived 
from resolved images are generally 2--4 times larger than 
those inferred from SED analysis assuming blackbody grains
\citep[e.g.][]{marshall2011,booth2013,morales2013}.
   
Here we report on our {\sl Herschel} observations of 11 debris discs, where 
young age and an earlier estimate of disc size from previous infrared observations hinted for 
stirring mechanisms other than self-stirring.
We first review the target selection, 
observations, and data reduction aspects of the programme (Sect~\ref{obsanddatared}). 
In Sect~\ref{dataanalysis} we summarize the stellar properties, present a basic analysis 
of the {\sl Herschel} images and measure fluxes.
We then compile and model the SED of the targets and 
model the resolved images using a simple geometrical approach (Sect.~\ref{results}). 
The results are discussed in Sect.~\ref{discussion}. Our conclusions are presented in Sect.~\ref{conclusion}.

\section{Observations and data reduction} \label{obsanddatared}

\subsection{Sample selection} 
Nine of our targets were selected on the basis of their previous {\sl Spitzer}/MIPS observations. 
By analysing {\sl Spitzer} data of 27 debris systems around F-type stars we already identified 
two objects, HD~50571 and HD~170773, whose dust emission were 
marginally extended in 70\,{\micron} MIPS images \citep{moor2011a}. In order to search for 
further candidates we queried the {\sl Spitzer} archive for bright debris systems 
(with flux density $>$ 200~mJy at 70~{\micron}) located within 120\,pc, 
and downloaded their MIPS 70~{\micron} observations.
Using the same method as in \citet{moor2011a} we 
found several additional marginally extended discs and estimated their characteristic sizes. 
We assumed that the derived sizes correspond to the size of the dust 
ring and that they are co-located with the parent planetesimals. Then, by adopting age estimates 
from the literature for each system, we evaluated the feasibility of the self-stirring mechanism for 
each disc based on the formulae from \citet{mustill2009}. 
Selecting only those objects not yet included in 
other {\sl Herschel} programmes, we identified nine discs
(including HD~50571 and HD~170773), where self-stirring would 
require initial disc masses either at the extreme upper end of the
mass distibution 
of protoplanetary discs obtained from millimetre measurements, or so high 
that the disc would be gravitationally 
unstable. 
These 9 objects are our prime candidates where alternative stirring mechanisms may operate.
To this sample we added HD~16743 and HD~182681, where disc sizes inferred from our SED 
analysis of a sample of debris systems turned out to be unusually large.

The uncertainty of the  disc sizes derived from the {\sl Spitzer}/MIPS profiles 
turned out to be too high to confidently exclude the self-stirring scenario.
New higher spatial resolution {\sl Herschel} observations are 
indispensable to measure disc sizes with the required accuracy.

\subsection{Herschel measurements}

We obtained far-IR and submillimetre maps for our targets 
using the Photodetector Array Camera and Spectrometer \citep[PACS,][]{poglitsch2010} and 
the Spectral and Photometric Imaging Receiver \citep[SPIRE,][]{griffin2010} 
onboard the {\sl Herschel Space Observatory}. 
Apart from HD~16743, which was observed as part of the OT1\_ckiss\_1 programme (PI: Cs. Kiss), 
all other maps were obtained in the framework of OT1\_pabraham\_2
(PI: P. \'Abrah\'am). Table~\ref{herschellog} presents the log of our {\sl Herschel} observations.
\begin{table}                                                                   
\setlength{\tabcolsep}{1.6mm}                                                   
\begin{center}                                                                  
\scriptsize                                                                     
\caption{{\bf Observation log}  \label{herschellog}}                                      
\begin{tabular}{lccc}                                                           
\hline\hline                                                                    
TARGET          &  INSTR./FILTER    &    OBS. ID    &   DATE \\                 
\hline                                                                          
     HD 9672 &  PACS 70/160 &  1342224377/1342224378 &   2011-07-18 \\
     HD 9672 & PACS 100/160 &  1342224379/1342224380 &   2011-07-18 \\
     HD 9672 &        SPIRE &             1342236226 &   2012-01-02 \\
    HD 10939 &  PACS 70/160 &  1342225326/1342225327 &   2011-07-23 \\
    HD 10939 & PACS 100/160 &  1342225328/1342225329 &   2011-07-23 \\
    HD 10939 &        SPIRE &             1342236211 &   2012-01-02 \\
    HD 16743 &  PACS 70/160 &  1342261271/1342261272 &   2013-01-17 \\
    HD 16743 &        SPIRE &             1342268403 &   2013-03-25 \\
    HD 17848 &  PACS 70/160 &  1342225120/1342225121 &   2011-08-01 \\
    HD 17848 & PACS 100/160 &  1342225122/1342225123 &   2011-08-02 \\
    HD 17848 &        SPIRE &             1342230802 &   2011-10-11 \\
    HD 21997 &  PACS 70/160 &  1342237454/1342237455 &   2012-01-13 \\
    HD 21997 & PACS 100/160 &  1342237456/1342237457 &   2012-01-13 \\
    HD 21997 &        SPIRE &             1342238289 &   2012-01-28 \\
    HD 50571 &  PACS 70/160 &  1342233376/1342233377 &   2011-12-01 \\
    HD 50571 & PACS 100/160 &  1342233378/1342233379 &   2011-12-01 \\
    HD 50571 &        SPIRE &             1342226639 &   2011-08-16 \\
    HD 95086 &  PACS 70/160 &  1342225005/1342225006 &   2011-08-01 \\
    HD 95086 & PACS 100/160 &  1342225007/1342225008 &   2011-08-01 \\
    HD 95086 &        SPIRE &             1342226634 &   2011-08-16 \\
   HD 161868 &  PACS 70/160 &  1342231117/1342231118 &   2011-10-18 \\
   HD 161868 & PACS 100/160 &  1342231119/1342231120 &   2011-10-18 \\
   HD 161868 &        SPIRE &             1342229577 &   2011-09-23 \\
   HD 170773 &  PACS 70/160 &  1342241398/1342241399 &   2012-03-14 \\
   HD 170773 & PACS 100/160 &  1342241400/1342241401 &   2012-03-14 \\
   HD 170773 &        SPIRE &             1342216001 &   2011-03-13 \\
   HD 182681 &  PACS 70/160 &  1342231927/1342231928 &   2011-11-05 \\
   HD 182681 & PACS 100/160 &  1342231929/1342231930 &   2011-11-05 \\
   HD 182681 &        SPIRE &             1342230820 &   2011-10-11 \\
   HD 195627 &  PACS 70/160 &  1342232492/1342232493 &   2011-11-18 \\
   HD 195627 & PACS 100/160 &  1342232494/1342232495 &   2011-11-19 \\
   HD 195627 &        SPIRE &             1342216008 &   2011-03-13 \\
\hline                                                                          
\end{tabular}                                                                   
\end{center}                                                                    
\end{table}                                                                     


\subsubsection{PACS observations and data reduction} \label{pacsdatared}

PACS data were obtained in mini scan-map mode 
(PACS Observer's Manual v2.5\footnote{http://herschel.esac.esa.int/Docs/PACS/html/pacs\_om.html}) 
at a medium scan speed of 
20{\arcsec}~s$^{-1}$, with 10 scan-legs of 3{\arcmin} length separated by 
4{\arcsec}. For targets in the OT1\_pabraham\_2 programme we made maps with 
scan angles of 70{$\degr$} and 110{$\degr$} 
both in the 70\,{\micron} and in the 100\,{\micron} bands, repeated four 
times in each scan direction. This setup provided in total 8 separate scans both at 70 and 100\,{\micron}.
Since every PACS observation provides a 160\,{\micron} measurement as well, 
at this wavelength we obtained in total 16 separate scans 
for our targets. In the case of HD~16743 only 100\,{\micron} and 160\,{\micron} maps were taken with a 
single repetition. With the medium scan speed the PACS beam size is 
$\sim$5\farcs6, $\sim$6\farcs8 and $\sim$11\farcs4, at 70, 100 and 160\,{\micron}, respectively.

The PACS raw data were reduced with the Herschel Interactive Processing Environment 
\citep[HIPE,][]{ott2010} version 11.1 using PACS calibration file release version 56 and the standard HIPE scripts. 
We selected those data frames where the actual scan speed of the spacecraft 
was between 15 and 25{\arcsec}~s$^{-1}$.
PACS maps are subject to 1/f noise that was removed by applying highpass filtering 
with filter widths of 25, 30, and 40 frames for
the 70, 100, and 160\,{\micron} data, respectively. 
At a scan speed of 
20{\arcsec}~s$^{-1}$ and with 10~Hz data sampling, these frame numbers 
correspond to filter widths of 102{\arcsec}, 122{\arcsec} and 162{\arcsec} 
in the 70, 100, and 
160\,{\micron} bands.
In order to avoid flux loss caused by this process, the immediate vicinity of our targets 
was excluded from the filtering using a circular mask placed at the sources' positions. 
The mask radius was 25{\arcsec} in the case of the
70 and 100\,{\micron} maps, and 30{\arcsec} at 160\,{\micron}. 
The best filter widths and source mask radii were determined by 
processing the maps with different parameter values and measuring the target flux with aperture photometry. 
By increasing the filter width, the measured flux increased until a point 
where the flux no longer changed, pinpointing the filtering parameters 
where flux loss can be neglected.       
Glitches were identified and removed using the second-level deglitching HIPE task.

As a final step of reduction, we compiled individual scan 
maps (corresponding to the individual repetitions) with pixel sizes of 1\farcs1, 1\farcs4, and 
2\farcs1 at 70, 100, and 160\,{\micron} using the {\sc 'photProject'} task. 
Mosaics were also created in each band, 
by combining the individual scan maps using a weighted average.

In order to ensure that we did not filter out low level
extended emission from the outer regions of the discs, the maps were also 
processed using the JScanam tool which is a Java and Jython based implementation 
of the Scanamorphos code \citep{roussel2013} integrated in HIPE. Application of this algorithm with 
the "galactic" option ensures the preservation of extended emission 
at all scales. By comparing the JScanam maps with the previous ones we found that 
the radial profiles of our sources were identical in the two data sets and the new maps did not 
reveal additional faint extended emission around our sources. 
For the further analysis we always use the original high-pass filter maps.

\subsubsection{SPIRE observations and data reduction}
All SPIRE observations were obtained in Small Scan Map mode (see Spire Handbook 
v2.5\footnote{http://herschel.esac.esa.int/Docs/SPIRE/spire\_handbook.pdf}) 
 resulting in simultaneous 250, 350, and 
 500\,{\micron} maps. We made two repetitions of the small maps for each target.
Data were reduced with HIPE v11.1 following the standard pipeline processing steps.
The final maps were produced using the {\sc 'naiveMapper'} task. 
 The beam sizes were $\sim$17\farcs6, $\sim$23\farcs9, and $\sim$35\farcs2 at 250, 350, and 
 500\,{\micron}, respectively, and the maps were resampled to pixel sizes of 
 6{\arcsec}, 10{\arcsec}, and 14{\arcsec} 
 at these wavelengths.

\subsection{Spitzer archival data}
Apart from HD~182681, all of our targets were observed both with the Multiband Imaging
Photometer for Spitzer \citep[MIPS,][]{rieke2004} and the Infrared Spectrograph \citep[IRS,][]{houck2004}
onboard the {\sl Spitzer Space Telescope} \citep{werner2004}. 
All data were downloaded from the 
{\sl Spitzer Heritage Archive\footnote{http://irsa.ipac.caltech.edu/applications/Spitzer/SHA/}}. \\

MIPS observations were performed at 
 24 and 70\,{\micron}. In the case of HD~16743, HD~50571, and HD~170773, 160\,{\micron}
 maps are available as well. 
We downloaded all 70\,{\micron} MIPS data that were obtained in photometric imaging 
mode (default scale, small-field size). 
The basic calibrated data, produced by the pipeline version 18.12, were processed 
with the MOsaicking and Point source Extraction \citep[MOPEX]{makovoz2005} tool performing the same steps 
as described in \citet{moor2011a}. The final
mosaic images had 4{\arcsec} pixels.

IRS observations of HD~21997 and HD~95086 were performed in Spectral Mapping mode, their spectra were taken 
from the literature \citep{moor2013a,moor2013b}. 
For the other targets, where the IRS Staring Mode was used in the observations, the spectra were retrieved 
from the CASSIS\footnote{The Cornell Atlas of Spitzer/IRS Sources (CASSIS) is a product of the Infrared 
Science Center at Cornell University, supported by NASA and JPL.} database \citep{lebouteiller2011}. 
As post-processing, we discarded outliers
by fitting polynomials to the data of individual IRS modules using a robust method and then searching for 
data points outlying by more than 4$\sigma$ from these fits. 
The longest wavelength parts of the LL1 and SL1 modules at wavelength $>$38\,{\micron} and $>$14.5\,{\micron} were discarded 
automatically because of their generally bad quality. 
Then CASSIS spectra were scaled to the predicted photospheres (Sect.~\ref{stellarpropssect}) using the
shortest wavelength parts of the spectra which are not contaminated by 
the emission from the disc. The scaling factor was less than 5\% in all cases.  
Finally, those individual IRS modules that were not aligned with each other were stitched 
together using multiplicative correction factors estimated from the overlapping 
spectral regions. 

\section{Analysis} \label{dataanalysis}

\subsection{Stellar properties} \label{stellarpropssect}

The spectral types of the selected stars range between B9 and F7. All of them are 
included in the {\sl Hipparcos} catalogue \citep{vanleeuwen2007}, their 
trigonometric distances lie between 28 and 90\,pc. 
Our sample contains only one known multiple system. HD~16743 and HD~16699AB,  
the latter being a binary itself, form a wide multiple system with a minimum separation 
of $\sim$12700\,au \citep[$\sim$216{\arcsec},][]{moor2011a,tokovinin2012}. 
We searched for common-proper-motion (CPM) companions in the 0.1\,pc vicinity for the other targets. 
Using the astrometric data 
from the UCAC4 catalogue \citep{zacharias2013} and criteria presented by \citet{halbwachs1983}, 
we found no additional CPM candidates.

To characterize the disc excess emission, the contributions of the stellar photosphere 
 and of the circumstellar dust grains to the measured flux must be separated. 
 In order to predict the photospheric fluxes at relevant infrared and submillimetre 
 wavelengths, and to estimate fundamental stellar properties, the  
 photosphere was modelled by fitting an ATLAS9 atmosphere model 
 \citep{castelli2003} to the optical and near-IR observations.
 Photometric data were taken from the {\sl Tycho~2} \citep{hog2000}, 
 {\sl Hipparcos} \citep{perryman1997}, and Two Micron All Sky Survey catalogues 
 \citep[2MASS,][]{cutri2003}. The data were further supplemented by Wide-field Infrared Survey Explorer 
 ({\sl WISE}) $W1$ band (centred {at} 3.4\,{\micron}) 
photometry from the {\sl WISE} All-Sky Database \citep{wright2010}. 
WISE 4.6\,{\micron} measurements were not used because of the well known systematic overestimation 
effect of the fluxes of bright sources in this 
band (Sec. 6.3 of the Explanatory Supplement to the WISE All-Sky Data Release Products) 
In the case of HD~161868 and HD~195627, the 2MASS data were also discarded because of 
saturation.
The metallicity data were collected from the literature. 
In those cases where more than one [Fe/H] estimates were found, we used their average. 
If no metallicity data were available we adopted solar metallicity for the 
given target. 
The surface gravity values were determined via an iterative process. 
Initially we set $\log{g} = 4.5$ for all targets, performed the atmospheric model 
fitting, and estimated the luminosity and
mass of the star as described below. Using the latter parameters and the 
derived effective temperatures, the $\log{g}$ values were re-interpolated in a grid with 
a stepsize of 0.25, and the
fitting of the photometric data was repeated until $\log{g}$  converged at a 
grid point. The resulting values are listed in Table~\ref{stellarprops}.

Our targets are located within 90~pc of the Sun, i.e. inside the relatively dust-free 
Local Bubble suggesting that their interstellar reddening might be negligible \citep[e.g.][]{reis2011,lallement2014}.
Apart from HD~95086 all of our stars have Str\"omgren {\sl uvby} photometry and 
H$_\beta$ indices in the catalogue compiled by \citet{hauck1998}.
In order to further evaluate whether their reddening can be really neglected, 
we derived $E(B-V)$ 
values for the targets by applying the appropriate 
calibration processes \citep{crawford1975,crawford1979,olsen1984}. 
We found that none of our sources has interstellar extinction.
In the case of HD~95086, the good agreement between the photometrically and 
spectroscopically estimated effective temperatures \citep{moor2013a} 
supports the negligible reddening. 
The effective temperature values yielded by the fitting as well as the 
derived stellar luminosities are presented in Table~\ref{stellarprops}.

Among the 11 selected systems three can be assigned to 
young kinematic groups or associations. HD~9672 belongs to the $\sim$40~Myr old 
Argus moving group \citep{zuckerman2012}, and HD\,21997 is part of the $\sim$30~Myr old Columba \citep{moor2006,torres2008} 
moving group. HD~95086 is a member of the Lower Centaurus Crux 
association \citep{dezeeuw1999,moor2013a,meskhat2013}.
In these cases we adopted the age of the corresponding group for the star.
Based only on its astrometric data taken from the {\sl Hipparcos} catalogue, HD~182681
is a probable candidate member of the $\beta$~Pic moving group, however, 
the highly uncertain radial velocity measurement of the star makes this classification 
doubtful (see Appendix). Thus, for HD~182681, as well as two other early type stars, HD~10939 and HD~17848, the 
age information were taken from \citet{nielsen2013}, who performed isochrone fitting for these objects.
In order to indicate the possible younger age of HD~182681 we changed the lower limit of its age confidence 
interval quoted by \citet{nielsen2013} to 19~Myr 
\citep[taking into account the recent age estimate of $\beta$~Pic moving group from][]{mamajek2014}. 
For age-dating HD~161868 and HD~195627, we used metallicity, effective 
temperature, and luminosity data from Table~\ref{stellarprops}. We  
performed a Bayesian approach of age estimates following 
the method outlined by \citet{nielsen2013} and using  
the isochrones compiled by \citet{siess2000}.
For the remaining stars the age estimates were taken from the literature.
The stellar masses were also taken from the literature or were estimated based on 
the appropriate Siess isochrones.

\begin{table*}                                                                                                                                                                                               
\setlength{\tabcolsep}{1.6mm}                                                                                                                                                                                
\begin{center}                                                                                                                                                                                               
\scriptsize                                                                                                                                                                                                  
\caption{{\bf Basic stellar properties.} (1) SIMBAD compatible identifier of the star. (2) Spectral type.                                                                                                    
(3) Distance of the star based on trigonometric parallax taken from the {\sl Hipparcos} catalogue \citep{vanleeuwen2007}.                                                                                    
(4) Derived effective temperature. (5) Surface gravity values fixed in the course of fitting stellar atmospheric models.                                                                                     
(6) Metallicity data from the literature; if more than one observations are available the average of                                                                                                         
the [Fe/H] is quoted. (7) References for metallicity data. 1 -- \citet{casagrande2011}, 2 --                                                                                                                 
\citet{erspamer2003}, 3 -- \citet{gray2003}, 4 -- \citet{gray2006}, 5 -- \citet{montesinos2009},                                                                                                             
6 --  \citet{soubiran2010}, 7 --  \citet{wu2011}. (8) Derived stellar luminosity.                                                                                                                            
(9) Stellar mass. (10) References for stellar mass data. 1 -- \citet{meskhat2013}, 2 --                                                                                                                      
\citet{moor2011b}, 3 -- \citet{roberge2013}, 4 -- this work. (11) Estimated age of the star and its                                                                                                          
confidence interval.                                                                                                                                                                                         
(12) References for age estimates. 1 -- \citet{meskhat2013}, 2 --                                                                                                                                            
\citet{moor2011a}, 3 -- \citet{moor2011b}, 4 -- \citet{nielsen2013}, 5 -- \citet{rhee2007},                                                                                                                  
6 -- this work, 7 -- \citet{zuckerman2012}. (13) Group membership. (14) Measured projected rotational velocity.                                                                                              
(15) References for rotational velocity. 1 -- \citet{abt2002}, 2 --                                                                                                                                          
\citet{diaz2011}, 3 -- \citet{glebocki2005}, 4 -- \citet{moor2011a}, 5 -- \citet{moor2013a},                                                                                                                 
6 -- \citet{nordstrom2004}, 7 -- \citet{royer2007}.  \label{stellarprops}}                                                                                                                                   
\begin{tabular}{lcccccccccccccc}                                                                                                                                                                             
\hline\hline                                                                                                                                                                                                 
Target          &   Sp.T    &    Dist.    &   $T_{eff}$ & $\log{g}$ & [Fe/H] & Ref. &  $L_*$         & $M_*$         &   Ref. &  Age   & Ref.& Membership & $v\sin{i}$    & Ref. \\                          
            &           &   [pc]      &    [K]      &           &        &      &  $[L_{\odot}]$ & $[M_{\odot}]$ &        &  [Myr] &     &            & [km~s$^{-1}$] &  \\                                  
                                                                                                                                                                                                             
\hline                                                                                                                                                                                                       
     HD 9672 &              A1V &    59.4 &  8900 &    4.25 &    0.10 &        5 &    16.4 &    2.00 &        3 &     40 [30,50] &        7 &      Argus &   196 &     7 \\
    HD 10939 &              A1V &    62.0 &  9100 &    4.00 &    0.00 &        - &    30.9 &    2.25 &        4 &  346 [305,379] &        4 &          - &    73 &     3 \\
    HD 16743 &      F0/F2III/IV &    58.9 &  6950 &    4.25 &   -0.06 &        1 &     5.2 &    1.50 &        4 &     30 [10,50] &        2 &          - &   100 &     4 \\
    HD 17848 &              A2V &    50.5 &  8450 &    4.25 &    0.00 &        - &    15.7 &    1.93 &        4 &  372 [269,467] &        4 &          - &   144 &     2 \\
    HD 21997 &           A3IV/V &    71.9 &  8300 &    4.25 &    0.00 &        - &    11.2 &    1.85 &        2 &     30 [20,40] &        3 &    Columba &    70 &     7 \\
    HD 50571 &         F7III-IV &    33.6 &  6550 &    4.25 &    0.00 &      1,4 &     3.2 &    1.30 &        4 &  300 [180,420] &        2 &          - &    60 &     6 \\
    HD 95086 &            A8III &    90.4 &  7550 &    4.25 &    0.00 &        - &     7.0 &    1.70 &        1 &     17 [13,21] &        1 &        LCC &    20 &     5 \\
   HD 161868 &              A0V &    31.5 &  8950 &    4.00 &   -0.54 &    3,6,7 &    26.5 &    2.10 &        4 &  450 [312,532] &        6 &          - &   210 &     7 \\
   HD 170773 &              F5V &    37.0 &  6650 &    4.25 &   -0.05 &      1,4 &     3.5 &    1.30 &        4 &      200 [-,-] &        5 &          - &    50 &     4 \\
   HD 182681 &           B8/B9V &    69.9 &  9650 &    4.25 &    0.00 &        - &    24.7 &    2.18 &        4 &   144 [19,208] &        4 &          - &   275 &     1 \\
   HD 195627 &            F1III &    27.8 &  7300 &    4.00 &   -0.11 &      2,4 &     7.4 &    1.57 &        4 & 805 [462,1100] &        6 &          - &   114 &     3 \\
\hline                                                                          
\end{tabular}                                                                   
\end{center}                                                                    
\end{table*} 

\subsection{Basic image analysis} \label{basicanalysis}
\begin{figure} 
\includegraphics[scale=.23,angle=0]{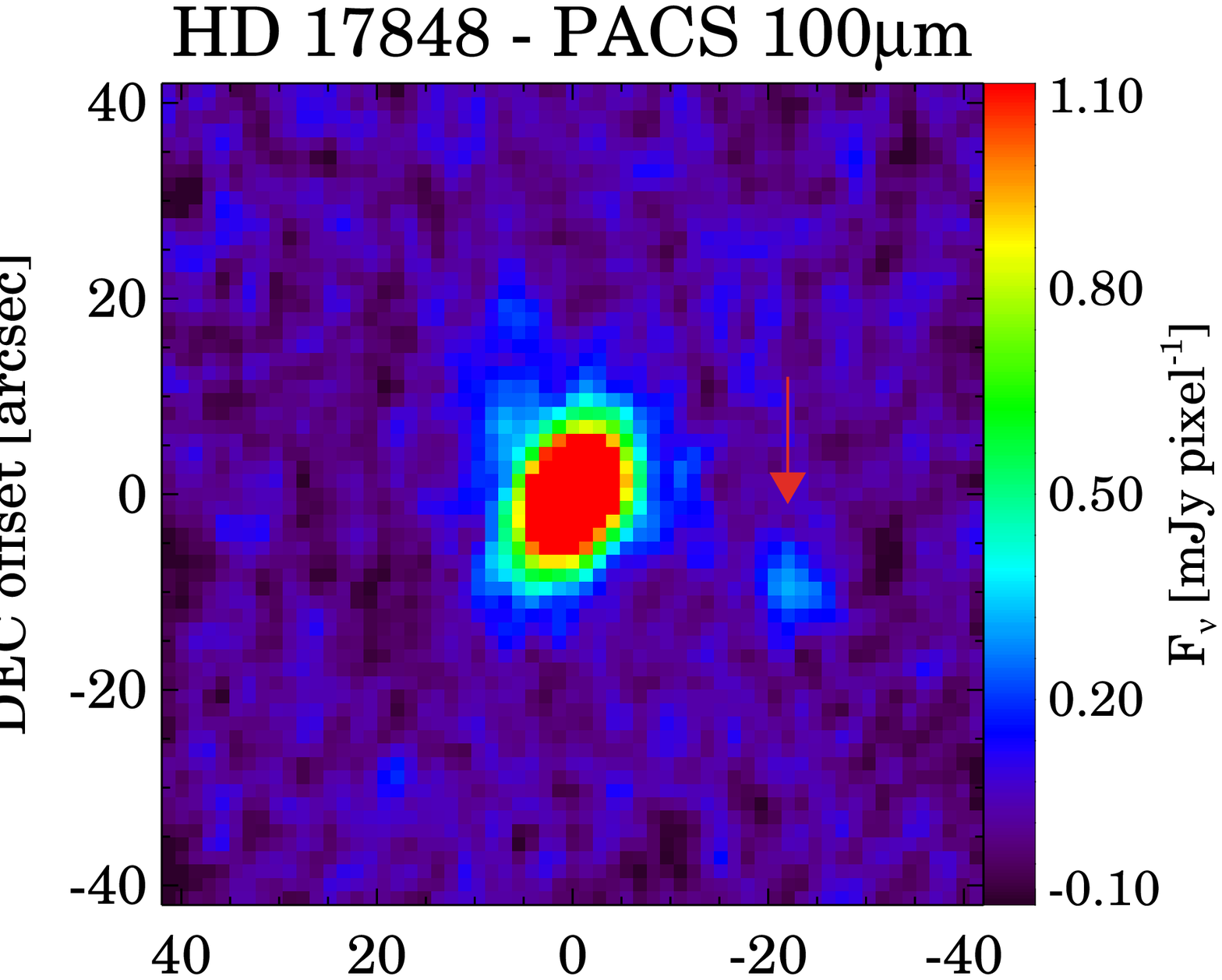}
\includegraphics[scale=.23,angle=0]{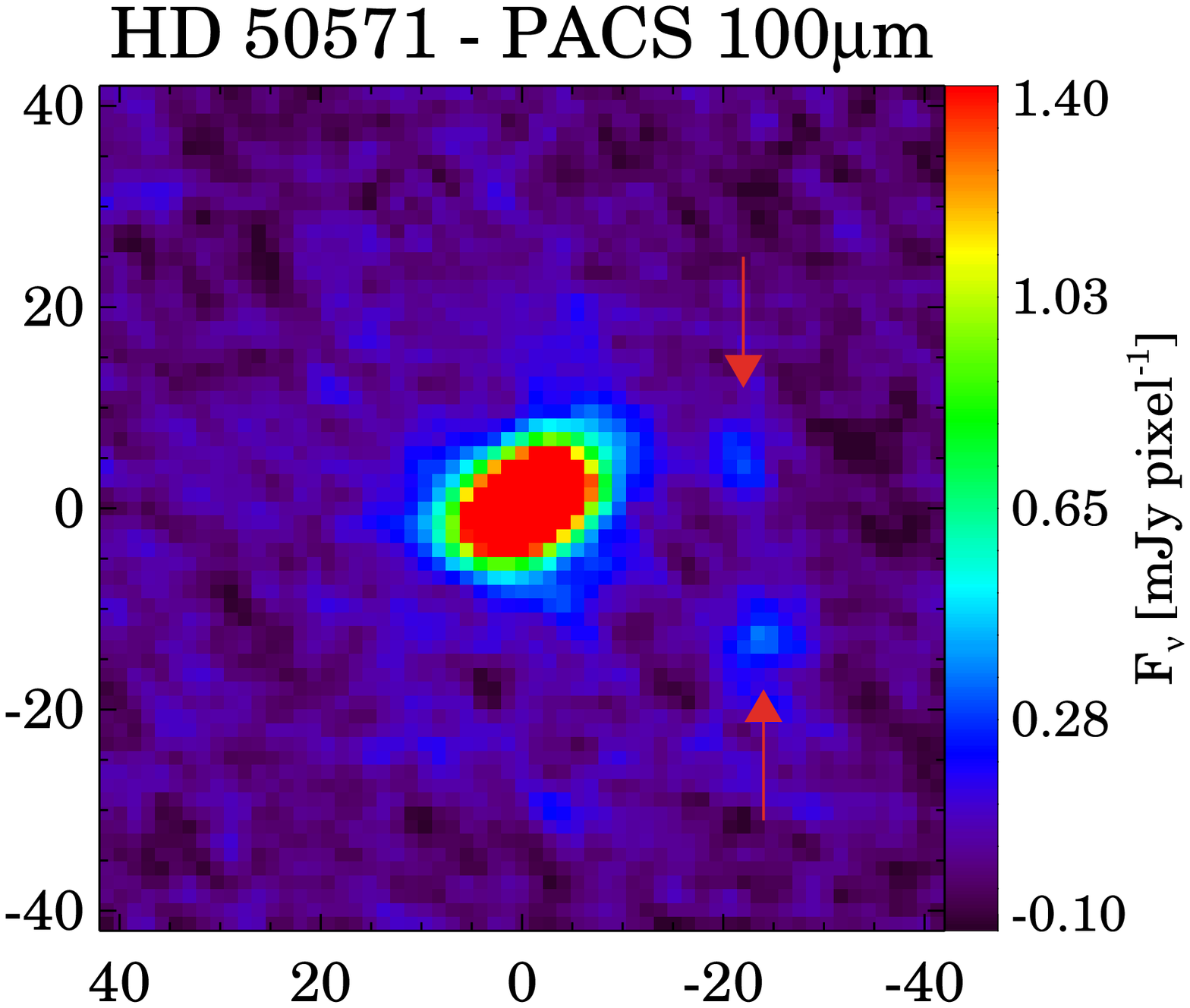} \\
\includegraphics[scale=.23,angle=0]{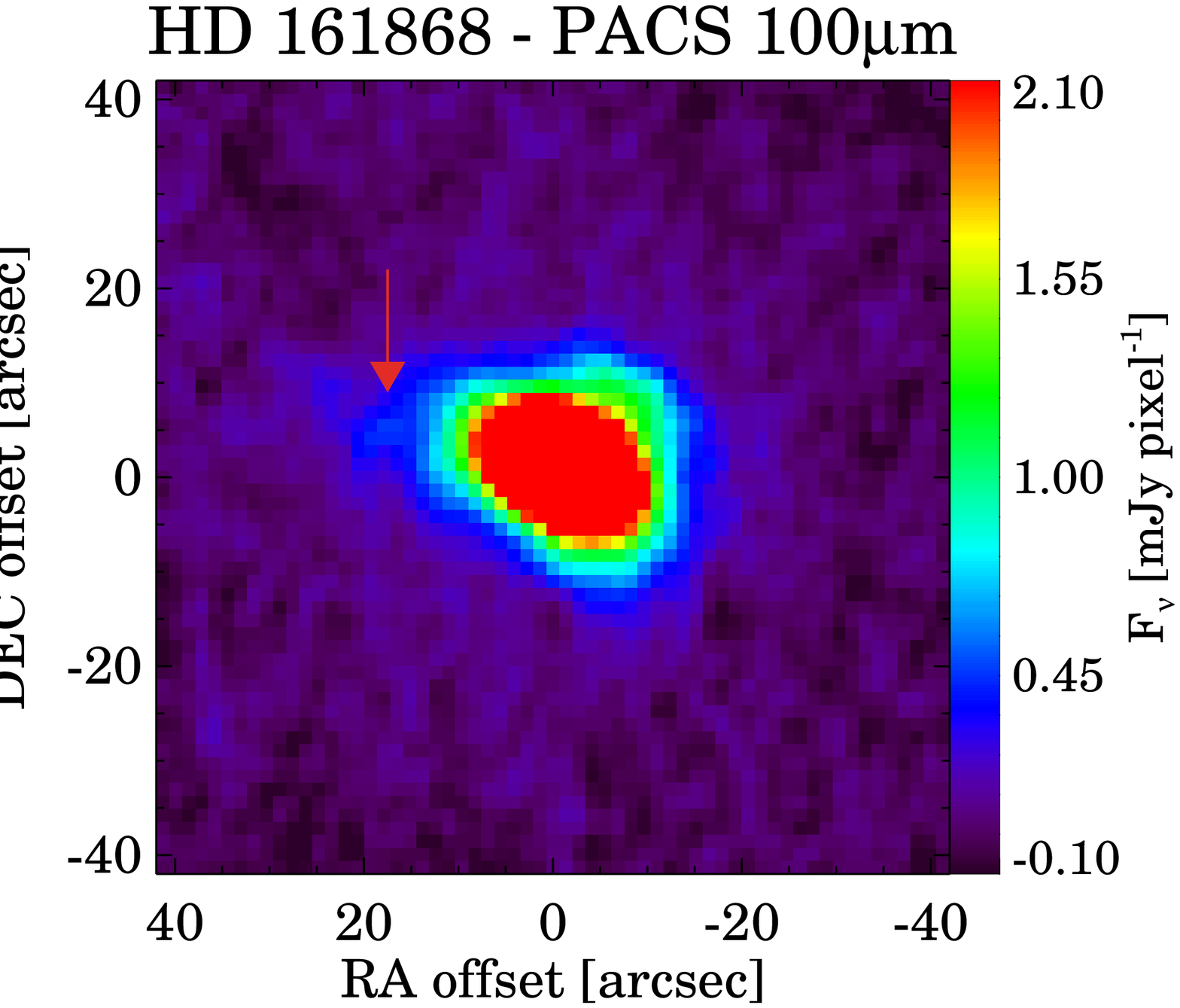} 
\includegraphics[scale=.23,angle=0]{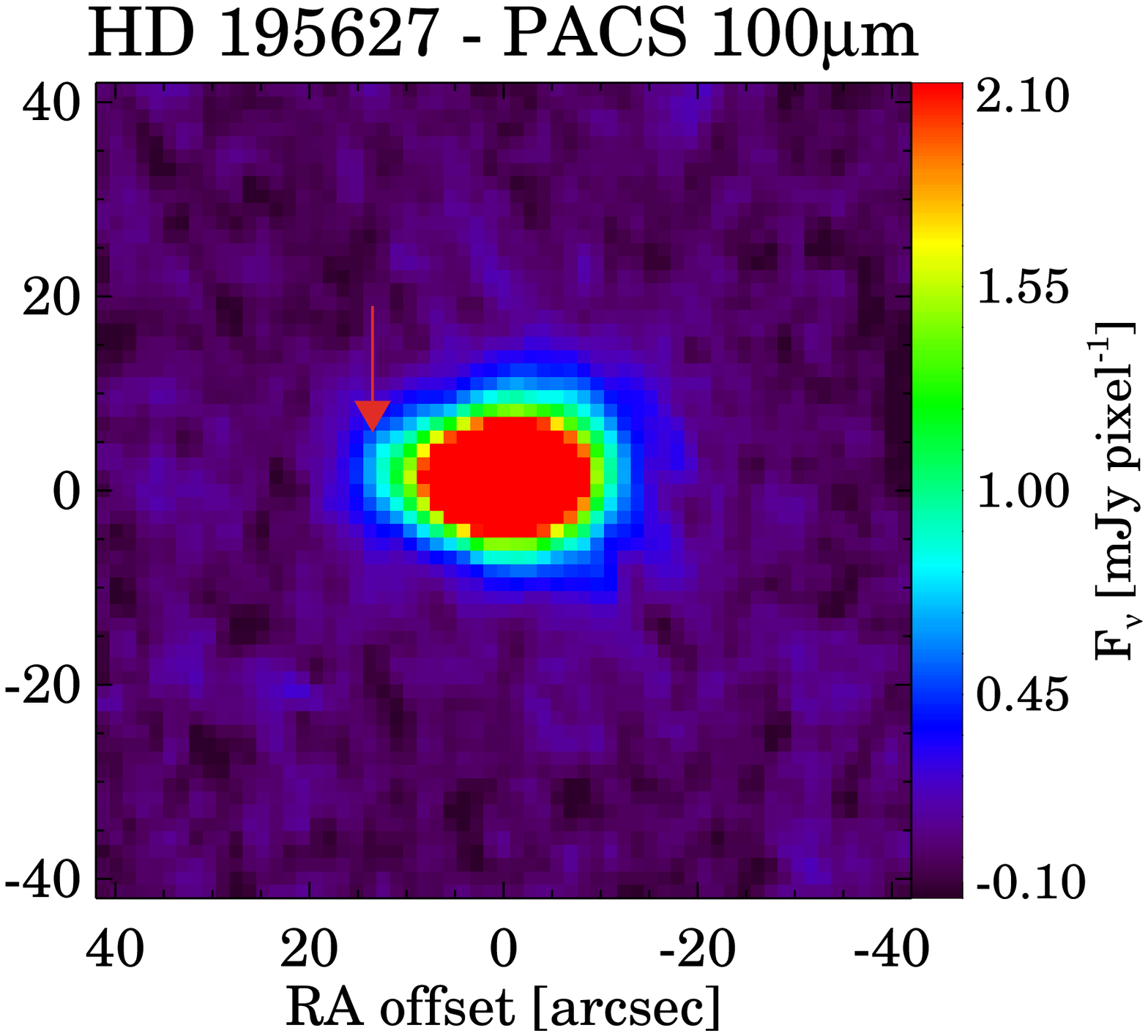} \\
\caption{ {\sl 
PACS 100\,{\micron} images of those objects where nearby 
background sources are present (Sect.~\ref{basicanalysis}). In the case of HD\,195627 the background source cannot be 
separated from the disc by eye, however the distortion of the eastern part of the disc image indicates its presence. 
Red arrows show the positions of the background sources.}
\label{fig1}
}
\end{figure}

All of our targets appeared as very bright sources on PACS images in all bands. 
We searched the PACS maps for nearby ($<$30{\arcsec}) background sources. 
Figure~\ref{fig1} shows the PACS 100\,{\micron} images for those four targets where background sources 
were detected.
At HD~17848 a background source is visible on the 100\,{\micron} and 160\,{\micron} 
images at $\sim$25{\arcsec} 
southwest 
from the target. On the maps of HD~50571 two additional sources can be identified at separations of 
22{\arcsec} and 
28{\arcsec}. In the cases of HD~161868 and HD~195627 there are background objects to east 
with
separations of $\sim$21{\arcsec} and $\sim$15{\arcsec}, respectively. All of these background sources were
found to be point-like, making it possible to subtract them after fitting appropriate point spread 
functions (PSFs) to them. 
We note that there is a bright source 33{\arcsec} away from HD~182681, but due to the large separation, it
does not have a significant effect on our analysis.

\begin{figure*} 
\includegraphics[scale=.24,angle=0]{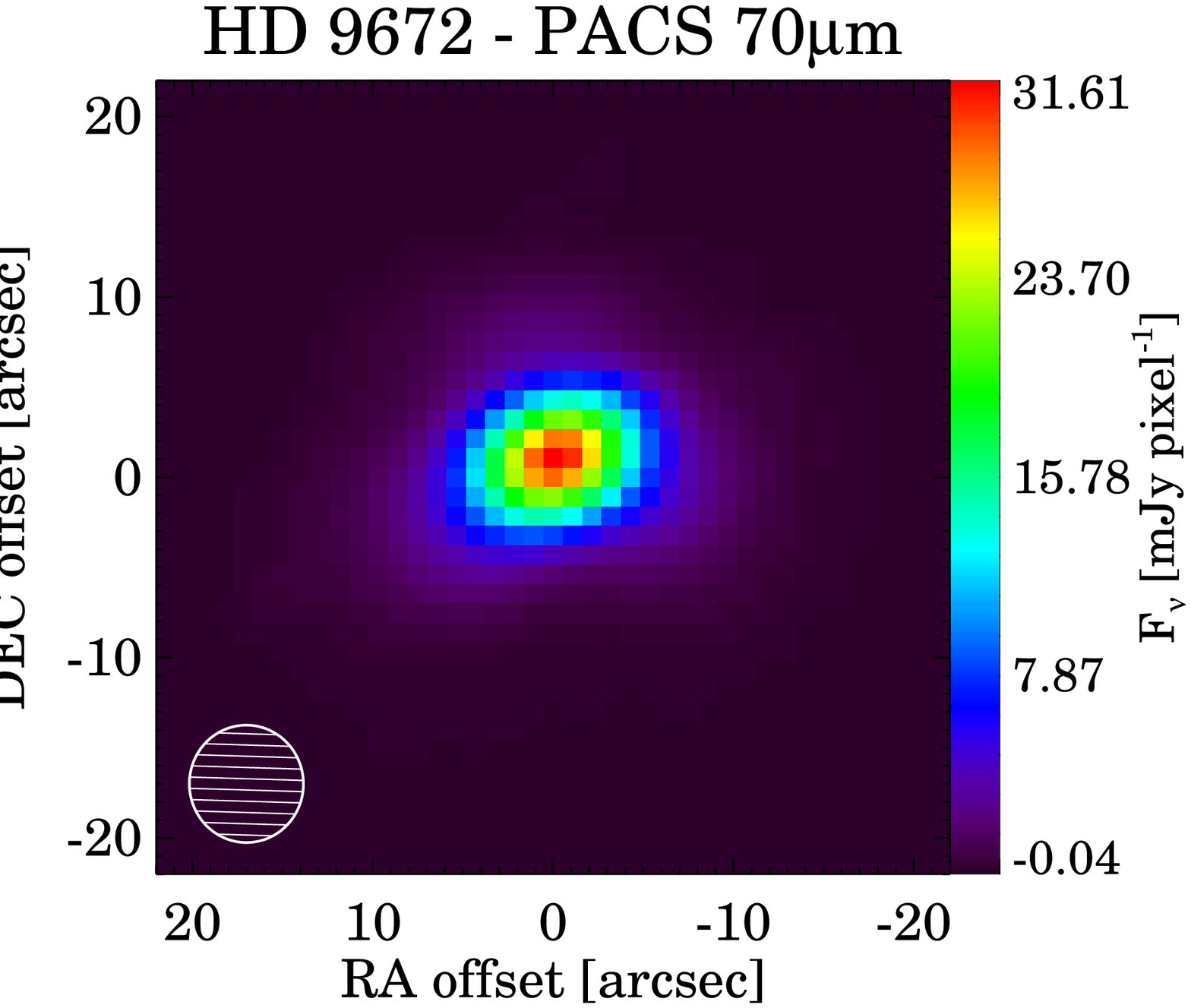}
\includegraphics[scale=.24,angle=0]{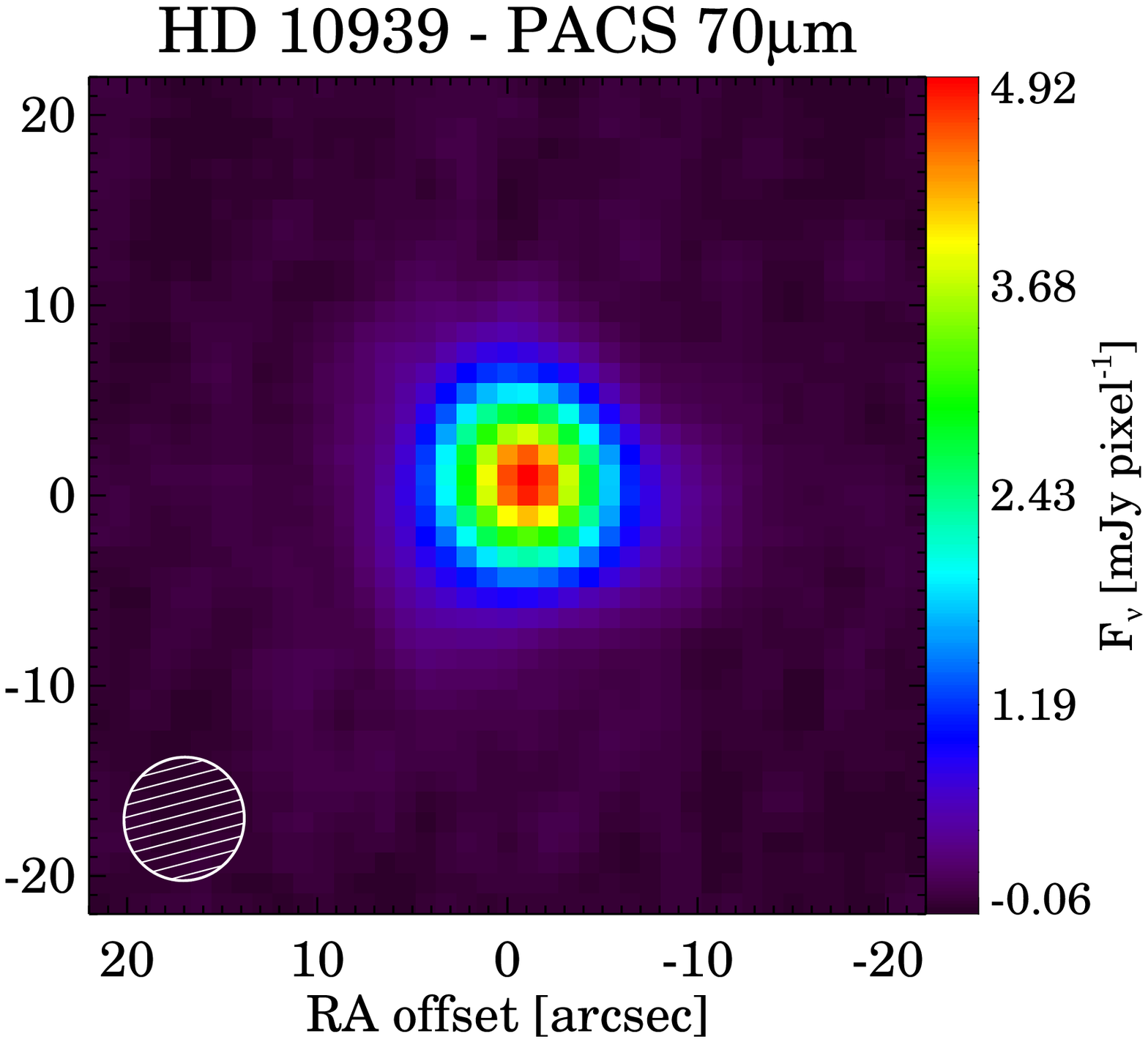}
\includegraphics[scale=.24,angle=0]{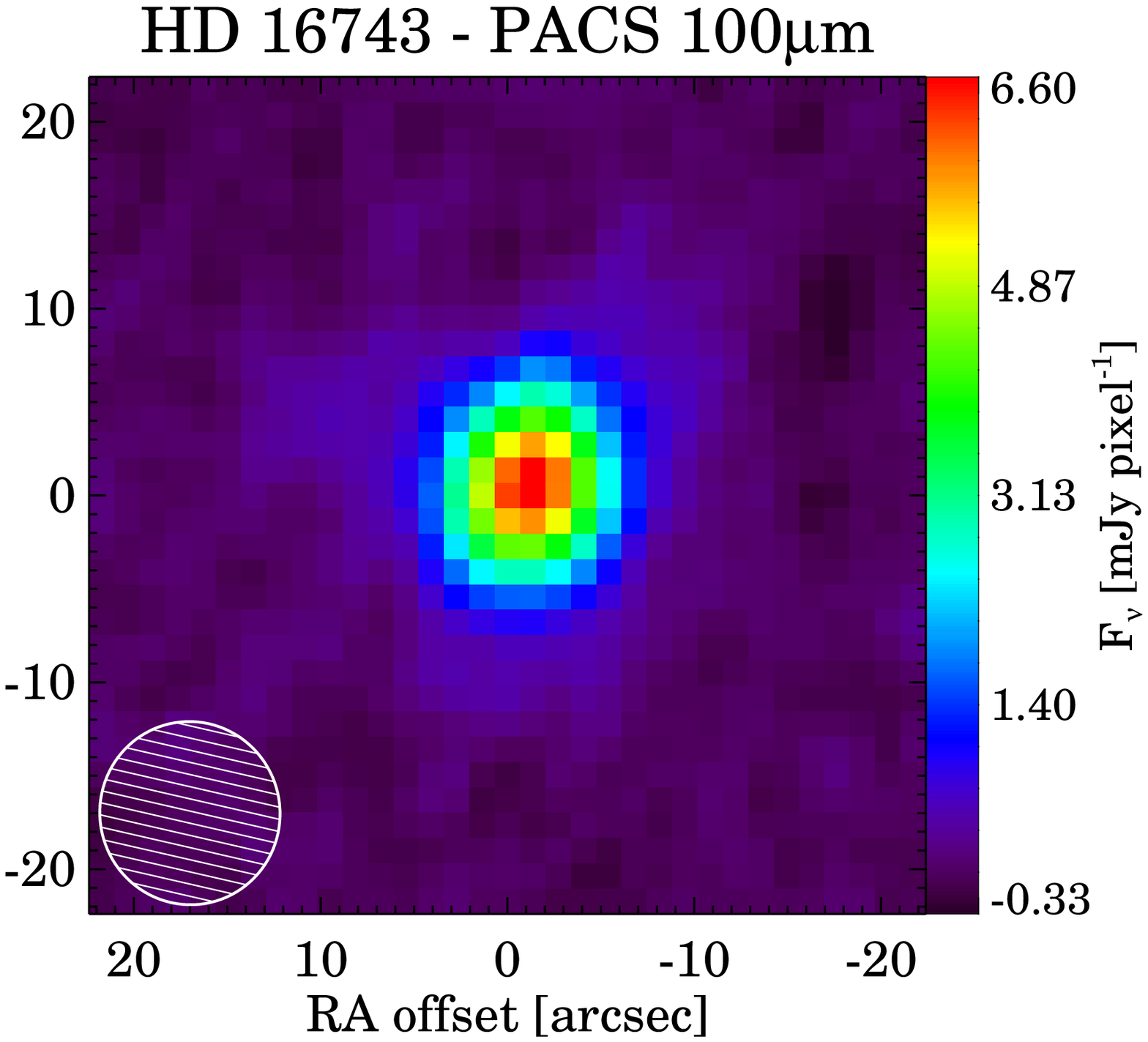} 
\includegraphics[scale=.24,angle=0]{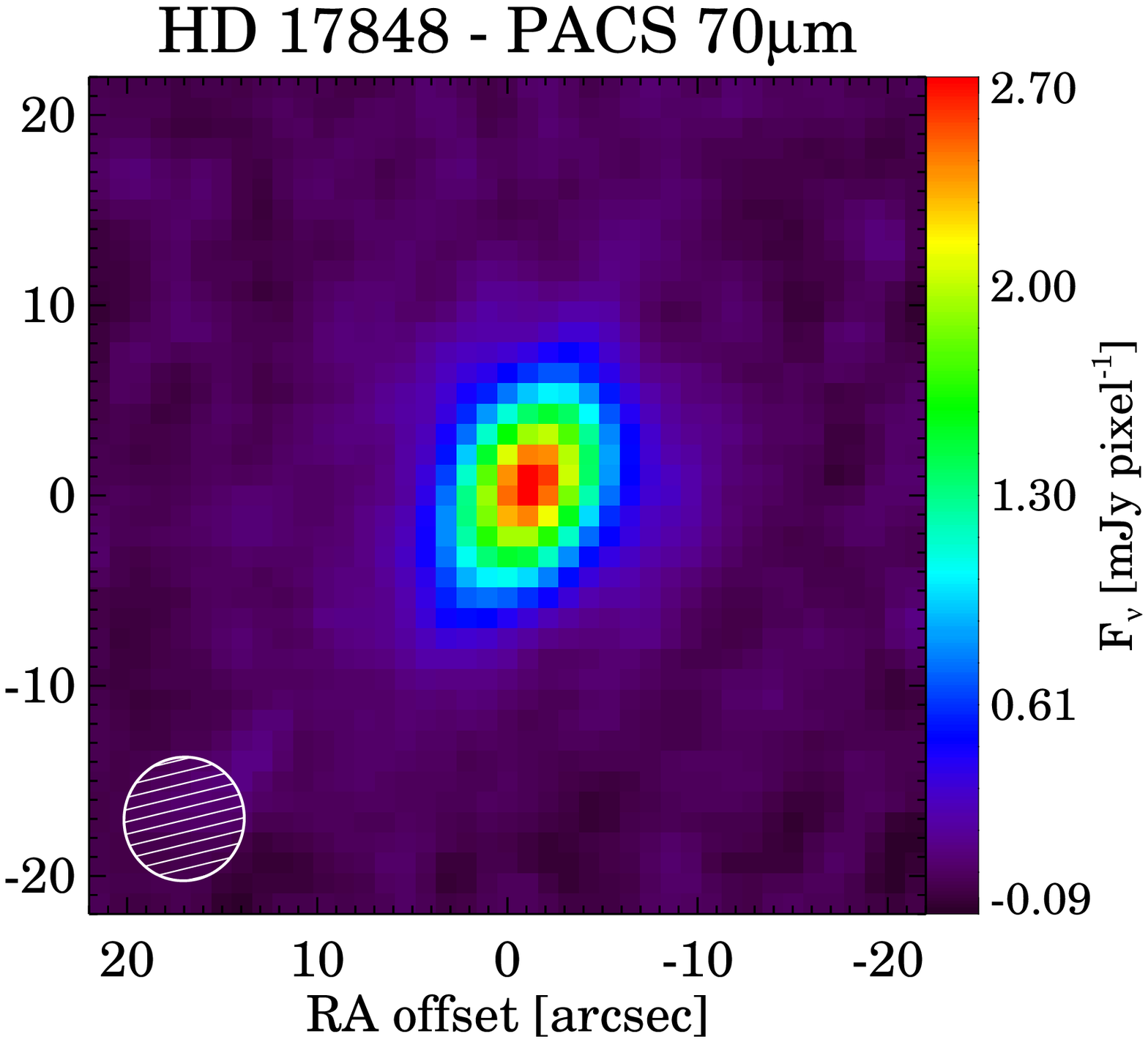} \\
\includegraphics[scale=.24,angle=0]{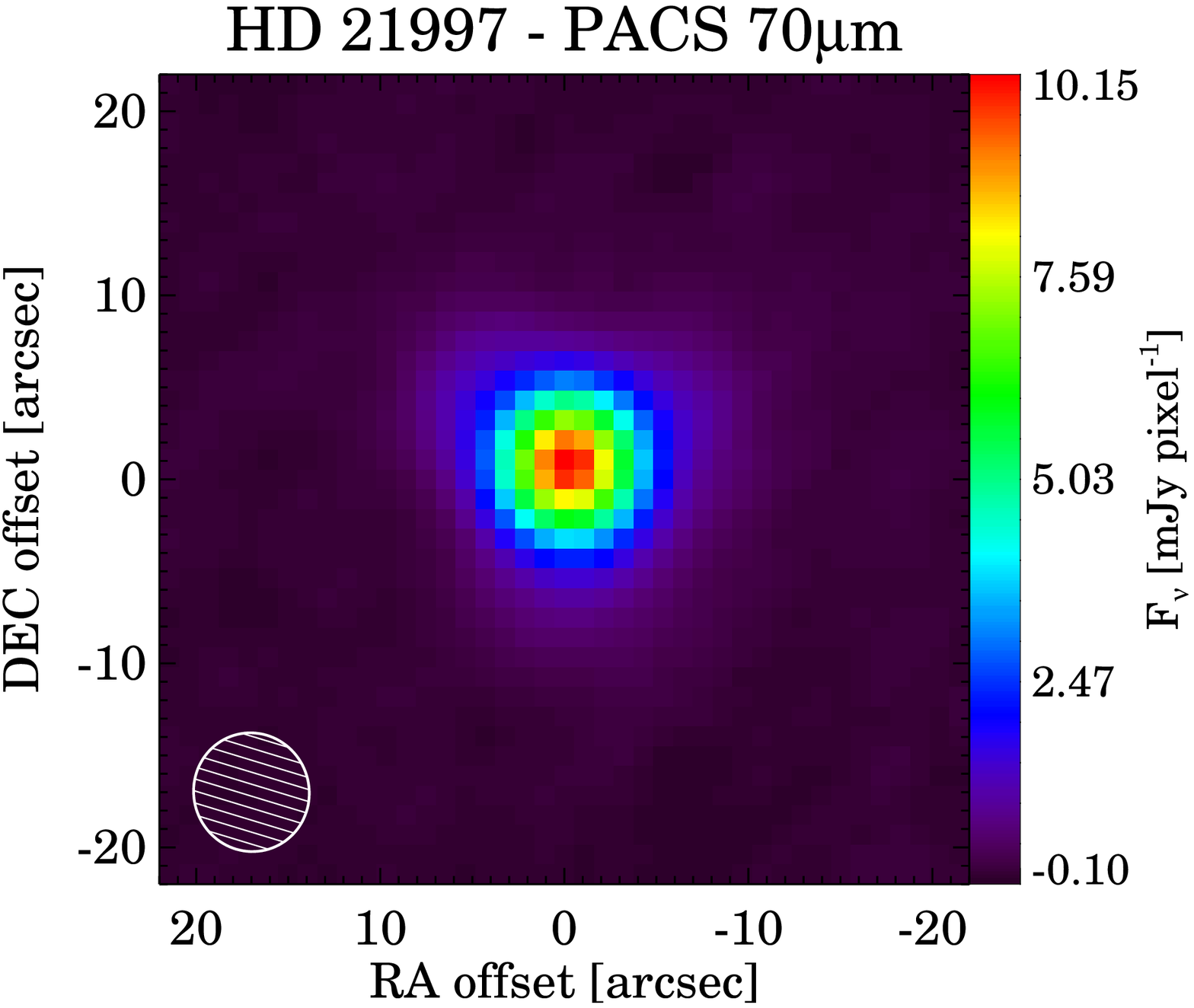}
\includegraphics[scale=.24,angle=0]{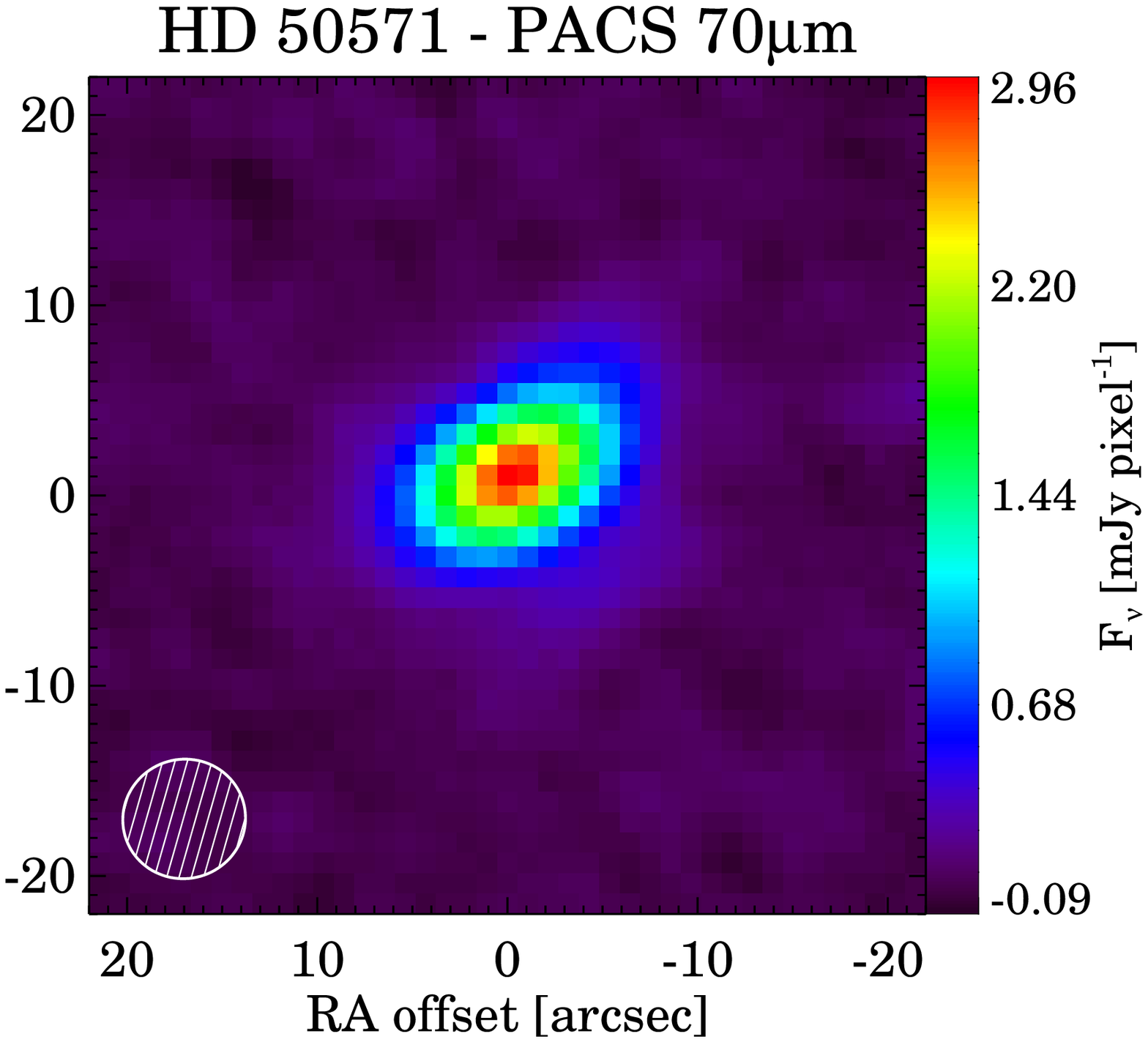}
\includegraphics[scale=.24,angle=0]{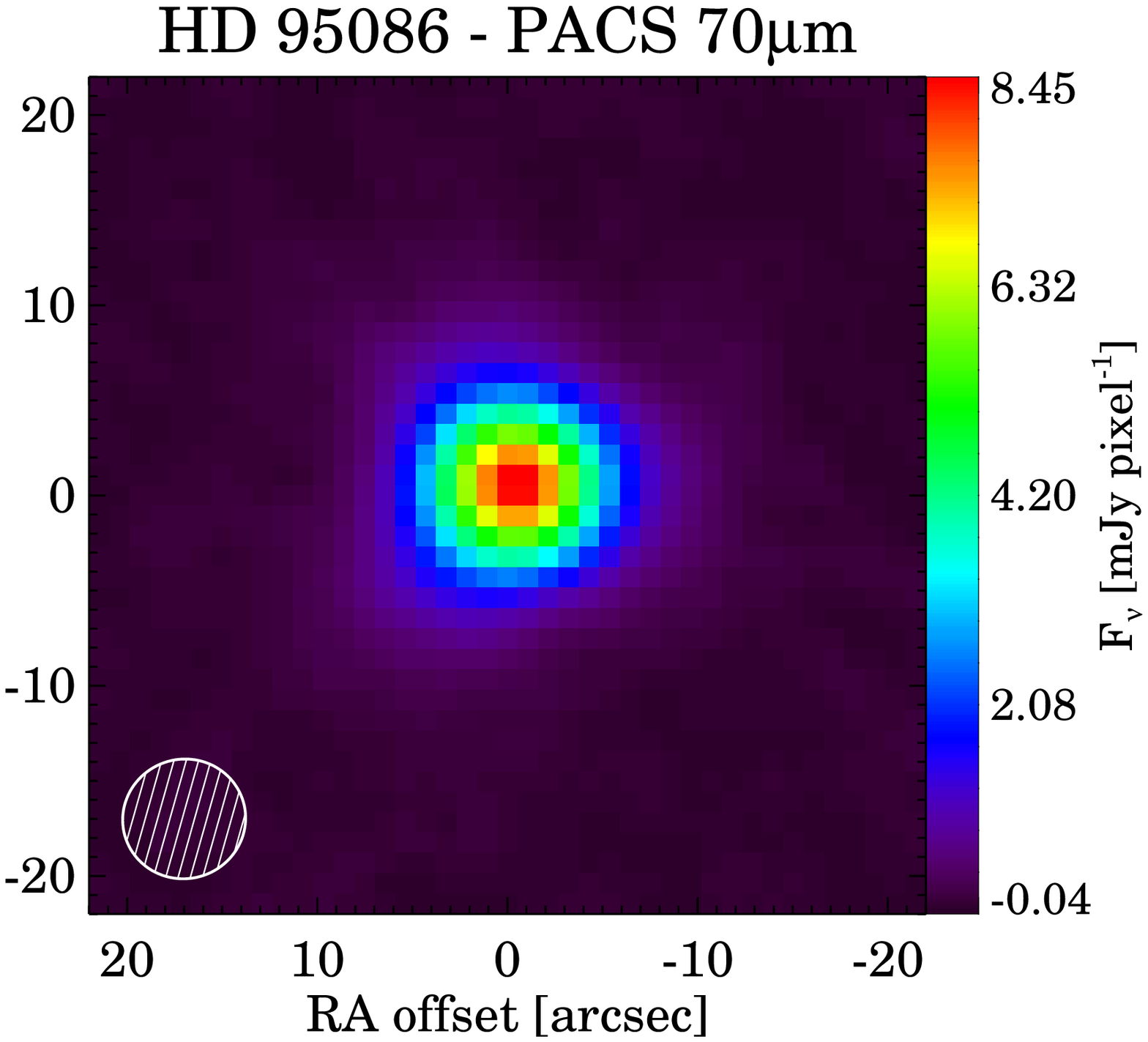} 
\includegraphics[scale=.24,angle=0]{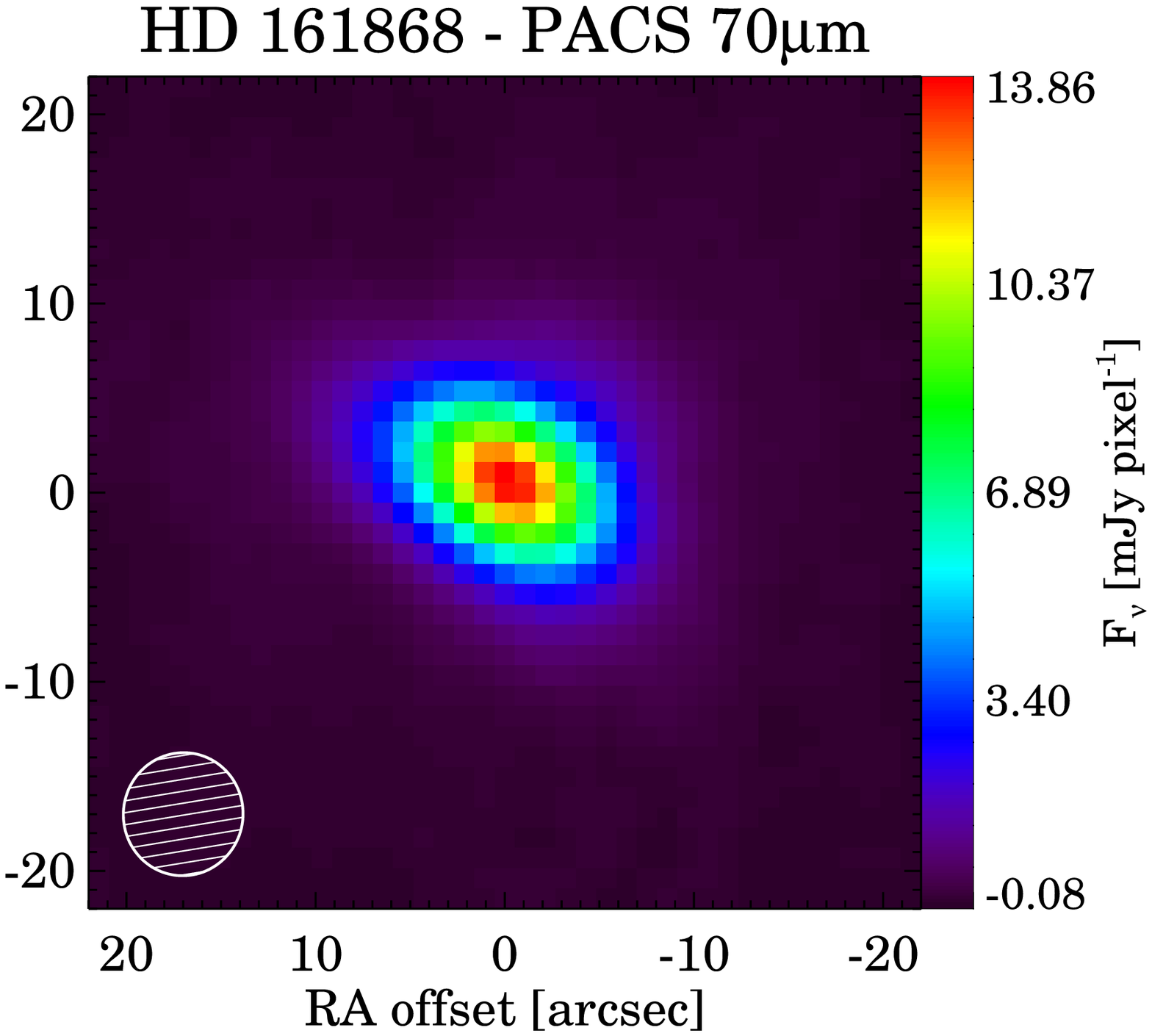} \\
\includegraphics[scale=.24,angle=0]{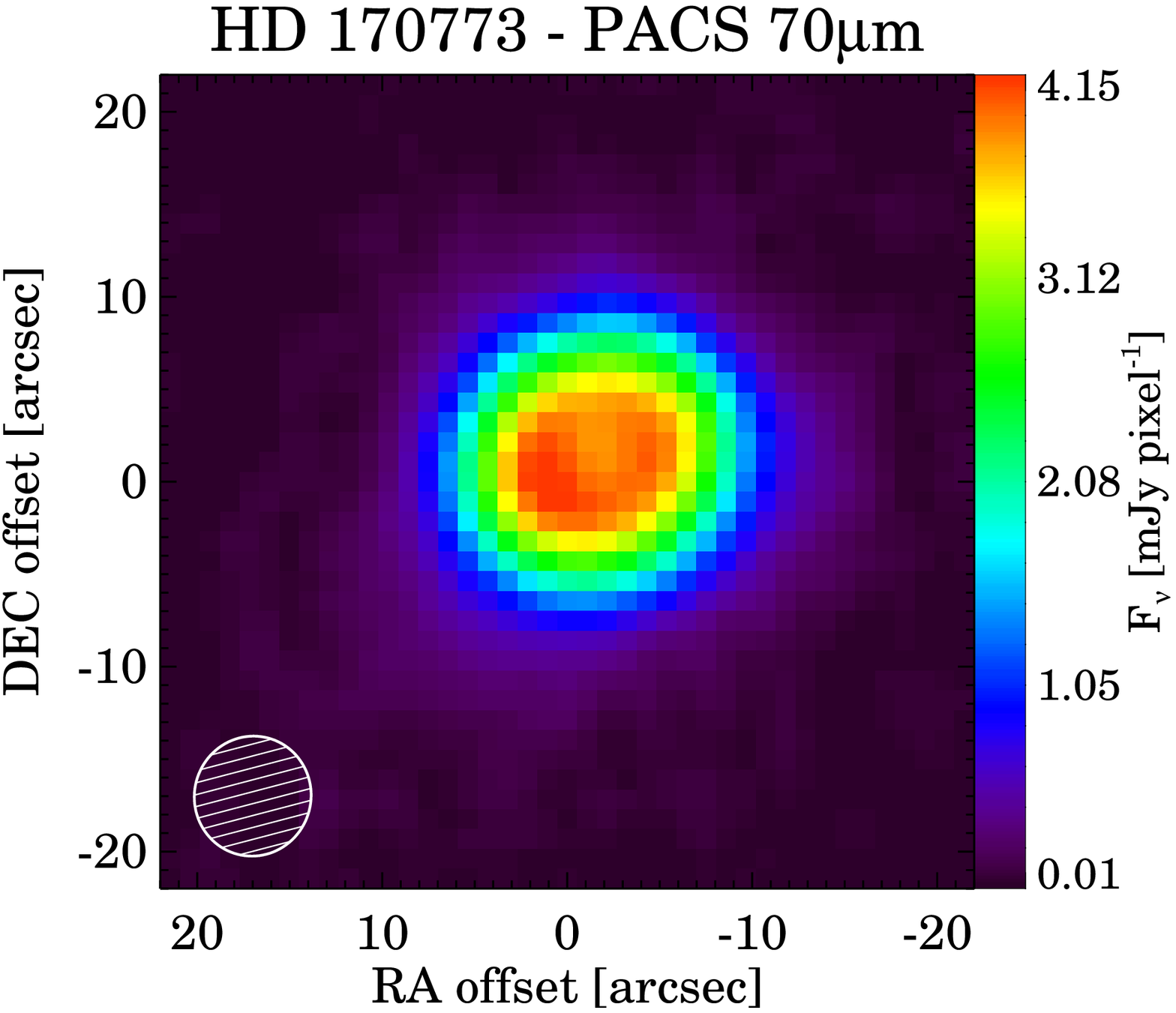}
\includegraphics[scale=.24,angle=0]{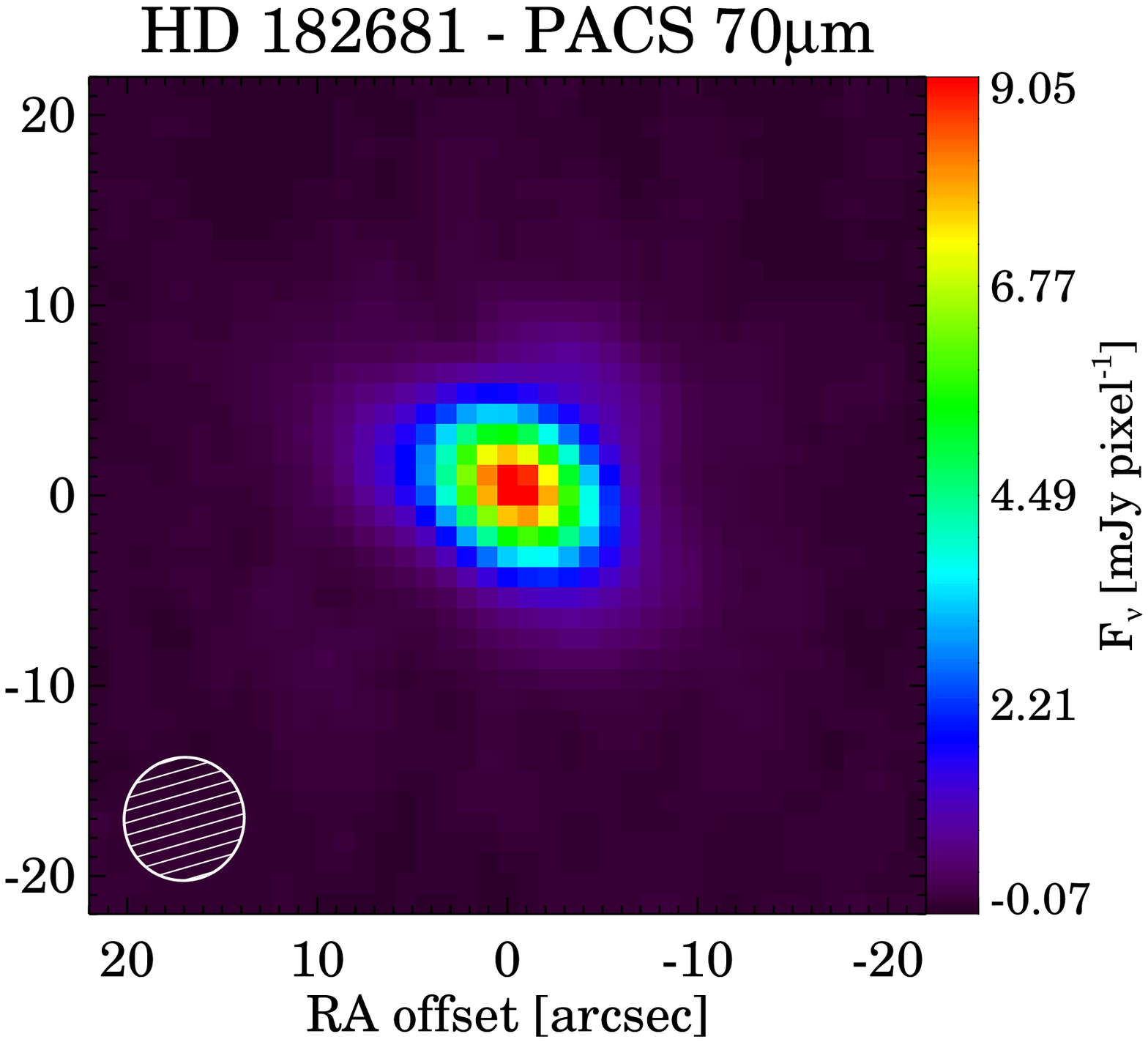}
\includegraphics[scale=.24,angle=0]{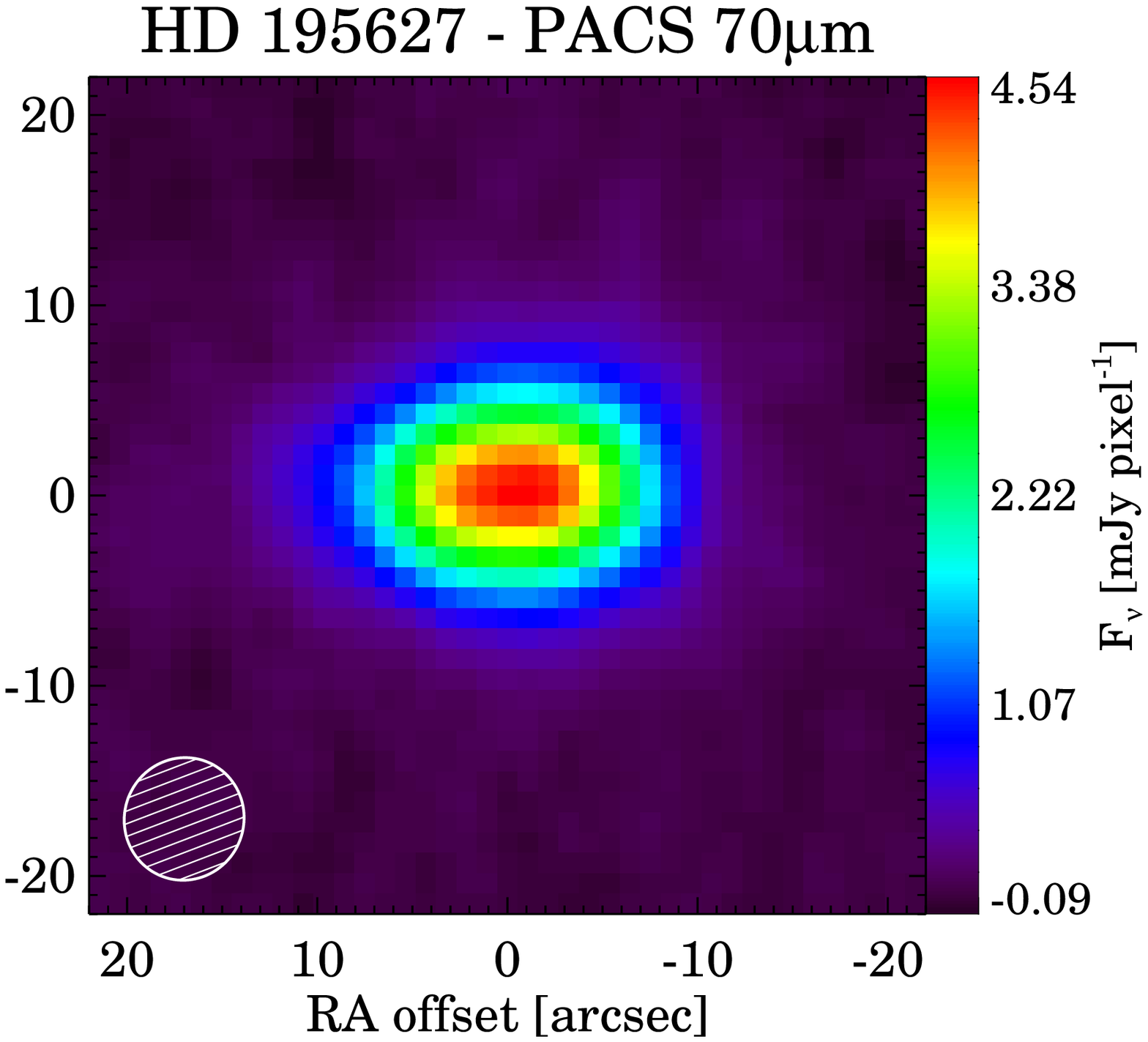} \\
\caption{ {\sl Resolved PACS images of our 11 targets at 70\,{\micron} (100\,{\micron} for HD\,16743 
where no 70\,{\micron}
PACS image is available). The hatched ellipses in the lower left corners show the average FWHMs of the 
rotated PSFs. 
}
\label{fig2}
}
\end{figure*} 
 
As Figure~\ref{fig2} demonstrates, at 70\,{\micron} most of our sources look elongated and spatially extended compared 
with the PACS beam. Actually, in the case of HD\,170773 our PACS 70\,{\micron} (and 100\,{\micron}) 
measurements resolve a broad 
ring of emission, i.e. even the inner edge of the dust ring was outlined.  
In order to evaluate whether the discs are indeed spatially extended 
we produced a first estimate of their position angles (PA, measured north through east), 
sizes, and inclinations ($i$, measured from face-on = 0{\degr}), 
which will be further refined in Sect.~\ref{ringmodel}. 
Following a common procedure in the literature,  
we fitted 2D Gaussians to 
the images and compared the resulting parameters 
to those we derived for PSF observations. This approach is widely used 
in studies of {\sl Herschel} resolved debris discs, therefore our results can be directly compared 
with similar results in the literature \citep[e.g.][]{bonsor2013,marshall2013,matthews2014a}.

PSFs for the different PACS bands were constructed using measurements of 
four photometric calibration stars ($\alpha$~Boo, $\alpha$~Tau, $\alpha$~Cet, $\beta$~And) that 
do not exhibit infrared excesses. We collected all those observations of these targets (14, 15 and 29 
measurements at 70, 100, and 160\,{\micron}, respectively) that were 
obtained in the same observing mode and same period of the mission (between Operational Days 795 and 1035) 
as our targets, and processed them with HIPE using identical data reduction steps and parameters 
as described in Sect.~\ref{pacsdatared}. The PSFs then were rotated to match the roll angle of the telescope at
the time of observing the specific source.
This last step was needed because the shape of the {\sl Herschel}/PACS PSFs 
were affected by the telescope's tripod resulting in an asymmetric beam pattern that rotated 
with the satellite roll angle. 

We found that apart from HD~16743 -- which is spatially extended  
only at 100\,{\micron}, but consistent with a point-source at 160\,{\micron} -- all of the targets 
are at least marginally extended in all bands. To estimate the discs' sizes we made a quadratic deconvolution.
We fitted 2D Gaussians to the appropriately rotated PSFs, then we derived the FWHM of the Gaussian along cuts 
parallel to the discs' major and minor axes. 
By fitting all available PSFs we could compute the average of the  
corresponding FWHMs and estimate the uncertainties related to 
the possible beam variation \citep[see also in][]{kennedy2012a}.
Then the disc angular sizes were derived as 
$\theta_{\rm disc} = \sqrt{\theta_{\rm image}^2 - <\theta_{\rm PSF}>^2 }$, 
where $\theta_{image}$ is the derived FWHM of the image along the major or minor axis, 
$<\theta_{\rm PSF}>$ is the average size of the PSF in the appropriate direction, while
$\theta_{disc}$ is the disc size along the major or minor axis.
Assuming that
the intrinsic structure of the disc is azimuthally symmetric, the inclination of the disc 
was also computed as the arc cosine of the axis ratio. Table~\ref{gausstable} lists the disc parameters based on 
this quadratic deconvolution.  
\begin{table*}                                                                                                                                                                                               
\setlength{\tabcolsep}{2.0mm}                                                                                                                                                                                
\begin{center}                                                                                                                                                                                               
\scriptsize                                                                                                                                                                                                  
\caption{{\bf Results of 2D Gaussian disc fitting}                                                                                                                                                           
 \label{gausstable}}                                                                                                                                                                                         
\begin{tabular}{lccccc}                                                                                                                                                                                      
\hline\hline                                                                                                                                                                                                 
Target          &   PACS band    &   \multicolumn{2}{c}{Disc size ($\theta_{maj} \times \theta_{min}$)} & $i$ & PA \\                                                                                        
            &               &   [\arcsec] & [au] & [$^\circ$] & [$^\circ$] \\                                                                                                                                
\hline                                                                                                                                                                                                       
     HD 9672 &     70{\micron} &      6.5$\pm$0.2$\times$2.5$\pm$0.3 &         383$\pm$9$\times$147$\pm$16 &                        67.4$\pm$2.7 &                       109.0$\pm$3.9 \\
             &    100{\micron} &      6.6$\pm$0.2$\times$2.6$\pm$0.3 &         392$\pm$9$\times$152$\pm$15 &                        67.2$\pm$2.5 &                       109.4$\pm$4.8 \\
             &    160{\micron} &      7.0$\pm$0.9$\times$3.9$\pm$1.5 &        417$\pm$51$\times$229$\pm$90 &                       56.7$\pm$15.5 &                       93.4$\pm$13.9 \\
\hline
    HD 10939 &     70{\micron} &      6.1$\pm$0.3$\times$5.5$\pm$0.3 &        381$\pm$15$\times$340$\pm$16 &                        26.8$\pm$7.2 &                       25.1$\pm$15.8 \\
             &    100{\micron} &      6.7$\pm$0.3$\times$5.9$\pm$0.3 &        417$\pm$17$\times$364$\pm$18 &                        29.3$\pm$6.7 &                       19.9$\pm$16.2 \\
             &    160{\micron} &      7.7$\pm$0.9$\times$7.1$\pm$0.8 &        480$\pm$57$\times$439$\pm$52 &                       23.8$\pm$21.8 &                       70.4$\pm$24.0 \\
\hline
    HD 16743 &    100{\micron} &      6.3$\pm$0.5$\times$3.7$\pm$0.7 &        370$\pm$29$\times$216$\pm$38 &                        54.2$\pm$8.0 &                      165.4$\pm$17.4 \\
    \hline
    HD 17848 &     70{\micron} &      7.7$\pm$0.4$\times$3.0$\pm$0.5 &        386$\pm$20$\times$149$\pm$26 &                        67.3$\pm$4.4 &                       153.8$\pm$7.3 \\
             &    100{\micron} &      9.2$\pm$0.5$\times$3.9$\pm$0.5 &        465$\pm$23$\times$195$\pm$26 &                        65.2$\pm$3.8 &                       154.4$\pm$8.3 \\
             &    160{\micron} &     10.6$\pm$1.5$\times$5.6$\pm$2.6 &       535$\pm$77$\times$284$\pm$131 &                       58.0$\pm$17.3 &                      147.8$\pm$17.6 \\
\hline
    HD 21997 &     70{\micron} &      4.9$\pm$0.3$\times$4.4$\pm$0.3 &        349$\pm$23$\times$313$\pm$23 &                       26.1$\pm$11.9 &                       24.7$\pm$27.5 \\
             &    100{\micron} &      4.9$\pm$0.4$\times$4.3$\pm$0.4 &        350$\pm$26$\times$307$\pm$28 &                       28.7$\pm$12.7 &                       27.0$\pm$29.2 \\
             &    160{\micron} &      6.1$\pm$1.2$\times$5.0$\pm$1.9 &       440$\pm$86$\times$360$\pm$137 &                       35.1$\pm$35.0 &                       56.4$\pm$37.8 \\
\hline
    HD 50571 &     70{\micron} &      7.8$\pm$0.5$\times$2.9$\pm$0.6 &         260$\pm$15$\times$96$\pm$20 &                        68.2$\pm$5.0 &                       121.2$\pm$8.5 \\
             &    100{\micron} &      9.1$\pm$0.3$\times$3.3$\pm$0.7 &         306$\pm$9$\times$109$\pm$24 &                        69.0$\pm$5.0 &                       120.7$\pm$7.3 \\
             &    160{\micron} &     10.6$\pm$1.1$\times$4.8$\pm$2.2 &        357$\pm$36$\times$161$\pm$73 &                       63.1$\pm$13.6 &                      133.3$\pm$13.0 \\
\hline
    HD 95086 &     70{\micron} &      6.0$\pm$0.2$\times$5.5$\pm$0.2 &        543$\pm$15$\times$499$\pm$15 &                        23.3$\pm$5.6 &                      100.6$\pm$15.6 \\
             &    100{\micron} &      6.4$\pm$0.2$\times$5.6$\pm$0.2 &        575$\pm$16$\times$508$\pm$17 &                        27.9$\pm$4.9 &                      114.0$\pm$15.9 \\
             &    160{\micron} &      7.2$\pm$0.8$\times$6.8$\pm$0.8 &        655$\pm$73$\times$614$\pm$75 &                       20.3$\pm$25.9 &                      139.1$\pm$15.5 \\
\hline
   HD 161868 &     70{\micron} &      8.5$\pm$0.1$\times$4.7$\pm$0.1 &          269$\pm$1$\times$148$\pm$1 &                        56.6$\pm$0.6 &                        59.2$\pm$1.0 \\
             &    100{\micron} &      9.7$\pm$0.1$\times$5.0$\pm$0.2 &          304$\pm$2$\times$158$\pm$5 &                        58.7$\pm$1.3 &                        61.1$\pm$2.6 \\
             &    160{\micron} &     10.9$\pm$0.7$\times$5.4$\pm$0.8 &        344$\pm$21$\times$171$\pm$25 &                        60.1$\pm$5.3 &                        66.1$\pm$6.5 \\
\hline
   HD 170773 &     70{\micron} &    11.8$\pm$1.1$\times$10.1$\pm$0.8 &        435$\pm$39$\times$372$\pm$29 &                       31.2$\pm$11.4 &                      115.3$\pm$24.0 \\
             &    100{\micron} &    11.8$\pm$0.4$\times$10.1$\pm$0.4 &        435$\pm$14$\times$373$\pm$13 &                        30.9$\pm$4.6 &                      116.9$\pm$18.7 \\
             &    160{\micron} &     11.4$\pm$1.2$\times$9.8$\pm$1.1 &        422$\pm$46$\times$362$\pm$41 &                       31.0$\pm$15.2 &                       99.5$\pm$21.7 \\
\hline
   HD 182681 &     70{\micron} &      5.8$\pm$0.1$\times$2.2$\pm$0.3 &         403$\pm$8$\times$156$\pm$20 &                        67.2$\pm$3.2 &                        56.0$\pm$3.9 \\
             &    100{\micron} &      5.7$\pm$0.2$\times$2.1$\pm$0.4 &        401$\pm$14$\times$144$\pm$25 &                        68.9$\pm$4.0 &                        55.7$\pm$7.3 \\
             &    160{\micron} &      6.6$\pm$1.0$\times$2.6$\pm$2.3 &       459$\pm$69$\times$184$\pm$161 &                       66.3$\pm$22.3 &                       63.5$\pm$19.3 \\
\hline
   HD 195627 &     70{\micron} &     11.6$\pm$0.2$\times$6.9$\pm$0.3 &          323$\pm$5$\times$190$\pm$8 &                        53.9$\pm$1.9 &                        92.4$\pm$3.7 \\
             &    100{\micron} &     12.9$\pm$0.3$\times$7.3$\pm$0.2 &          358$\pm$7$\times$202$\pm$6 &                        55.7$\pm$1.5 &                        91.0$\pm$4.3 \\
             &    160{\micron} &     15.6$\pm$0.5$\times$7.7$\pm$1.0 &        432$\pm$15$\times$212$\pm$28 &                        60.5$\pm$4.4 &                        89.0$\pm$5.2 \\
\hline
\hline                                                                          
\end{tabular}                                                                   
\end{center}                                                                    
\end{table*}                                                                    


All of the targets are detected in the 250 and 350\,{\micron} SPIRE maps. 
HD~9672, HD~161868, HD~170773 and HD 195627 are visible at 500\,{\micron} 
as well. Four of our targets -- HD~10939, HD~16743, HD~17848, and HD~50571 -- are detected 
at submillimetre wavelengths for the first time. 
HD~170773 and HD 195627 are found to be marginally extended 
at 250\,{\micron} and 350\,{\micron} when compared to the corresponding SPIRE beam 
(taken from the SPIRE calibration context). 
In the latter case 
the emission is extended with the same position angle of $\sim$90$\degr$ as in the PACS images.
We note, however, that the very nearby background source is also located along 
 the same PA.
Considering the low inclination of HD\,170773, the object is not particularly elongated in the 
SPIRE images, and the PA can only be determined with a large uncertainty. 
HD~161868 lies on the top of background emission originated from an 
extended ridge that is especially bright at 500{\micron} thus, we do not attempt to 
derive geometrical parameters from the SPIRE images for this source.

\subsection{PACS and SPIRE photometry}

The source flux determination was performed on the final {\sl Herschel} PACS mosaics 
after the nearest background sources were removed from the affected images (Sect.~\ref{basicanalysis}).
We used aperture photometry by placing the apertures at the centroid position.
We found that the offsets between the sources' optical positions (corrected for the proper motion using the 
epochs of PACS observations) and the derived centroids in the PACS images are $\leq$2\farcs4 for most of our targets, 
i.e. within the 1$\sigma$ pointing accuracy of {\sl Herschel}.
Even the largest offset of 3\farcs2, measured for HD~10939 
is well within the 2$\sigma$ pointing uncertainty of the telescope.
The aperture radius was chosen to cover the resolved disc emission: 24{\arcsec} for HD\,170773 
and HD\,195627, and 20{\arcsec} for the other sources.
We used the same aperture size at all wavelengths.
The background was always computed in a sky annulus between 50{\arcsec} and 60{\arcsec}.  
To remove the possible contamination of any remaining background objects,
we used an iterative sigma-clipping method with a 3$\sigma$ clipping threshold in the sky annulus.
In order to estimate the sky noise in each PACS band, we distributed
sixteen apertures, with the same size as the source aperture, randomly along
the background annulus. We performed aperture photometry without background subtraction 
in each aperture and computed the sky noise as the standard deviation 
of derived background flux values. 
Finally we applied aperture correction to account for the flux outside the aperture using correction 
factors taken from the appropriate calibration file.
According to our tests, in the case of the selected aperture configurations 
the nominal correction factors for point sources can be  
safely applied. 
The total uncertainty of the photometry was derived as the quadratic sum of the 
measurement errors and the absolute calibration uncertainty of
 7\% \citep{balog2013}.
We note that in most cases, especially at 70 and 100\,{\micron}, the uncertainty of the absolute calibration 
dominates the error budget. 

Background sources discovered in PACS images close to some targets might also be present 
in SPIRE maps. However, because of the coarser spatial resolution, in most cases they cannot be  
clearly deblended from our main targets. In order to minimize the possible contamination of these 
sources (and possible additional ones) and the extended background emission in the case of 
HD~161868, 
we used PSF fitting in the flux determination. 
Apart from HD~170773 and HD~195627, which were found to be marginally extended at 250~{\micron} and 350~{\micron},
 we used the SPIRE beam profiles taken from the calibration context in the course of 
 flux extraction. 
For HD~170773 and HD~195627 at 250~{\micron} and 350~{\micron} the original SPIRE beam was convolved with a 2D 
Gaussian whose parameters (FWHM of the major and minor axes, position angle) were 
taken from the Gaussian fitting performed at 100~{\micron} (Sect.~\ref{basicanalysis}, Table~\ref{gausstable}).
The offsets between the stellar positions corrected for the proper motion and 
the sources' centroids are generally smaller than 3\farcs1, corresponding to
1.3$\sigma$ pointing uncertainty of {\sl Herschel}.  The only exception is HD 195627, where the 
flux peak is shifted by 5--6\arcsec east of the nominal stellar position raising a complicated issue 
since the nearby source is also located to the east, and the disc is elongated in the east-west direction too. 
In this case we placed the centre of the aperture at the nominal stellar position, but 
some contamination from the background source cannot be excluded.
The final uncertainties
were derived as the quadratic sum of the measurement errors and the
overall calibration uncertainty of 5.5\% for the SPIRE
photometer \citep{bendo2013}.

The derived PACS and SPIRE flux densities of the targets
are listed in Table~\ref{phottable}. 
We note 
that PACS and SPIRE photometry of HD\,21997 and HD\,95086 
were taken from \citet{moor2013b} and 
\citet{moor2013a}, respectively.

\subsection{MIPS photometry}
When performing photometry in MIPS 70~{\micron} maps we utilized the spatial information 
 from the higher resolution PACS 70~{\micron} images. 
As a first step, a PSF profile valid for sources with blackbody temperatures of 60~K was constructed using 
the method described by \citet{gordon2007}. The compiled PSF was convolved with a two-dimensional 
elliptical Gaussian constructed
for each source by using the 
parameters derived from the Gaussian fitting in the PACS 70~{\micron} image. 
These spatially extended profiles were used to determine the 
flux density of the targets via PSF fitting. 
The final photometric uncertainty was computed by quadratically adding 
the calibration uncertainty of 7\% \citep{gordon2007} and the measurement error. 
The MIPS 70~{\micron} band photometry (with an effective 
filter wavelength of 71.42~{\micron}) shows a good agreement with the 
PACS 70~{\micron} photometry. If we take into account the typical color correction factors 
of 0.9 and 0.98 for MIPS and PACS photometry, respectively, the flux densities  
obtained with the two instruments at different epochs are consistent with each other 
 within their 1~$\sigma$ uncertainty. 

From {\sl Herschel} we know that the discs are spatially extended at 160~{\micron} as well. 
However, 
the MIPS 160~{\micron} beam is significantly 
larger than the size of the discs. Therefore the sources can be considered as point-like, and
the 160~{\micron} photometry was simply taken from \citet{moor2011a} when available.
Photometry in MIPS 24~{\micron} filter were also taken from the literature.
MIPS photometric data for our sources are also listed in Table~\ref{phottable}.

\begin{table*}                                                                                                                                                                                               
\setlength{\tabcolsep}{1.2mm}                                                                                                                                                                                
\begin{center}                                                                                                                                                                                               
\scriptsize                                                                                                                                                                                                  
\caption{{\bf Photometric data used in SED compilation.} (1) SIMBAD compatible identifier of the star. (2-10)                                                                                                
Measured flux densities in PACS, SPIRE, and MIPS bands. The quoted fluxes are in mJy and are not colour-corrected.                                                                                           
For HD\,21997 and HD\,95086 these photometric data were taken from \citet{moor2013b} and                                                                                                                     
\citet{moor2013a}, respectively.                                                                                                                                                                             
  (11) References for MIPS photometry: 1 -- \citet{ballering2013}, 2 --                                                                                                                                      
\citet{moor2011b}, 3 -- \citet{roberge2013}, 4 -- this work. (12) References for additional infrared and                                                                                                     
ground-based submillimetre photometric data:  1 -- IRAS PSC, 2 --                                                                                                                                            
IRAS FSC, 3 -- \citet[AKARI IRC][]{ishihara2010}, 4 -- \citet{moor2011b}, 5 -- \citet{nilsson2010},                                                                                                          
6 -- \citet{panic2013}, 7 -- \citet{su2008}, 8 -- \citet{williams2006}, 9 -- \citet[WISE][]{wright2010}, 10 -- \citet[AKARI FIS][]{yamamura2010}                                                             
\label{phottable}}.                                                                                                                                                                                          
\begin{tabular}{lccccccccccc}                                                                                                                                                                                
\hline\hline                                                                                                                                                                                                 
Target          &  \multicolumn{3}{c}{PACS} & \multicolumn{3}{c}{SPIRE} & \multicolumn{3}{c}{MIPS} & Refs. & Additional photometry \\                                                                        
            &  70{\micron} & 100{\micron} & 160{\micron} & 250{\micron} & 350{\micron} & 500{\micron} & 23.675{\micron} & 71.42{\micron} & 155.9{\micron} &   &                                              
	      \\                                                                                                                                                                                                    
\hline                                                                                                                                                                                                       
     HD 9672 &         2163$\pm$151 &         1919$\pm$134 &          1066$\pm$75 &           363$\pm$20 &           166$\pm$11 &             76$\pm$8 &           259$\pm$10 &         1749$\pm$123 &                    - &        3,4 &       2,3,9,10 \\
    HD 10939 &           396$\pm$28 &           403$\pm$28 &           277$\pm$20 &             94$\pm$7 &             43$\pm$6 &              3$\pm$6 &            108$\pm$4 &           379$\pm$27 &                    - &        1,4 &          2,3,9 \\
    HD 16743 &                    - &           369$\pm$27 &           226$\pm$32 &             82$\pm$6 &             38$\pm$6 &              8$\pm$8 &             50$\pm$2 &           388$\pm$26 &           174$\pm$24 &        2,4 &          2,3,9 \\
    HD 17848 &           213$\pm$17 &           210$\pm$18 &           138$\pm$11 &             50$\pm$5 &             28$\pm$6 &             6$\pm$10 &             86$\pm$3 &           204$\pm$17 &                    - &        1,4 &          2,3,9 \\
    HD 21997 &           697$\pm$49 &           665$\pm$47 &           410$\pm$30 &           151$\pm$11 &             66$\pm$9 &             33$\pm$9 &             55$\pm$2 &           663$\pm$46 &                    - &        1,4 &        2,3,8,9 \\
    HD 50571 &           223$\pm$17 &           262$\pm$19 &           188$\pm$16 &             71$\pm$7 &             46$\pm$6 &             13$\pm$7 &             70$\pm$2 &           235$\pm$17 &           214$\pm$36 &        2,4 &          2,3,9 \\
    HD 95086 &           690$\pm$48 &           675$\pm$47 &           462$\pm$32 &           213$\pm$12 &            120$\pm$8 &            63$\pm$10 &             45$\pm$2 &           655$\pm$44 &                    - &        1,4 &          1,3,9 \\
   HD 161868 &          1222$\pm$85 &          1051$\pm$73 &           587$\pm$44 &           177$\pm$12 &            98$\pm$10 &            56$\pm$11 &           434$\pm$18 &          1118$\pm$78 &                    - &        1,4 &      1,3,6,7,9 \\
   HD 170773 &           806$\pm$56 &          1109$\pm$78 &           875$\pm$61 &           379$\pm$21 &           167$\pm$11 &             73$\pm$7 &             65$\pm$2 &           785$\pm$55 &           692$\pm$83 &        2,4 &      1,3,4,5,9 \\
   HD 182681 &           607$\pm$42 &           463$\pm$33 &           243$\pm$18 &             84$\pm$7 &             30$\pm$5 &              8$\pm$7 &                    - &                    - &                    - &        1,4 &        2,3,6,9 \\
   HD 195627 &           629$\pm$44 &           607$\pm$43 &           405$\pm$29 &           145$\pm$14 &             70$\pm$7 &             34$\pm$7 &            204$\pm$8 &           644$\pm$45 &                    - &        1,4 &     1,3,5,9,10 \\
\hline                                                                          
\end{tabular}                                                                   
\end{center}                                                                    
\end{table*}                                                                    


\section{Results} \label{results}

\subsection{Disc properties derived from SEDs} \label{sedmodelling}

\begin{figure*} 
\includegraphics[width=170mm]{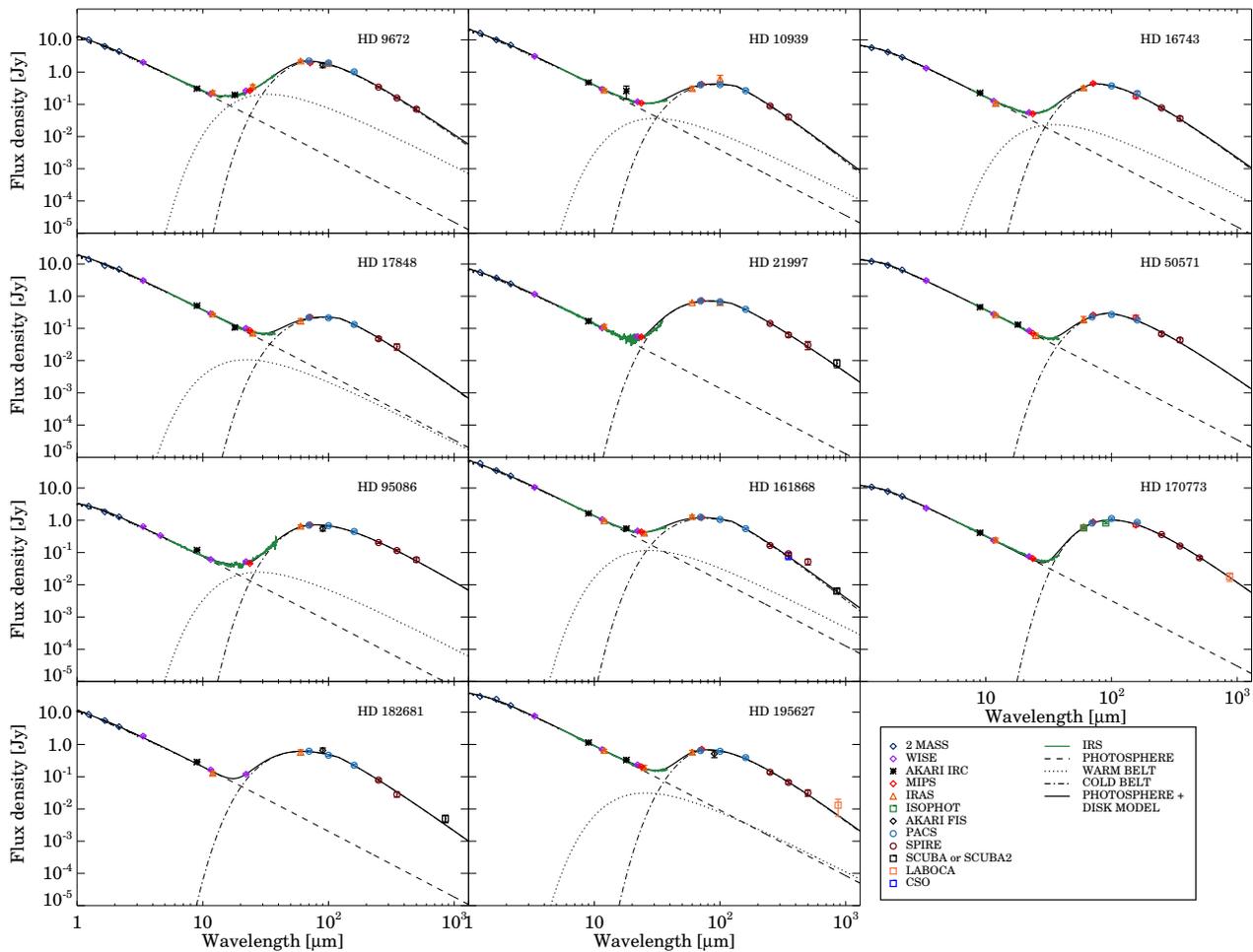}
\caption{{\sl SEDs of the studied systems. Stellar photospheres and the best fit 
models (Sect.~\ref{sedmodelling}) are also displayed.}
}
\label{sedplot}
\end{figure*}

We compiled the SED of each object by combining the optical and 
near-IR data points (see Sect.~\ref{stellarpropssect}) with the new 
PACS, SPIRE and MIPS flux densities and photometry obtained by 
different infrared space missions and ground-based submillimetre 
observations.
IRAS photometry with moderate quality (quality index of 2) was re-evaluated 
by utilizing the SCANPI tool\footnote{http://scanpiops.ipac.caltech.edu:9000}. In most 
cases the results were consistent with the values quoted in the IRAS 
catalogues, except for HD~182681 where at 60~{\micron} 
we adopted the photometry derived from SCANPI (0.549$\pm$0.100\,Jy). 
References for the utilized catalogues and literature data are 
summarized in Table~\ref{phottable}. 
{\sl Spitzer} IRS spectra of the targets are found to be featureless. For the subsequent SED modelling process 
the IRS spectra were resampled into the following 11 adjacent 
wavelength bins: 10--12, 12--14, 14--16, 16--18, 18--20, 20--23, 23--26, 26--29, 
29--32, and 32--35{\micron}.
In the computation of uncertainties of the derived flux densities in a given bin, 
we adopted 5\% absolute calibration uncertainty for IRS added in quadrature.
The SEDs for HD\,21997 and HD\,95086 were adopted from \citet{moor2013b} and 
\citet{moor2013a}, respectively. 
The constructed SEDs are plotted in Figure~\ref{sedplot}.

The infrared excess emission of debris discs is attributed to 
optically thin thermal emission of second generation circumstellar 
dust grains. 
In most cases the measured excess emission 
can be well fitted by a single temperature modified blackbody or a combination 
of two different temperature components that are thought to represent narrow, spatially separated 
warm and cold rings in which the bulk of the emitting dust grains is confined. 
This simple model/assumption can provide estimates of some 
basic disc  properties, e.g. the characteristic dust temperature(s) and 
the fractional luminosity (the ratio of the integrated luminosity
of the dust emission to the integrated luminosity of the host star). 

In the course of modelling we fitted the excess emission that was derived as a difference between the measured 
flux densities and the predicted photospheric fluxes. 
The average accuracy of the predicted photospheric fluxes is estimated to be around 3\%.
The final errors in the excess flux densities were computed as a quadratic sum of 
 uncertainties of the measured and the predicted photospheric fluxes. 
In the case of HD\,161868 the SPIRE measurement at 500\,{\micron} deviated significantly from the trend 
delineated by other submillimetre observations (see Fig.~\ref{sedplot}). 
This source coincides with bright background 
emission at this wavelength (Sect.~\ref{basicanalysis}) and notwithstanding the applied PSF photometry the 
measured flux density may be 
affected by background contamination. Therefore this data point was ignored in the 
SED fitting. The IRAS FSC catalogue includes a moderate quality 100\,{\micron} photometry for HD\,16743. 
This measurement was also omitted, because the quoted flux density is inconsistent with 
the much better quality PACS photometry.   

The excess SEDs were fitted by single temperature and two-temperature 
models. For the single temperature approach we used a modified blackbody 
where the emissivity is equal to 1 at $\lambda \leq \lambda_0$ and follows $(\lambda/\lambda_0)^{\rm -\beta}$
for longer wavelengths. Here we fitted four parameters, $A_{\rm c}$ a scaling factor that is proportional to the 
solid angle of the emitting
region, $T_{\rm bb,c}$, $\lambda_0$, and $\beta$.
In the two-temperature model, we added a warmer simple blackbody component (characterized by two parameters, $A_{\rm w}$ and $T_{\rm bb,w}$)
to the one defined above.
The best-fitting model was found by applying a Levenberg-Marquardt algorithm \citep{markwardt2009}. 
An iterative way was used to compute and apply colour corrections for the photometric data 
during the fitting process \citep[see e.g.][]{moor2006}.

In order to decide whether the single-temperature or the two-temperature model is better for a given source, 
we applied a variant of the 
Akaike Information Criterion, the so-called corrected $AIC_c$ \citep{burnham2002}. 
By penalizing the usage of 
unnecessary additional model parameters, this method provides a way to compare the relative quality 
of different models. The corrected $AIC_c$
was derived as: 
\begin{equation}
AIC_c = \chi^2 + 2k + \frac{2 k (k+1)}{n-k-1}
\end{equation}
where $n$ is the number of observations, and $k$ is the number of parameters in the model.
Better model quality gives a lower $AIC_c$.
Based on this test, we found the single-component model to be better only in four cases 
(HD\,21997, HD\,50571, HD\,170773, HD\,182681). 
The best-fit models for each of our targets are plotted in Figure~\ref{sedplot}. 

The fractional luminosity of the discs was computed as 
$f_{\rm dust} = {L_{\rm dust}}/{L_{\rm bol}}$. 
The integrated dust emission was derived using the fitted models.
Based on the derived characteristic temperatures and the 
stellar luminosities (Table~\ref{stellarprops}), dust belt radii 
valid for large blackbody grains, were also derived following \citet{backman1993}: 
\begin{equation}
\frac{R_{\rm bb}}{au} = \left(\frac{L_{\rm star}}{L_{\sun}}\right)^{0.5} \left(\frac{278\,K}{T_{\rm bb}}\right)^2.
\end{equation}
In the case of two-temperature models we assumed that the warm and cold components 
correspond to two spatially distinct dust belts and determined radii for the cold and warm component 
separately. 
$T_{\rm bb}$ is set equal to the temperature values ($T_{\rm bb,w}$ and $T_{\rm bb,c}$, for the warm and 
cold components, respectively) 
found in our fitting process.

The obtained fundamental disc properties are listed in Table~\ref{mbbtable}.
\begin{table*}                                                                                                                                                                                               
\setlength{\tabcolsep}{1.2mm}                                                                                                                                                                                
\begin{center}                                                                                                                                                                                               
\scriptsize                                                                                                                                                                                                  
\caption{{\bf Results of modified blackbody fitting.}                                                                                                                                                        
 \label{mbbtable}}                                                                                                                                                                                           
\begin{tabular}{lccccccccc}                                                                                                                                                                                  
\hline\hline                                                                                                                                                                                                 
Target          &  \multicolumn{3}{c}{Warm component} & \multicolumn{5}{c}{Cold component} &   \\                                                                                                            
            &  $T_{bb,w}$  & $f_{d,w}$  & $R_{bb,w}$  &  $T_{bb,c}$  & $\lambda_0$                                                                                                                           
	    &  $\beta$ & $f_{d,c}$  & $R_{bb,c}$ & $M_d$  \\                                                                                                                                                        
	    &   [K] &  [10$^{-5}$] &  [au] &   [K] &  [{\micron}] &   & [10$^{-5}$] &  [au] & [M$_\oplus$] \\                                                                                                       
\hline                                                                                                                                                                                                       
     HD 9672 &      155$\pm$13 &    20.0$\pm$2.2 &    12.9$\pm$2.2 &        56$\pm$4 &       62$\pm$31 &   0.83$\pm$0.11 &    90.2$\pm$3.3 &       99$\pm$17 & 0.267$\pm$0.041 \\
    HD 10939 &      167$\pm$41 &     2.2$\pm$0.5 &    15.4$\pm$7.7 &        54$\pm$2 &      135$\pm$42 &   1.18$\pm$0.24 &     9.9$\pm$1.0 &      143$\pm$13 & 0.051$\pm$0.014 \\
    HD 16743 &      146$\pm$35 &     6.7$\pm$2.4 &     8.2$\pm$4.0 &        50$\pm$3 &       71$\pm$22 &   0.95$\pm$0.21 &    44.7$\pm$3.1 &       70$\pm$10 & 0.058$\pm$0.015 \\
    HD 17848 &      228$\pm$38 &     1.2$\pm$0.3 &     5.9$\pm$2.0 &        55$\pm$5 &      123$\pm$24 &   0.92$\pm$0.35 &     6.5$\pm$0.5 &      101$\pm$21 & 0.028$\pm$0.011 \\
    HD 21997 &               - &               - &               - &        61$\pm$1 &      123$\pm$31 &   0.82$\pm$0.23 &     5.7$\pm$0.1 &        67$\pm$4 & 0.163$\pm$0.047 \\
    HD 50571 &               - &               - &               - &        46$\pm$2 &       93$\pm$35 &   0.81$\pm$0.21 &     1.3$\pm$0.1 &        64$\pm$6 & 0.029$\pm$0.007 \\
    HD 95086 &      184$\pm$31 &    15.4$\pm$1.8 &     6.1$\pm$2.1 &        54$\pm$2 &       60$\pm$30 &   0.37$\pm$0.09 &   150.1$\pm$5.5 &        67$\pm$6 & 0.688$\pm$0.130 \\
   HD 161868 &      186$\pm$51 &     2.4$\pm$0.6 &    11.5$\pm$6.3 &        66$\pm$3 &      125$\pm$44 &   1.10$\pm$0.13 &    10.3$\pm$0.9 &        89$\pm$9 & 0.022$\pm$0.004 \\
   HD 170773 &               - &               - &               - &        37$\pm$2 &       35$\pm$30 &   0.92$\pm$0.11 &     5.1$\pm$0.1 &      100$\pm$11 & 0.178$\pm$0.053 \\
   HD 182681 &               - &               - &               - &        82$\pm$1 &      120$\pm$28 &   0.83$\pm$0.18 &     2.8$\pm$0.1 &        56$\pm$2 & 0.067$\pm$0.018 \\
   HD 195627 &      199$\pm$81 &     1.9$\pm$0.8 &     5.3$\pm$4.3 &        45$\pm$2 &       60$\pm$41 &   0.97$\pm$0.15 &    11.5$\pm$0.8 &      102$\pm$13 & 0.032$\pm$0.008 \\
\hline                                                                          
\end{tabular}                                                                   
\end{center}                                                                    
\end{table*}                                                                    

All of our targets were successfully detected at submillimetre wavelength that allowed 
us to estimate the dust mass in the system.
We followed the standard approach
assuming optically thin emission characterized by a single temperature (in our case, the 
temperature of the cold component): 
\begin{equation}
M_{d} = \frac{F_{\nu, \rm excess} d^2}{ B_{\nu}(T_{bb,c}) \kappa_{\nu}}, 
\end{equation}
where $F_{\nu,excess}$ is the measured excess at the longest wavelength 
where submillimetre data is available (except in the case of HD\,195627 where 
we used the 500\,{micron} SPIRE data point instead of the 870\,{micron} 
LABOCA measurement that has low signal-to-noise ratio), $d$ is the
distance to the source, $\kappa_{\nu} = \kappa_{0}
(\frac{\nu}{\nu_0})^{\beta}$ is the mass absorption coefficient, and
$B_{\nu}$ is the Planck function. 
We adopted $\kappa_{0} = $2\,cm$^2$\,g$^{-1}$ at $\nu_0 = 345$~GHz \citep[e.g.][]{nilsson2010}, the
$\beta$ and $T_{bb,c}$ values 
were taken from Table~\ref{mbbtable}.
The derived dust masses and their uncertainties are also listed in Table~\ref{mbbtable}.

\subsection{Geometrical disc model} \label{ringmodel}

In order to precisely derive geometrical parameters for the discs
around our targets, we fitted the PACS images using a simple,
non-physical disc model grid. In our first approach, we assumed that the
emitting dust is located in a narrow ring around the central star.
Such a geometry could be
expected from our SED modeling, where we found that the
long-wavelength excess could be fitted with a single temperature
modified blackbody, the cold component in Table~\ref{mbbtable}. Our simple model
has three free parameters, the average disc radius ($R_{\rm avg}$),
the position angle ($PA$) and the inclination ($i$) of the disc. This
disc is assumed to be 0.1$R_{\rm avg}$ wide, i.e.~it extends from 0.95
$R_{\rm avg}$ to 1.05 $R_{\rm avg}$ in radial direction. We calculated 6048 models in a
regular grid. We used 18 different position angles centred on the
value determined from the 2D Gaussian fitting (Sect.~\ref{basicanalysis}), with a step between 0.5 and
10$^{\circ}$ (larger step for longer wavelengths and for more face-on
discs, and smaller step for shorter wavelengths and more edge-on
discs). We took 21 different inclinations from 0 to 90$^{\circ}$,
using a regular grid for $\cos i$. For the average disc radius, we
used 16 different values, centred on those obtained from the Gaussian
fitting, and a step between 2 and 8\,au (smaller for smaller discs and
shorter wavelengths, larger for larger discs and longer
wavelengths). The surface brightness of the disc was assumed to be
homogeneous. In each model, we included a central point source as
well, meant to represent the emission from the stellar photosphere and
from the warm inner dust ring where exist. The flux ratio of the
central point source to the disc in our model was adjusted to match
the respective flux ratio of each target from the SED modeling. These
model images were then convolved with a PSF appropriately rotated to
match the PSF angle of each observation (for the PSF we used the
observation of $\alpha$ Boo with OBS ID 1342247705 at 70\,{\micron} and
with OBS ID 1342223348 at 100 and 160\,{\micron}). The convolved model
images were then downsampled and shifted to match the pixel size and
astrometry of the actual observations.

In order to select the best-fitting model from our grid, we used
Bayesian analysis. We added in quadrature pixel values of the residual
image (the difference of the model image and the observed image) in an
aperture centred on the source, did the same in the error image, and
took the ratio of the two numbers as $\chi^2$. The Bayesian
probability assigned to a certain model is $\exp(-\chi^2/2a)$, where
$a$ is the area of the aperture. We then marginalized the Bayesian
probabilities and obtained 1D probability distributions as a function
of the three free parameters in the model. We fitted the distributions
with Gaussians to determine the optimal value of $R_{\rm avg}$, $PA$,
and $i$. In cases where the probability distribution for the position
angle was too wide, we fitted it with a von Mises distribution, the
circular equivalent of the Gaussian. With the optimal parameters, we
calculated a model image and residuals. The residual images showed
that our optimal models do not fit the observed images very well. For
all targets, bright residuals at the stellar position and at large
radial distances, and dark residuals in-between could be seen. This
effect is strongest at the best resolution 70{\micron} images, but
still noticeable for some sources at 100{\micron}. At 160{\micron}, the
residuals were in all cases below 3$\sigma$.

\begin{table}                                                                                                                                                                                                
\setlength{\tabcolsep}{0.8mm}                                                                                                                                                                                
\begin{center}                                                                                                                                                                                               
\scriptsize                                                                                                                                                                                                  
\caption{{\bf Disc properties derived from the geometrical model.                                                                                                                                            
\label{disktable}}}                                                                                                                                                                                          
\begin{tabular}{lcccccc}                                                                                                                                                                                     
\hline\hline                                                                                                                                                                                                 
Target         &   PACS band      & R$_{\rm avg}$  & R$_{\rm in}$  & R$_{\rm out}$ & $i$        & PA  \\                                                                                                     
            & ({\micron})  &   (au)        & (au)        &    (au)       & ({\degr})  & ({\degr}) \\                                                                                                         
\hline                                                                                                                                                                                                       
     HD 9672 &    70.0 &     253.0$\pm$13.5 &          46$\pm$28 &         459$\pm$28 &              84.1$\pm$5.9 &      109.4$\pm$1.6 \\
             &   100.0 &     264.4$\pm$21.0 &          52$\pm$41 &         476$\pm$41 &              79.7$\pm$6.0 &      109.2$\pm$1.8 \\
             &   160.0 &     227.1$\pm$24.5 &          83$\pm$58 &         371$\pm$58 &             77.7$\pm$10.8 &      110.5$\pm$4.5 \\
             &   WMEAN &     251.2$\pm$10.3 &          52$\pm$21 &         452$\pm$21 &              81.4$\pm$3.9 &      109.4$\pm$1.2 \\
\hline
    HD 10939 &    70.0 &     153.8$\pm$16.5 &          47$\pm$31 &         260$\pm$31 &             35.8$\pm$10.0 &      16.5$\pm$14.7 \\
             &   100.0 &     171.3$\pm$20.0 &          69$\pm$36 &         273$\pm$36 &             32.5$\pm$10.0 &      20.1$\pm$13.6 \\
             &   160.0 &     185.3$\pm$33.9 &         115$\pm$78 &         290$\pm$78 &              23.1$\pm$34.0 &      20.8$\pm$51.9 \\
             &   WMEAN &     163.9$\pm$11.9 &          61$\pm$23 &         267$\pm$23 &              34.1$\pm$7.1 &       18.5$\pm$9.8 \\
\hline
    HD 16743 &   100.0 &     156.6$\pm$19.7 &          72$\pm$39 &         240$\pm$39 &              58.5$\pm$8.3 &      165.0$\pm$6.7 \\
\hline
    HD 17848 &    70.0 &     156.1$\pm$17.3 &          45$\pm$42 &         266$\pm$42 &              77.1$\pm$9.7 &      154.6$\pm$5.0 \\
             &   100.0 &     182.6$\pm$21.6 &          63$\pm$52 &         301$\pm$52 &              69.3$\pm$9.6 &      159.3$\pm$6.4 \\
             &   160.0 &     215.8$\pm$47.5 &         47$\pm$173 &        384$\pm$173 &             75.6$\pm$22.6 &     154.6$\pm$14.6 \\
             &   WMEAN &     170.1$\pm$13.0 &          52$\pm$32 &         283$\pm$32 &              73.4$\pm$6.5 &      156.3$\pm$3.8 \\
\hline
    HD 21997 &    70.0 &     133.6$\pm$14.8 &          42$\pm$28 &         224$\pm$28 &             28.6$\pm$15.7 &       3.1$\pm$19.9 \\
             &   100.0 &     139.3$\pm$17.2 &          72$\pm$35 &         206$\pm$35 &             19.0$\pm$15.3 &       4.7$\pm$36.8 \\
             &   160.0 &     168.6$\pm$30.7 &          91$\pm$75 &         245$\pm$75 &             25.8$\pm$25.4 &      30.8$\pm$43.7 \\
             &   WMEAN &     139.9$\pm$10.5 &          57$\pm$21 &         219$\pm$21 &             24.0$\pm$10.1 &       7.2$\pm$16.2 \\
\hline
    HD 50571 &    70.0 &      99.9$\pm$11.5 &          30$\pm$29 &         168$\pm$29 &             80.7$\pm$10.4 &      119.2$\pm$5.2 \\
             &   100.0 &     123.4$\pm$13.7 &          46$\pm$29 &         200$\pm$29 &              76.1$\pm$8.7 &      120.8$\pm$4.3 \\
             &   160.0 &     135.1$\pm$26.6 &          39$\pm$85 &         230$\pm$85 &             78.1$\pm$18.3 &     120.0$\pm$11.1 \\
             &   WMEAN &      112.1$\pm$8.4 &          38$\pm$20 &         187$\pm$20 &              78.0$\pm$6.3 &      120.1$\pm$3.2 \\
\hline
    HD 95086 &    70.0 &     205.5$\pm$17.4 &          44$\pm$33 &         366$\pm$33 &              29.1$\pm$9.6 &      91.4$\pm$12.6 \\
             &   100.0 &     217.8$\pm$25.1 &          72$\pm$45 &         363$\pm$45 &             24.2$\pm$13.0 &     106.0$\pm$17.6 \\
             &   160.0 &     268.3$\pm$36.2 &         170$\pm$79 &         366$\pm$79 &             17.5$\pm$18.2 &     129.2$\pm$41.0 \\
             &   WMEAN &     217.4$\pm$13.3 &          66$\pm$25 &         365$\pm$25 &              25.9$\pm$7.1 &       98.3$\pm$9.9 \\
\hline
   HD 161868 &    70.0 &      156.2$\pm$7.6 &          50$\pm$14 &         261$\pm$14 &              64.9$\pm$1.9 &       57.8$\pm$1.8 \\
             &   100.0 &     177.8$\pm$10.0 &          54$\pm$19 &         300$\pm$19 &              63.7$\pm$2.0 &       60.6$\pm$1.6 \\
             &   160.0 &     164.4$\pm$12.5 &          35$\pm$26 &         292$\pm$26 &              64.3$\pm$5.3 &       60.8$\pm$4.2 \\
             &   WMEAN &      164.2$\pm$5.4 &          49$\pm$10 &         278$\pm$10 &              64.3$\pm$1.3 &       59.5$\pm$1.2 \\
\hline
   HD 170773 &    70.0 &      171.5$\pm$3.7 &           74$\pm$8 &          268$\pm$8 &              30.7$\pm$3.0 &      118.1$\pm$5.4 \\
             &   100.0 &      175.9$\pm$4.6 &          89$\pm$10 &         262$\pm$10 &              31.7$\pm$2.3 &      117.4$\pm$4.3 \\
             &   160.0 &     176.7$\pm$14.7 &          93$\pm$27 &         259$\pm$27 &              31.0$\pm$5.3 &      123.1$\pm$9.4 \\
             &   WMEAN &      173.4$\pm$2.8 &           81$\pm$6 &          265$\pm$6 &              31.3$\pm$1.7 &      118.3$\pm$3.2 \\
\hline
   HD 182681 &    70.0 &     154.3$\pm$15.8 &          44$\pm$31 &         263$\pm$31 &              79.5$\pm$4.8 &       54.1$\pm$3.3 \\
             &   100.0 &     164.5$\pm$21.4 &          73$\pm$44 &         255$\pm$44 &              80.9$\pm$9.6 &       56.2$\pm$5.3 \\
             &   160.0 &     175.7$\pm$54.1 &         59$\pm$198 &        410$\pm$198 &             84.2$\pm$21.5 &      59.0$\pm$17.8 \\
             &   WMEAN &     158.8$\pm$12.4 &          52$\pm$25 &         263$\pm$25 &              79.9$\pm$4.2 &       54.8$\pm$2.8 \\
\hline
   HD 195627 &    70.0 &      113.7$\pm$6.1 &          15$\pm$12 &         212$\pm$12 &              54.4$\pm$1.9 &       93.1$\pm$2.2 \\
             &   100.0 &      124.7$\pm$8.4 &          25$\pm$17 &         224$\pm$17 &              55.9$\pm$2.2 &       92.1$\pm$2.7 \\
             &   160.0 &     151.4$\pm$15.5 &          65$\pm$35 &         237$\pm$35 &              59.0$\pm$4.9 &       89.1$\pm$5.4 \\
             &   WMEAN &      120.6$\pm$4.7 &           22$\pm$9 &          217$\pm$9 &              55.4$\pm$1.4 &       92.4$\pm$1.6 \\
\hline
\hline                                                                          
\end{tabular}                                                                   
\end{center}                                                                    
\end{table}                                                                     

\begin{figure} 
\includegraphics[width=90mm]{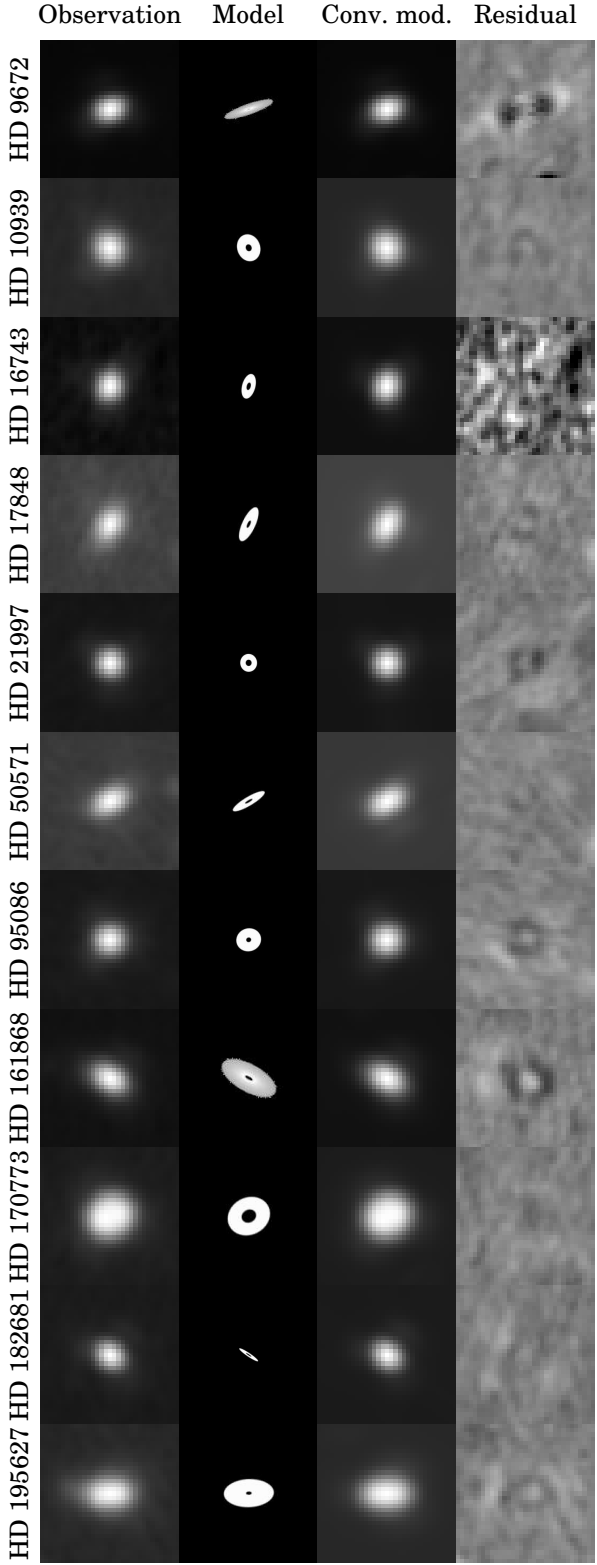}
\caption{{\sl Measured and model images, as well as the residuals for 
the 100\,{\micron} band observations of our 11 targets.}
}
\label{residual}
\end{figure}

The structure of the residual images at 70{\micron} strongly suggests
that the adopted model geometry -- a narrow ring -- is not appropriate for
our targets. For this reason, as a next step, we used an annulus, i.e. a disc with an inner hole. 
It was characterized by five free parameters:
the average
radius ($R_{\rm avg}$), the position angle ($PA$), the inclination
($i$), the power law exponent of the brightness profile (p), and the
thickness of the disc ($t$), defined so that the inner radius is at
$R_{\rm avg} (1-t/2)$, while the outer radius is at $R_{\rm avg} (1+t/2)$. 
We calculated a grid of 848\,160 models, with 18 different $PA$,
31 different $i$, 20 different $R_{\rm avg}$, 19 different $t$, and 4
different $p$ values. The disc thickness $t$ was varied from 0.1 to
1.9 with a regular step of 0.1, while the range and step of the other
parameters were the same as above. 
We found that the $p$ values are not well constrained from the images. Moreover, 
they are degenerated with the inner and outer disc radii: steeper disc 
radial profiles are accompanied with tendentiously larger inner/outer radii.
For the further analyses we adopted p = 0 for all 
sources
except HD\,9672 and HD\,161868, 
where a steeper profile with p = -1 provided significantly better fits.
Bayesian analysis, fitting of the
optimal parameters (listed in Table~\ref{disktable}), calculating the best-fit
model image and residuals (shown in Figure~\ref{residual} for observations at 100\,{\micron}) 
were done as described
above. 

We note that the disc position angles and inclinations derived in 
the narrow and broad geometrical models were fully consistent with 
each other, and the $PA$ values were also in good agreement with the 
ones obtained from 2D Gaussian fittings. 
As for disc inclinations, the 2D Gaussian and the geometrical models 
gave similar results below $i = 60${\degr}, while in more edge-on 
systems the Gaussian model provided somewhat lower inclination.
Apart from the case of HD\,170773, the inner edge of the outer dust belt 
has not been resolved, and its relative uncertainty is usually comparable to the derived $R_{\rm in}$ values.
The average and the outer radii, however 
are better determined, and for any given object the deduced  
 70, 100, and 160\,{\micron} outer radii agree within 
 their uncertainties.

For each object we calculated the weighted averages of the different disc parameters 
obtained in different PACS bands (Table~\ref{disktable}).

\subsection{Comparison of our results to previous resolved data} \label{ringmodel}

Five of our targets have been successfully 
resolved by other instruments. 
The disc around HD\,9672 has been
spatially resolved in mid-IR thermal emission using the Keck telescope 
by \citet{wahhaj2007}. They derived a position angle of 125$\pm$10{\degr}
and an inclination of 60$\pm$15{\degr} for the disc, both parameters 
are consistent with our results. 
\citet{wahhaj2007} argued that the observed mid-IR emission 
dominantly comes from an inner disc component.
\citet{roberge2013} found this disc to be significantly extended
at a position angle of $\sim$114{\degr}
based on a previous 70\,{\micron} {\sl Herschel} image 
(we used a later deeper observation in our analysis) of the 
source. 
They derived a disc radius of 200\,AU, and constrained its 
inclination to be $\geq 44${\degr}.

In the case of HD\,21997, our ALMA observation \citep{moor2013b} clearly resolved a broad ring of emission between 55 and 
150\,au, with a position angle of 21\fdg5$\pm$3\fdg5 and an inclination of 32\fdg9$\pm$2\fdg6. 
Apart from the outer radius all of these parameters  
 match well the PACS-based values within the uncertainties. 
The outer edge of the disc is somewhat smaller in the ALMA image, indicating that the
large grains may have a more confined distribution than the smaller grains emitting in
the PACS bands. Moreover, ALMA is an interferometer, and may have filtered out some
disc emission at the largest spatial scales \citep[see more details in][]{moor2013b}.

The disc around HD\,161868 was resolved 
both at 24 and 70\,{\micron} with the {\sl Spitzer Space Telescope} 
\citep{su2008}. They derived a disc position angle and inclination 
of 55$\pm$2{\degr} and 50$\pm$5{\degr}, respectively, which are broadly consistent with our values 
($PA$=59\fdg5$\pm$1\fdg2 and $i$=64\fdg3$\pm$1\fdg3). With a radius of 520\,au at 70\,{\micron}, 
the disc was found to be somewhat more extended than in our analysis based on the PACS images, although this comparison is hampered 
by the coarser spatial resolution of {\sl Spitzer}. 

Analyzing the {\sl Spitzer} MIPS 70\,{\micron} images 
of HD\,50571 and HD\,170773, \citet{moor2011a} revealed marginally extended emission at these sources.
The disc around HD\,50571 was found to be extended only in one direction along
a position angle of 91{\degr} with a characteristic radius of $\sim$160\,au, which deviate 
from the $PA$ of 120{\degr} and $R_{\rm avg}$ of 112\,au derived from our PACS images.
This discrepancy may partly be related to those nearby background sources, which are 
blended with the disc on {\sl Spitzer} images. 
The characteristic radius of 180\,au and position angle of 110{\degr} derived for  
HD\,170773 are in good agreement with our new results based on PACS data (see Table~\ref{disktable}).

\section{Discussion} \label{discussion}

\subsection{General points concerning the disc parameters}

\paragraph*{Systems with warm inner belts.}
All of our targets harbour cold, extended outer dust component.  
In 
seven cases (64\%) our SED analysis implies the existence of a warmer 
component as well (Sect.~\ref{sedmodelling}). 
This fraction is consistent with the results of \citet{chen2014} who found that in a large sample 
of 499 debris discs exhibiting {\sl Spitzer}/IRS excess, two-temperature models provided 
better fit in 66\% of the systems.
That the SED analysis indicates two components with different temperatures does not
necessarily mean that the system harbours a spatially distinct 
inner dust belt as well.
\citet{kennedy2014} pointed out the possibility that the emission from a single 
narrow dust belt including grains with different sizes and thereby different temperatures 
can reproduce a two-component SED. However, this model may not be feasible for our seven discs: 
by placing them in Figures 13--15 of \citet{kennedy2014} these systems are clearly 
out of the parameter space covered by the single narrow disc models, although 
the fact that our discs are radially extended might slightly change this conclusion.

Supposing that in most of our cases the warm and cold components are associated  
with spatally distinct dust belts, what could be the origin of the warm 
dust?
In our case the temperature of the warm dust component
ranges between 145 and 230\,K, where sublimation is an active process in comets \citep[T$>$110\,K,][]{wyatt2008}.  
Therefore, the production of warm grains is not limited to collisions in the inner dust belt, but 
sublimation of icy bodies, originated from the outer planetesimal belt and entering 
the inner part of the system, can also 
contribute.    
We know a smaller number of warm debris discs where the dust production may be 
due to a recent transient event, e.g. collisions between large planetesimals \citep{wyatt2008,telesco2005,stark2014}.
In order to evaluate whether the warm components discovered around our targets 
might be related to similar transient events, we utilized 
 the analytical steady state
model of \citet{wyatt2007}.
Their model describes  the collisional evolution of a planetesimal belt, and argues 
that at any given age there is a maximum dust fractional luminosity ($f_{max}$), since 
initially more massive discs consume their mass faster. \citet{wyatt2007} proposed that dust belts 
with fractional luminosity of $\gg 100 \times f_{max}$ can be considered as a result of a transient 
event that increases the dust production for a short period. Using their formulae, 
 the maximum possible fractional luminosity of the warm belt 
can be estimated as $f_{max} = 0.16\times 10^{-3} R_{bb,w} M_{*}^{-5/6} L_{*}^{-1/2} t_{\rm age}^{-1}$, where 
the radius of the belt ($R_{bb,w}$), stellar mass, luminosity and age ($M_*, L_*, t_{age}$) were taken from 
Tables~\ref{mbbtable} and \ref{stellarprops}. We found that the measured fractional luminosities ($f_{d,w}$) 
are always less than 0.1 $f_{\rm max}$.
Therefore no transient events have to be invoked to explain the amount of observed warm dust, and 
a steady-state evolution of an inner planetesimal belt can be consistent with the observations. 
The fractional luminosity of the cold disc component ($f_{d,c}$) exceeds the
 fractional luminosity measured for the warm component of the same system 
 ($f_{d,w}$) 
 by a factor of four at least.
 Since the relative dust masses in the inner and outer dust belts are proportional 
 to $\propto f_d R_d^2 $ \citep{wyatt2008}, 
 the difference is even significantly higher in terms of dust mass.

\paragraph*{Basic characteristics of the outer cold belts.}
The characteristic temperatures of the cold outer belts in our targets are in most cases well 
within the range found for large
samples of cold debris discs \citep{ballering2013,chen2014}. The only exception 
is HD\,170773, which is one of the
coldest known debris discs with its $T_{bb,c} = 37$\,K.
The derived $\beta$ values range between 0.4 and 1.2, consistently with earlier results indicating 
that more processed debris dust grains typically have $\beta$ values between 0 and 1.2 
\citep[e.g.][]{roccatagliatta2009,nilsson2010,booth2013}, as opposed to the 
$\beta \sim 2$ of interstellar dust grains \citep{li2001}.
The $\lambda_0$ parameter is related to characteristic grain radius in the disc \citep[e.g.,][]{backman1993}.
For our sample $\lambda_0$ is between 35 and 135\,{\micron}, in most cases  
the obtained parameter is close to 100\,{\micron}, a typical value
in debris discs \citep{williams2006}.
Thus, although due to our selection criteria, our sample represents unusually large, extended structures, 
the properties of
dust grains seem to be very similar to those in smaller, more typical debris discs.

\paragraph*{Evolution of dust mass.}
\begin{figure} 
\includegraphics[width=90mm]{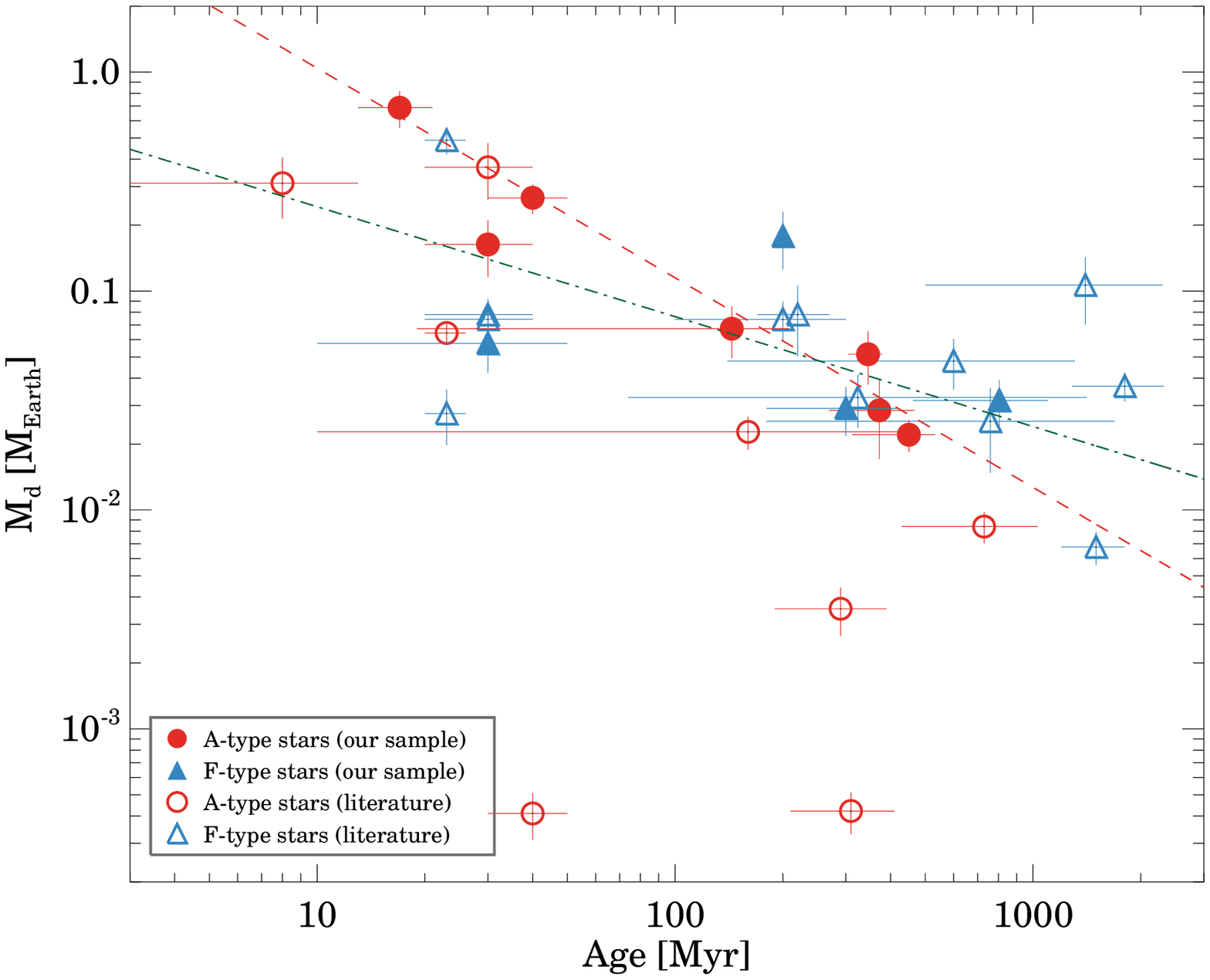}
\caption{{\sl Measured dust masses as a function of age for our discs and for other 
debris systems detected at submillimetre/millimetre
wavelengths. Literature data are 
from the following sources: \citet{sheret2004,najita2005,williams2006,nilsson2009,nilsson2010,churcher2011,
kennedy2012b,meuss2012,booth2013,eiroa2013,panic2013}. The red dashed line shows our fit 
to the upper envelope of data points corresponding to A-type stars. Our fitting to 
the whole sample are displayed by a green dashed-dotted line.} 
}
\label{mdust}
\end{figure}
In Figure~\ref{mdust} we plotted the dust masses inferred from 
submillimetre observations as a function of system age.
For comparison, we collected additional submillimetre/millimetre 
detections for debris discs around A- and F-type stars from the literature. 
The dust masses of the latter systems were recalculated using 
the same $\kappa_0$ mass absorption coefficient and technique 
adopted for our targets (Sect.~\ref{sedmodelling}).
We note that the dust mass estimates are usually quite uncertain because 1) 
the submillimetre detections have generally low signal-to-noise ratio; 2) the errors of 
such parameters as e.g. dust temperature, $\beta$, distance
are propagated to the final uncertainty. Moreover, the mass absorption coefficient is
currently not well known at submillimetre
wavelengths \citep[e.g.,][]{nilsson2010}. 
Figure~\ref{mdust} shows that nearly all of our sources are located close to the upper envelope 
of the distribution, i.e. at a specific age they are among the most massive debris discs. 
Since the plotted data are already biased towards the most massive discs -- 
low-mass discs can be detected only around nearby, older stars \citep{panic2013} -- 
our objects belong to the top range of debris discs in terms of mass. 
By analyzing dust masses of debris discs around Sun-like 
stars, \citet{roccatagliatta2009} found a moderate 
correlation with age. On a larger sample, including also intermediate mass stars, \citet{panic2013} reported 
that the upper envelope of the dust mass distribution remains relatively flat at all ages.
Our survey provides new or improved dust mass estimates for several 
systems, and in particular around A-type stars. Focusing only 
 on discs around A-type stars, the data points
hint at a decrease of the upper envelope of the distribution with increasing age.
Using an analytic steady state evolutionary model \citet{wyatt2007} 
predicted that the decay of dust mass is inversely proportional to time ($M_{\rm dust} \propto t^{-1}$), 
whereas more detailed analytic models suggest
a gentler slope of t$^{-0.3...-0.4}$ \citep{loehne2008}.
Modelling the collisional evolution of debris discs \citet{gaspar2013} argued that the decay rate 
varies with time and $M_{\rm dust} \propto t^{-0.8}$ at the fastest point of this process.
We performed a Bayesian linear regression method developed by \citet{kelly2007} to 
derive the best fit line (in log--log scale) to data points corresponding to A-type stars and 
obtained a slope of $-$0.95$\pm$0.16 for the upper envelope. By fitting the whole sample 
including both discs around A- and F-type stars
using the same method we obtained a slope of $-$0.50$\pm$0.12.
These results are consistent with the predictions of above-mentioned models.

\subsection{Location of the cold planetesimal belt} \label{location}

In Sect.~\ref{ringmodel} we used a geometrical model to estimate 
the apparent extent of the IR emitting region in the PACS images. 
For each object we calculated the weighted averages of the $R_{\rm in}$, $R_{\rm avg}$, and 
$R_{\rm out}$ parameters obtained in different PACS bands.
The inferred averaged outer radii $\langle R_{\rm out} \rangle$ range from 190 to 450~au 
(Table~\ref{disktable}), thus 
these discs are significantly larger than our Kuiper-belt in which most objects 
  are located between 39 and 48~au \citep{jewitt2009}. 
The ratio of the average disc radii inferred from 
the IR images to the radii derived from the dust temperature assuming 
blackbody grains ($\Gamma = \langle R_{\rm avg} \rangle / R_{bb,c}$ ) is higher than 1 for all our systems,   
ranging between 1.1 and 3.1. It is consistent 
with previous results derived for other spatially resolved debris discs 
 and can be interpreted as the signature 
of small, ineffectively emitting grains in the system \citep{booth2013,morales2013,rodriguez2012}.
  
We found that the observed far-IR brightness distributions 
can be better fitted by broad dust rings in all of our discs. 
Indeed, \citet{booth2013} also suggested that at least 4 among their 9 resolved discs 
are relatively broad. 
Such results do not inevitably mean that the parent planetesimal belt is also broad since 
collisional processes can naturally lead to a dust distribution more extended than 
the birth ring itself \citep{strubbe2006}. In debris discs mutual collisions gradually grind large
bodies into smaller ones that are expelled by radiation pressure, producing 
an outwardly extended dust population \citep{krivov2006}.
In principle the Poynting-Robertson (PR) drag can also contribute to a wider dust distribution 
by drifting grains towards the central star. However, by computing both the dust removal timescale 
due to PR-drag \citep[$\tau_{PR}$, derived using the formula of][]{wyatt2005} and due to collisions 
 \citep[$\tau_{coll}$, derived following][]{zuckerman2012} in our systems,
 we found that $\tau_{coll} \ll \tau_{PR}$ for all relevant grain sizes. It implies 
 that these discs are collisionally dominated, i.e. PR drag 
 can be neglected and the width of the dust distribution is mainly 
 affected by processes related to the radial component of the stellar radiation pressure. 
 
 For investigating the feasibility of different stirring mechanisms in our targets, 
 we have to estimate the location of the dust-producing planetesimals.
 Therefore it would be important to know how our PACS observations are affected by small grains 
 expelled from the parent belt, i.e. how well we can trace the birth ring at these wavelengths? 
 Those small grains where 
 $\beta_{gr}$, the ratio of the radiation pressure to gravity forces, 
is higher than 0.5 are blown away 
on hyperbolic orbits on a short timescale, while larger grains are moved onto more eccentric 
orbit \citep{krivov2006}. It results in a radial grain size segregation beyond the parent belt. 
Using a numerical model, \citet{thebault2014} investigated the grain size distribution 
beyond a narrow birth ring of colliding planetesimals. They found that at a specific radial distance
the geometrical cross section of dust is dominated by the largest grains that can reach the given region, 
i.e. those particles whose apoastron is located at that radial position. 
Because of their decreasing emissivity, grains with a size of $< \frac{\lambda}{2 \pi}$ have typically 
little contribution to the  
emission observed at wavelength $\lambda$. As a consequence, at longer and longer wavelengths we 
can detect smaller and smaller part of these extended outskirt of dust. 
Using equation 19 from \citet{burns1979} 
we computed the $\beta_{gr}$ 
value for an compact, spherical astrosilicate grain \citep{draine2003} with a size of $\frac{70}{2 \pi}$, $\frac{100}{2 \pi}$, and 
$\frac{160}{2 \pi}$\,{\micron} and with a density of 2.7\,g~cm$^{-3}$ 
at all our targets. Assuming a narrow belt of parent bodies in circular orbit at a radial distance of $r_{br}$ 
 we also derived the orbital eccentricity ($ e = \frac{\beta_{gr}}{1-\beta_{gr}}$) and 
 the apoastron radial distance ($r_{ap}$) of these grains. 
 We found that for eight out of our eleven objects
 $r_{ap} / r_{br} < 1.3$ at 100{\micron}. For the six fainter stars this result holds also at 70\,{micron}. 
 In the case of the three most luminous objects (HD\,161868, HD\,182681, and HD\,10939) 
 at 70 and 100\,{\micron} 
 the observable region could be more extended, however at 160\,{\micron} the effectively emitting 
 grains are also expected to be confined within $1.3 r_{br}$. 
This suggests that even in these cases 
at least at 160\,{\micron} we can trace the birth ring and its immediate vicinity with a reliability 
of 30\%. 
 
The calculations above contain several simplifications, e.g. 1) narrow birth rings 
were assumed; 2) when computing the grain sizes we adopted 
compact homogenous spheres, while in real discs the dust particles may be porous and larger 
for the same $\beta_{gr}$ \citep{kirchschlager2013}. 
In order to further evaluate the possible difference between the location of dust grains effectively emitting 
at PACS wavelengths and the location of the parent planetesimal belt, we made an additional empirical 
comparison. 
Emission at (sub)millimetre wavelengths ($>$300~{\micron}) predominantly comes from large
grains with size of typically $>$50~{\micron}. Such large grains are little affected by radiative
forces, thus, being close to their birth area they trace the distribution of the parent planetesimals 
\citep{thebault2014}. 
From the literature, we selected four debris discs ($\beta$\,Pic, HD\,107146, HD\,109085, HR\,8799) that were resolved 
at millimetre wavelengths \citep{dent2014,hughes2011,wyatt2005,patience2011} and for which {\sl Herschel} PACS observations are also 
available in the Archive and whose angular size is comparable with that of our sources. 
We processed the {\sl Herschel} data
with HIPE utilizing the same method, 
as described in Sect.~\ref{pacsdatared}, and then disc sizes were derived by applying our broad disc 
models (see Sect.~\ref{ringmodel}). HD\,21997, one of our targets was also resolved at submillimetre \citep{moor2013b}
wavelength, thus we added it to this sample.  
We found the PACS-based average disc radii ($\langle R_{\rm avg} \rangle$) always to be smaller than the outer radii measured on 
millimetre images. 

Our theoretical considerations predict that the radius of the planetesimal belt 
may not differ from $R_{\rm out}$ by more than 30\%, where $R_{\rm out}$ is the disc radius 
determined from the PACS images. Since $\langle R_{\rm avg} \rangle$ 
was always found to be smaller than $\langle R_{\rm out} \rangle$/1.5 in our sample (Table~\ref{disktable}), 
it can be considered as a lower limit for the extent of the planetesimal belt.
Combining this result with our conclusions from discs resolved at submillimetre wavelengths,  
in the following analyses we will adopt the derived $\langle R_{\rm avg} \rangle$ values (Table~\ref{disktable})
as a proxy for the outer edge of the planetesimal belts.

\subsection{Stirring mechanism}

In the literature three different mechanisms have been invoked 
to explain the dynamical excitation of planetesimals
in debris discs: self-stirring, planetary stirring, and stirring 
initiated by stellar encounters (Sect.~\ref{intro}).
In the following, we examine which stirring scenarios 
could work in our disc sample.

\subsubsection{Self-stirring}

Using a hybrid multiannulus coagulation code, \citet{kb2008} followed the size and orbital 
evolution of initially small (1\,m--1\,km) planetesimals in a broad belt between 30 and 150\,au.
In the self-stirred model, bodies of the initial planetesimal population have 
low random velocities, thus their collisions lead to coagulation. 
After certain time, the largest planetesimals start accreting smaller 
bodies very effectively due to gravitational focusing, resulting in their runaway 
growth. In this phase the system becomes more and more dominated by a few big 
planetesimals whose growth slows down when they become large enough to increase
the velocity dispersion of neighbouring small planetesimals and thereby  
decreasing the gravitational focusing factor.
Then the growth of the large planetesimals switches to the much slower oligarch 
growth stage. According to the model of \citet{kb2008} the growing oligarchs 
can excite the motion of neighbouring small bodies so 
efficiently that their collisions produce
debris instead of mergers. 
Once fragmentation begins, continued stirring
leads to a collisional cascade, where a primordial body is ground
down into smaller and smaller fragments until all its mass ends
up in grains small enough to be removed from the system by radiation pressure.
After a peak in the dust production, as the cascade removes a large fraction of
planetesimals from the given region, the collisions become less frequent leading 
to a decline in dust replenishment. 
Since the growth time is proportional to $\frac{P}{\Sigma}$, where $P$ 
is the orbital period and $\Sigma$ is the surface density, the formation 
of large planetesimals requires more time at larger
radii, i.e. it is an inside-out process, the collisional cascade is ignited in 
the inner disc first and then the active
dust production propagates outward.
In their model, \citet{kb2008} adopted 
a disc with an initial surface density distribution of  
\begin{equation}
\Sigma(a) = \Sigma_0(M_*) x_m (a/a_0)^{-3/2},
\end{equation} where $\Sigma_0$ is the reference surface density at a radius of $a_0$ = 30\,au, while $x_m$ is a scaling factor. 
The reference surface density was scaled with the stellar mass as $\Sigma_0(M_*) = 0.18(M_*/M_{\sun})$\,g\,cm$^{-2}$ 
($\Sigma_0=0.18$\,g\,cm$^{-2}$ corresponds to the surface densitiy of the minimum mass solar nebula, MMSN, 
at the radius of 30\,au).
\citet{kb2008} found that the formation of the first 1000\,km icy planetesimals at a radius $a$ occurs 
at: 
\begin{equation} 
t_{\rm 1000} = 145 x_{\rm m}^{-1.15} (a/80\,{\rm au})^3 (2M_{\sun}/M_*)^{3/2} [\rm Myr]. \label{t1000}
\end{equation}
Thus the more massive was the disc initially the faster is the initalization of the cascade  
at a specific radius. Since the initial disc mass (or surface density) is limited, 
the front of stirring -- the region where the collisional cascade is ignited at a specific 
age -- is also limited. Adopting a protoplanetary disc with a surface density distribution 
of $\Sigma(a) \propto a^{-3/2}$ and with a gas-to-dust mass ratio of 100:1, 
\citet{mustill2009} proposed that $x_m$ have to be $<$10 since a more massive disc 
would become gravitational unstable at radii $>$100\,au.
Indeed, the mass of a disc that extends from 1\,au to 200\,au (typical outer radius for our systems) 
and have a surface density ten times higher than that of MMSN would be $\sim$ 0.55\,M$_{\odot}$.
(Sub)millimetre observations indicate that isolated Herbig Ae stars with stellar masses similar 
to our objects typically harbour less massive discs \citep{williams2011,sandell2011}. 
Thus, even if we take into account the well known uncertainties in disc mass estimates \citep{williams2011}, 
the suggested
$x_{m,max} = 10$ limit seems to be rather conservative and secure.

\begin{figure} 
\includegraphics[scale=0.45]{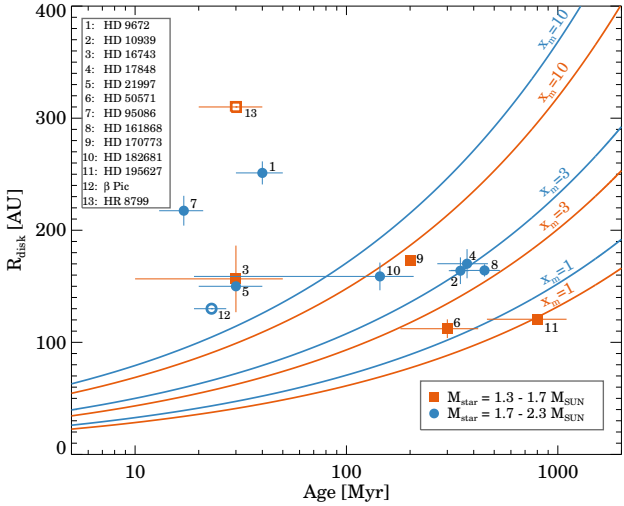}
\caption{{\sl Derived disc radii as a function of stellar ages overplotted by 
self-stirring models computed for host stars with masses of 1.5 and 2.0\,M$_{\odot}$ (orange and blue lines) 
and for three different initial $x_m$ values. Our targets are marked by filled symbols, the two additional discs 
are indicated by empty symbols. We note that for members of different kinematic groups,  
HD\,9672, HD\,21997, $\beta$\,Pic, HD\,95086 and HR\,8799 
the age estimates are more reliable.}
}
\label{stirring}
\end{figure}

In Figure~\ref{stirring} we plotted the obtained average disc radii ($\langle R_{\rm avg} \rangle$) as a function 
of system age for our 
targets. 
The disc around HD\,21997 was also resolved with ALMA at 
886\,{\micron}, in its case we adopted the outer radius derived from the submillimetre 
interferometric data \citep{moor2013b}. 
We also plotted two additional young debris discs
around stars with masses similar to our targets: $\beta$\,Pic and HR\,8799. 
$\beta$\,Pic belongs to the $\sim$23\,Myr old \citep{mamajek2014} 
$\beta$\,Pic moving group, while HR\,8799 is a member of the $\sim$30\,Myr old Columba association \citep{zuckerman2011}, 
thus they have reliable age estimates. Information for the outer edges of their discs 
were taken from resolved images published in \citet{dent2014,patience2011,matthews2014a}.

Using Eqn.\,5, in Figure~\ref{stirring} we overplotted the disc radius where 
1000\,km size planetary embryos has just formed, for a small grid of discs 
with $x_{\rm m}$ = 1, 3, 10 and for $M_* = 1.5, 2.0$\,M$_\odot$. The figure suggests 
that the size of discs around our older targets (HD\,10939, HD\,17848, 
HD\,50571, HD\,161868, HD\,170773, HD\,182681 and HD\,195627) can be well 
explained within the framework of the self-stirring model. In these cases 
there was enough time for the formation of large embryos even at the outer 
edge of the disc, and the initial surface density of the original 
protoplanetary disc did not exceed 10 times that of the MMSN. 

The figure also suggests that the initial disc masses in our sample were rather
high, in almost all cases higher than the MMSN by a factor of 1--10.
This conclusion is in accordance with our earlier findings that 
in Figure~\ref{mdust} the present-day dust masses in our 
systems are among the most massive debris discs for their age. 
Large dust mass suggests a massive underlying planetesimal belt, whose 
formation probably required a high mass protoplanetary disc as well. 

While constructing Fig.~\ref{stirring} we assumed that the timescale of the 
outward
propagation is equal to the age of the system. This might not be true, however,
because the expanding ring of planetesimal stirring could have reached in the
meantime the outer edge of the protoplanetary disc where it stopped. The typical
size of protoplanetary discs -- measured from silhouettes of 22 proplyds in the
Orion region -- range between 50 and 200\,au \citep{vicente2005}, very similar
to the typical sizes of our debris discs. Thus, it is possible that in some of
our systems the expansion of the collisional cascade has already finished some
time ago, and we now observe a late phase evolution of the debris disc. Note
that in this case, due to the reduced evolutionary timescale, the initial
protoplanetary disc had to be even more massive than suggested in 
Fig.~\ref{stirring}.

Our results do not mean that self-stirring is the only possible explanation for
the seven mentioned objects in our sample. Since these discs are rather massive,
giant planets might have also been formed, and they could contribute to the
stirring of planetesimals (for more details see Sect.~\ref{planetstirring}). 

All of our objects with age $\leq$40\,Myr, HD\,9672, HD\,16743, HD\,21997, and HD\,95086 as well 
as $\beta$\,Pic and 
HR\,8799
are located above the model curves 
predicted by the self-stirred model, implying that they would require 
 $x_m > x_{m,max} = 10$. Age estimates of these targets  
are reliable, in most cases based on moving group membership.
Since the differences between the derived and predicted sizes are larger than the 
uncertainties related to estimates of the location of dust producing planetesimals, these objects 
can be considered as prime candidates where the self-stirring scenario may not work.

In order to check how robust are the conclusions concerning these debris
discs, one can test how sensitive are the outcomes of the self-stirring model on
its basic assumptions. \citet{kb2010} investigated some modifications of the
self-stirring model. In one, they derived an equation similar to Eqn.~5 for a 
disc with a flatter surface density profile. In the other, they started their simulations using 
 uniform initial planetesimal sizes of 1\,km or 10\,km or 100\,km instead of  
an ensemble of 1\,m to 1\,km sized bodies. In the third one, they considered different
fragmentation properties of the colliding planetesimals.
We recalculated the model curves in Fig.~\ref{stirring} considering these modifications, 
and found that HD\,9672, HD\,21997, HD\,16743 and HD\,95086, as well as, $\beta$\,Pic and 
HR\,8799 are still outliers. In the previous analyses the initiation of dust production at a given radius was 
linked to the formation of $\sim$1000\,km size bodies. In fact, \citet{kb2008} 
found that the cascade is already initiated at smaller sizes, once the largest 
oligarchs reach sizes of 500\,km. Unfortunately, they offered no formulae 
for the formation of bodies smaller than 1000\,km. 
Their Fig.~8 and 9 suggest that the outer front where the largest bodies are $\sim$500 km
is at most 30-40\% larger than the site where 1000\,km sized planetesimals are formed.
Considering our conservative approach in determining the outer size of the planetesimal belt (Sect.~\ref{location}), 
the above arguments probably will not change our conclusions.

The initial size distribution of
the first planetesimals is very little constrained
by current models in the literature \citep{johansen2014}.
The recently proposed turbulent concentration and gravitational clumping models \citep[e.g.][]{johansen2007,cuzzi2008} predict 
significantly larger initial planetesimal
 sizes ($>$100\,km) than the classical coagulation model utilized by 
\citet{kb2008}. 
As mentioned above, \citet{kb2008} tested different intial planetesimal sizes, however, 
they used uniform size distribution and not an ensemble.
 Moreover, the maximum planetesimal size was only
100\,km, while graviturbulent processes can produce
planetesimals with a size of 1000\,km or even larger bodies especially in
discs with large initial mass \citep{johansen2012}. 
If the latter processes produce an ensemble of smaller planetesimals that can be stirred by the largest ones, 
then the collisional cascade can be started 
earlier than in the standard coagulation model. 
This phenomenon might help to produce very extended, young discs within the framework 
of self-stirring scenario. The detailed investigation of this case is out of the scope of the present paper.

In the following we will investigate whether alternative stirring mechanisms could explain 
the formation of the six outlier debris discs.

\subsubsection{Planetary stirring scenario} \label{planetstirring}
If the debris system harbours also a stellar companion or giant planet(s) -- which is 
not unreasonable in our massive discs -- then this large body may
interact with planetesimals formed in the same system.
By investigating the effects of the secular perturbations of a 
planet on a planetesimal disc, \citet{mustill2009} found that they can drive planetesimals on new 
intersecting orbits, increasing the frequency and velocity of their collisions. 
Due to this dynamical excitation, 
the collisions become destructive and produce dust even at large radial distances.
The planet's secular perturbation first initiates a collisional cascade at the closest 
region of the planetesimal belt, but the excitation 
 shifts to outer regions with time.   
Thus planetary stirring -- similarly to self-stirring -- 
is an inside-out process.
\citet{mustill2009} showed that in the case of a perturber located closer to the star than 
the minor bodies (internal perturber), the crossing time -- the time when the initially 
non-intersecting planetesimals'
orbits begin to cross -- is   
\begin{eqnarray} 
t_{\rm cross} = 1.53\times10^3 \frac{(1-e_{\rm pl}^2)^{3/2}}{e_{\rm pl}} \left( \frac{a}{\rm 10au} \right)^{9/2} \times \nonumber \\
\left( \frac{M_*}{M_\odot} \right)^{1/2} \left( \frac{M_{\rm pl}}{M_\odot} \right)^{-1} 
\left( \frac{a_{\rm pl}}{\rm 1au} \right)^{-3} [{\rm yr}],  \label{tcross}
\end{eqnarray}
where $a_{\rm pl}$ and $e_{\rm pl},$ are the semi-major axis and eccentricity of the planet's orbit, 
$M_{\rm pl}$ is the mass of the planet, while $a$ is the semi-major axis of the planetesimals. 
Depending on the planet's orbital parameters and mass, its secular perturbation could    
excite planetesimal eccentricities, starting a collisional fragmentation at a given radius faster than 
the time needed for the growth of 1000\,km size bodies. Thus, this stirring model can explain the presence of dust-producing 
 planetesimals at large stellocentric radius even if the self-stirring 
 scenario would require too high initial disc masses.  
 
 We used Eq.~\ref{tcross} to evaluate the feasibility of planetary stirring in our 
systems. In the calculations we used the age of the system as $t_{\rm cross}$,  
assuming that the timescale of planet formation is negligible. 
Moreover, we assumed that the planet formed in situ and did not migrate during system's lifetime.
Whether in a given system the hypothetical planet can affect the planetesimals even at the observed outer 
disc radius 
via secular perturbations depends on its mass ($M_{\rm pl}$), and semi-major axis ($a_{\rm pl}$), 
and eccentricity. 
In our seven older systems with ages of $>$50\,Myr, even a Jupiter-mass planet orbiting at 
$>$35\,au with a moderate eccentricity of 0.1 can excite the motion 
of planetesimals located at $\langle R_{\rm avg} \rangle$ via its 
secular perturbation. In two cases out of these seven (HD\,50571 and 
 HD\,195627) the perturbing planet's semi-major axis can be even lower, $\gtrsim$15\,au. 
The current high-contrast direct imaging observations are not sensitive enough to 
detect such a relatively low-mass planet at the given separations.   
These results suggest that although these seven systems can be explained via self-stirring, 
they can be equally well explained by the effect of a planet. Future, more refined planet 
search techniques will resolve this question.

\begin{figure*} 
\includegraphics[scale=.32,angle=0]{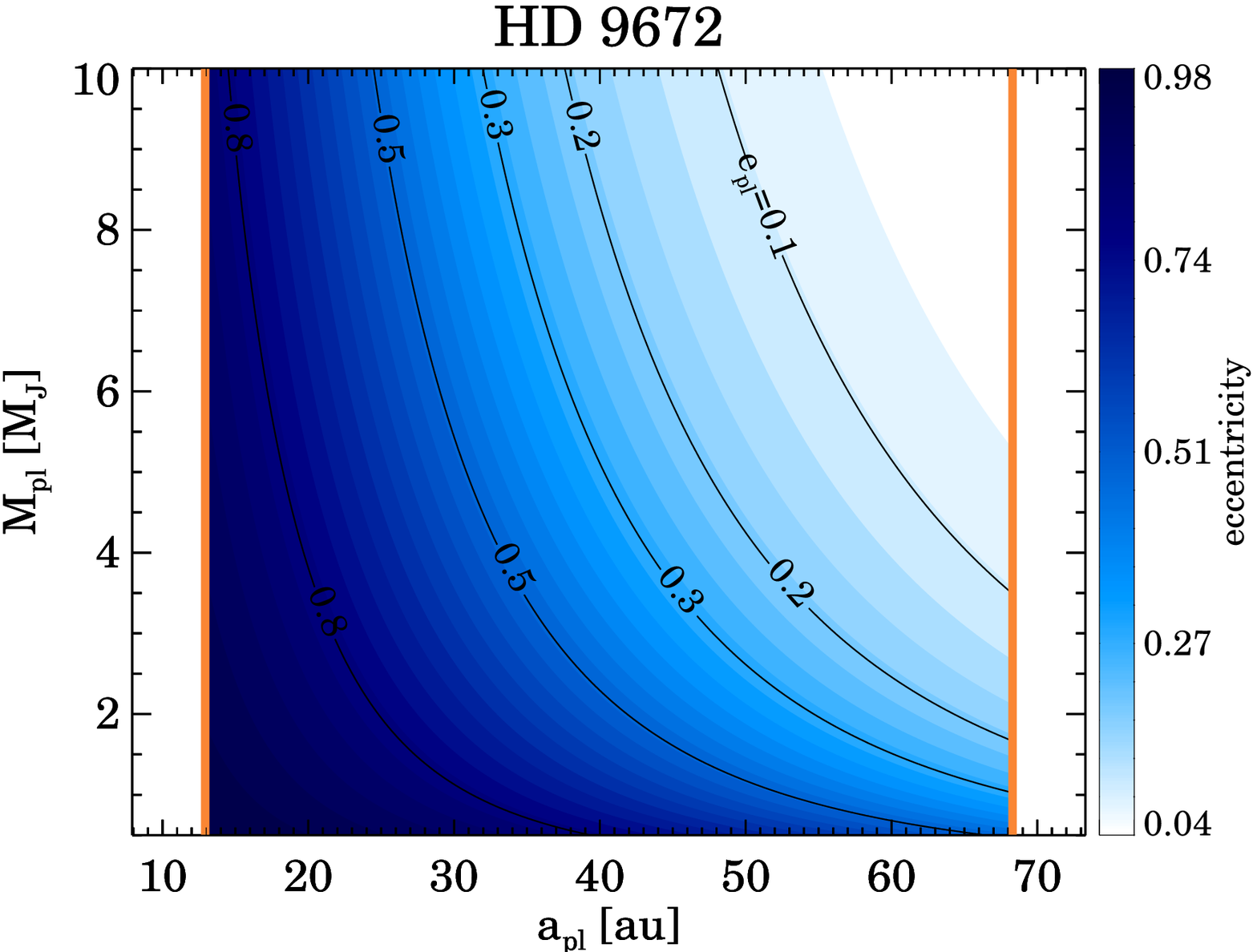}
\includegraphics[scale=.32,angle=0]{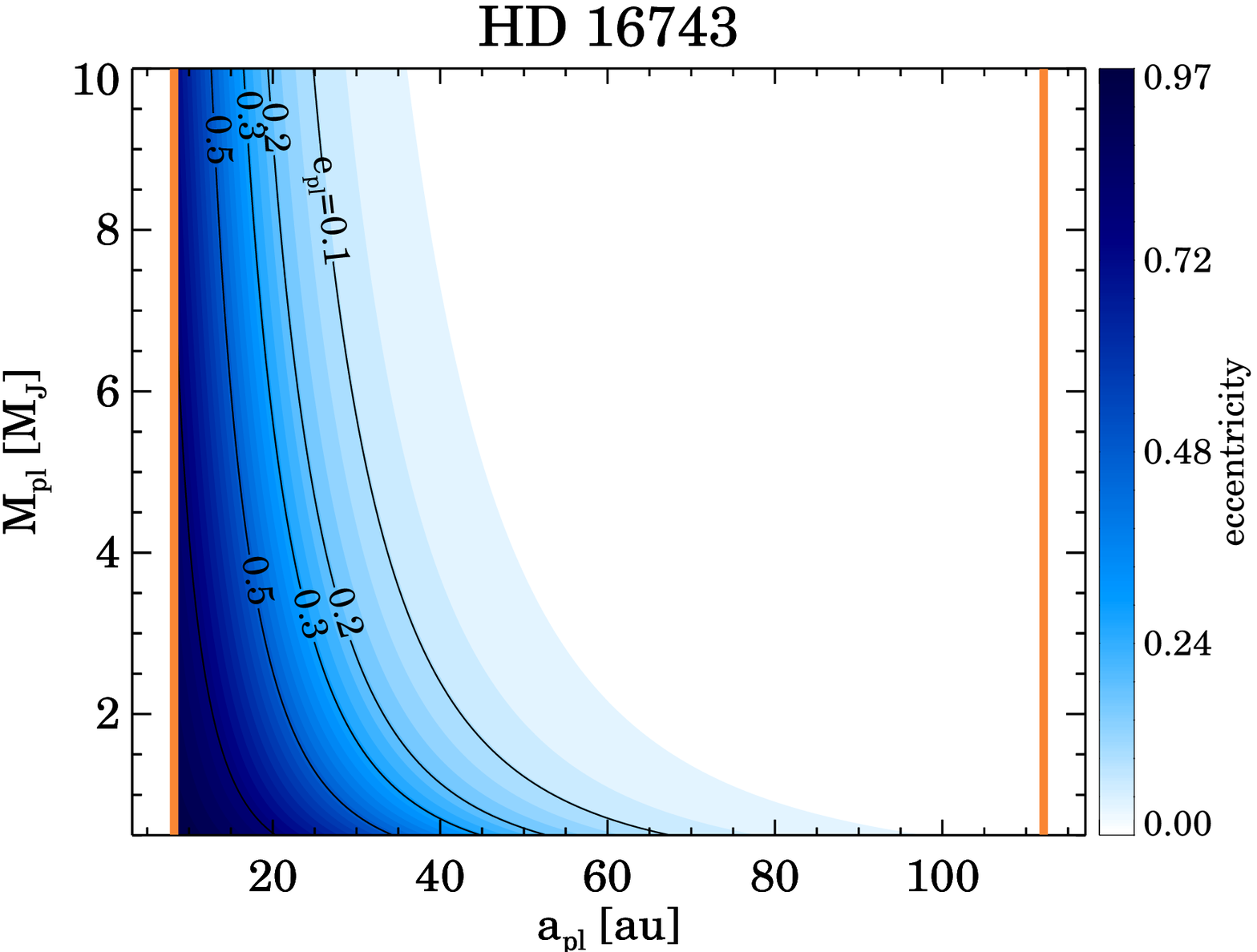}
\includegraphics[scale=.32,angle=0]{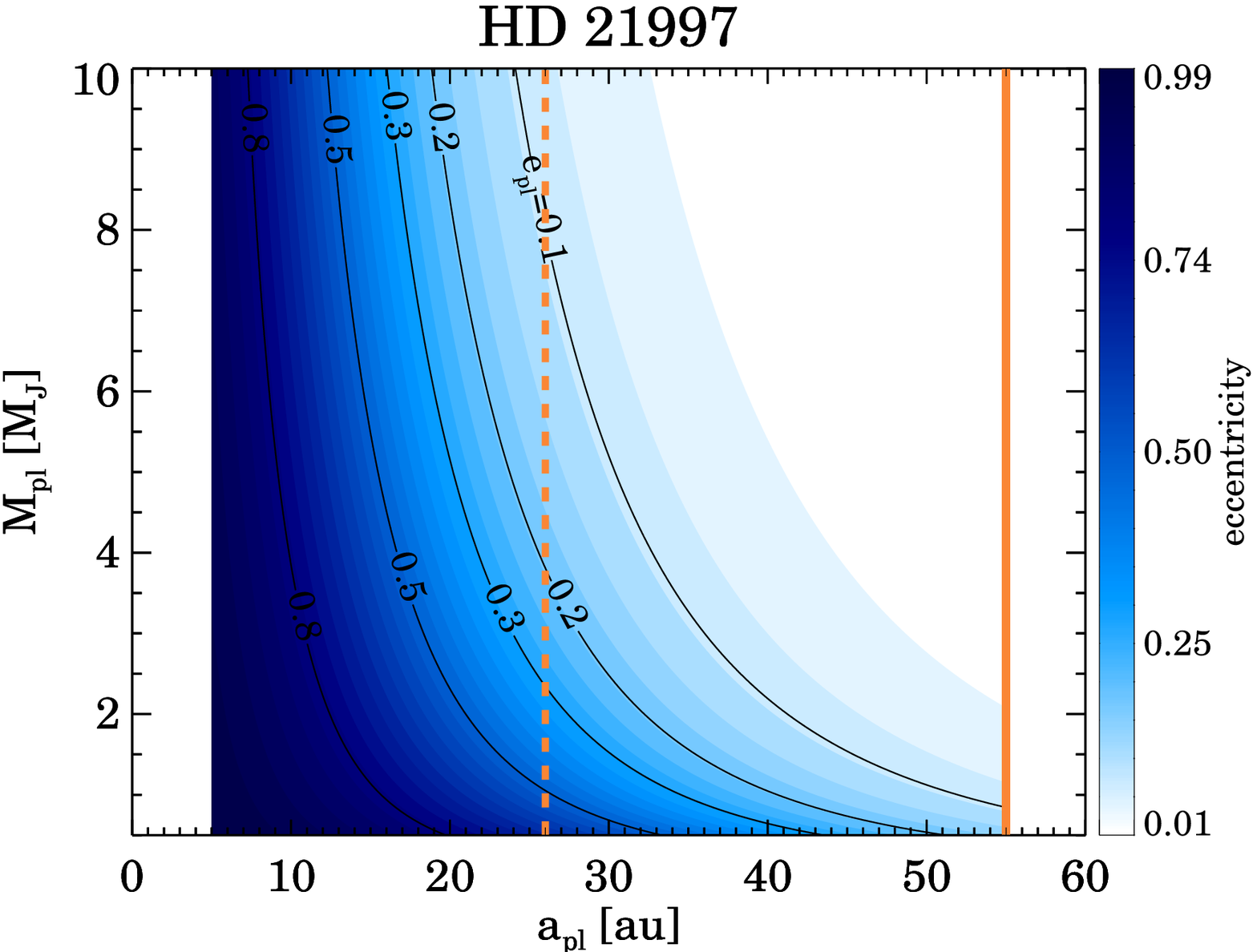} \\
\includegraphics[scale=.32,angle=0]{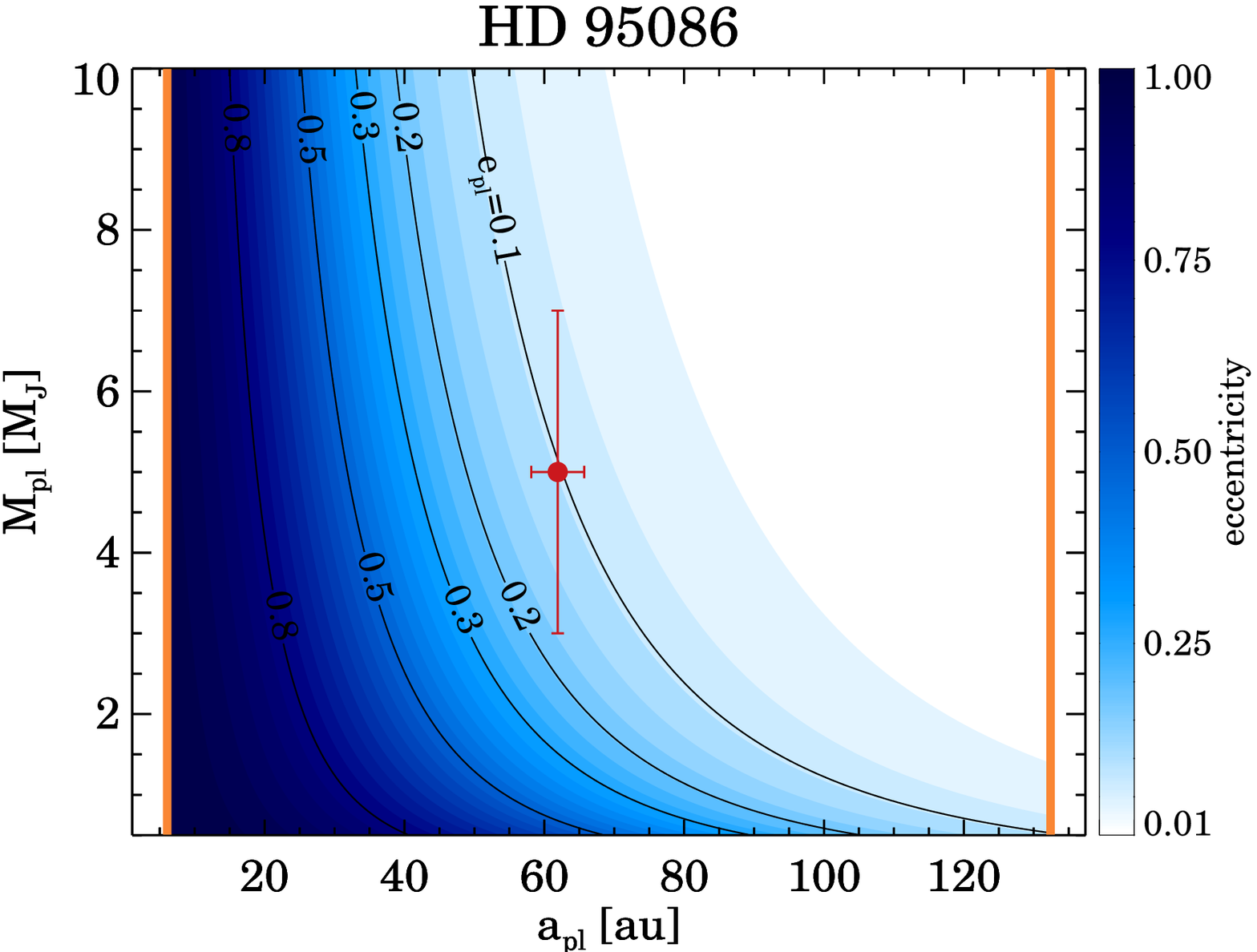}
\includegraphics[scale=.32,angle=0]{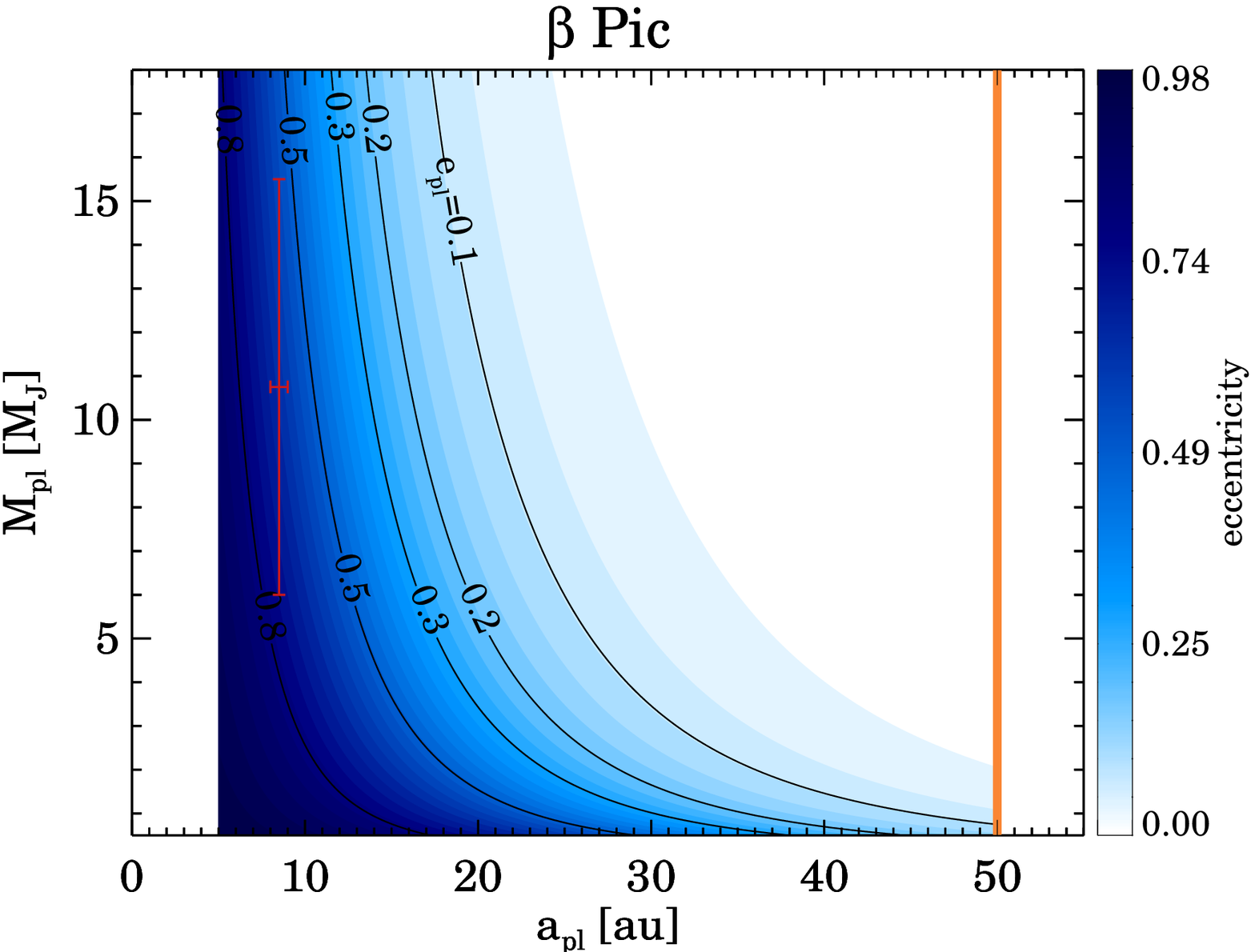}
\includegraphics[scale=.32,angle=0]{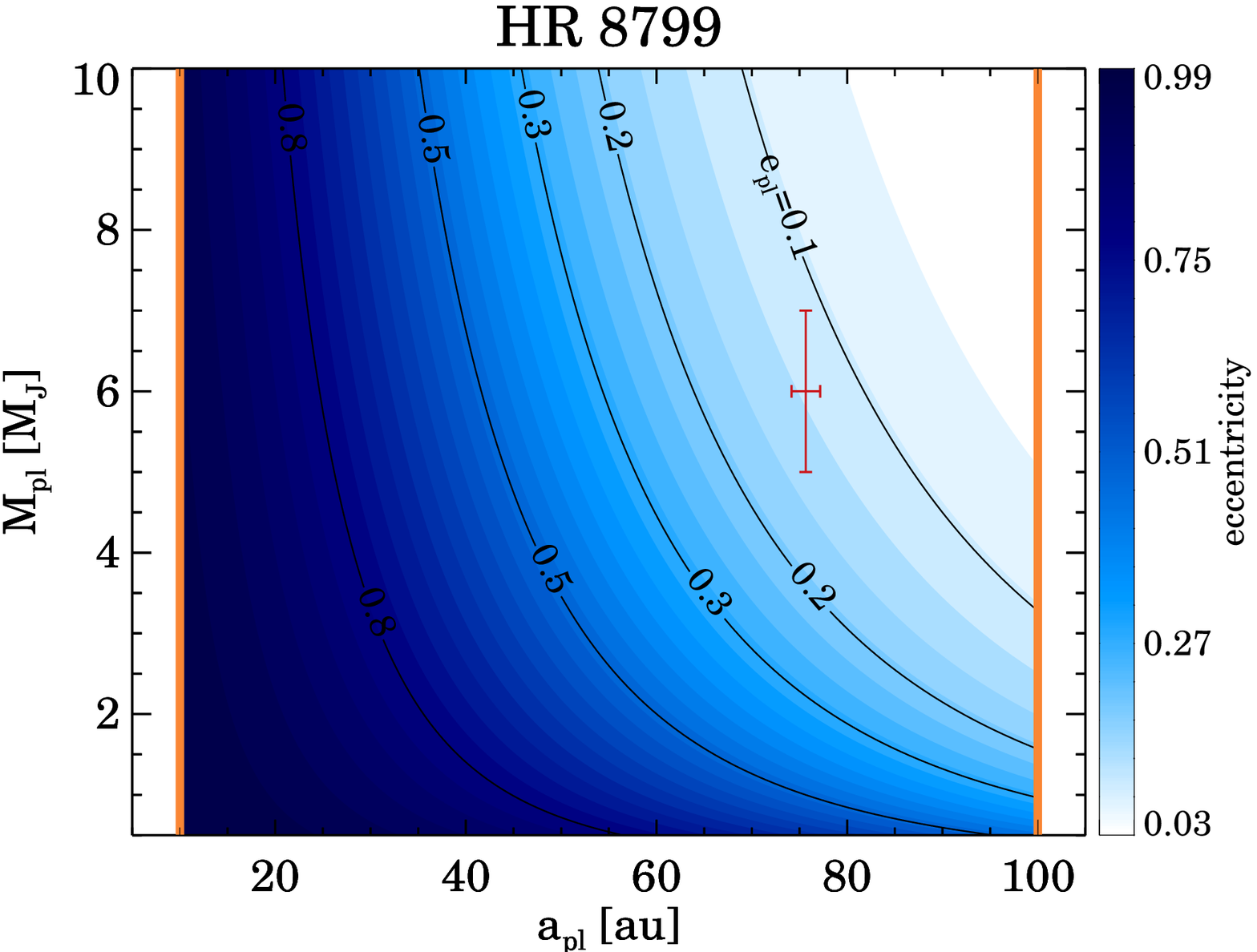}
\caption{ {\sl  Evaluation of the planetary stirring model for those targets where the self-stirring 
model turned out to be unfeasible. 
The figure shows the eccentricity values, in the $a_{\rm pl}, M_{\rm pl}$ parameter space, 
necessary to stir planetesimals located at 
the outer edge of the specific discs in less than the systems' age. 
The black contour lines correspond to eccentricity values of 0.1, 0.2, 0.3, 0.5 and 0.8. 
Vertical solid lines indicate the inner radius of the outer dust belt and the radius of the inner dust belt
(if it is present).
For HR\,8799, $\beta$\,Pic, and HD\,21997 the inner radii of the outer component 
were taken from the literature \citep{matthews2014a,dent2014,moor2013b}.
For the other targets this parameter was set based on our geometrical models (Sect.~\ref{planetstirring}). 
In the case of HD\,21997, where gas is also present, the upper limit for the inner 
radius of the gas disc \citep[taken from][]{kospal2013} was also marked by a dashed orange line. There are three systems (HR\,8799, $\beta$\,Pic, and HD\,95086) 
where a giant planet or planets have already been discovered. In these cases the planets were also displayed by red symbols 
(at HR\,8799, where four giant planets are known, we plotted only the outermost one).
\label{plstirring}}  
 }
\end{figure*}

For systems where the self-stirring model found to be unfeasible 
-- HD\,9672, HD\,16743, HD\,21997, HD\,95086 from our sample and HR\,8799 and $\beta$\,Pic -- 
we performed a more detailed analysis regarding the planetary stirring at the outer edge of 
the planetesimal belt. 
In Fig.~\ref{plstirring} we plotted the necessary eccentricity values as a 
function of 
$M_{\rm pl}$ and $a_{\rm pl}$ for these systems. 
We assume an internal perturber, therefore $a_{\rm pl}$ must be smaller than the inner radius 
of the outer disc. 
For HR\,8799, $\beta$\,Pic, and HD\,21997 we
used inner radii derived from resolved submillimetre images. 
For the other cases we used our Herschel-based geometrical model results.
Since $\langle R_{\rm in} \rangle$ was not constrained well, we explored the range of 
$a_{\rm pl} < \langle R_{\rm in} \rangle + \sigma_{\langle R_{\rm in} \rangle} $
for these cases.
When an inner belt is present we adopted its blackbody radius as a lower limit for $a_{\rm pl}$, otherwise 
five au, the orbital radius of Jupiter was used. 
The figure demonstrates that in all systems there is a large variety of $M_{\rm pl}$, 
$a_{\rm pl}$ pairs where a planet with a moderate eccentricity ($<$0.1) can stir the outer disc 
in less than the systems' age.  Thus, planetary stirring is a reasonable explanation for these sources. 
However, the planets are typically more massive than in the previous 7 cases.

Three systems (HR\,8799, $\beta$\,Pic and HD\,95086)  
are already known to harbour wide orbit giant planets discovered by direct imaging technique 
\citep{marois2008,lagrange2010,rameau2013}. For these objects we overplotted 
the planet's parameters in Fig.~\ref{plstirring} with red symbols. The planet of HD\,95086 has a mass of 
5$\pm$2\,M$_{\rm J}$ \citep{rameau2013} and orbital separation of $\sim$62\,au 
(computed from the projected distance, assuming that the orbital plane and the disc are coplanar), therefore it 
needs to have an eccentricity 
 of $\gtrsim$0.1 for stirring planetesimals at $\langle R_{\rm avg} \rangle \sim 220$\,au. 
The mass of $\beta$\,Pic\,b is in the range of 6.0--15.5\,M$_{\rm J}$ \citep{bonnefoy2013},
 while the semi-major axis is 8--9\,au and the eccentricity is $<$0.17 \citep{chauvin2012}.
Based on the formula of \citet{mustill2009} we found that $\beta$\,Pic\,b can force 
intersecting planetesimal orbits 
only at radial distance of $\sim$90\,au or less. 
Thus this planet alone cannot explain the presence 
of large dust grains at larger radii  
\citep[based on ALMA data the disc outer radius is $\sim$130\,au,][]{dent2014}.
At HR\,8799 we considered only 
the outermost planet in our calculations. HR\,8799\,b has a projected distance of $\sim$68\,au to the host-star, 
its mass is estimated to be 5--7\,M$_{\rm J}$ \citep{marois2010}. 
The inclination of the debris disc was derived to be 26$\pm3${\degr} based on resolved {\sl Herschel} PACS images 
\citep{matthews2014a}. Assuming the same inclination for planet's orbit, 
the orbital eccentricity must be $\gtrsim$0.1 to efficiently excite  
the motion of planetesimals even at $\sim$310\,au, the estimated outer edge of the planetesimal belt 
\citep{matthews2014a}. 

According to our current knowledge, among our targets only HD\,16743 have stellar companions. 
However, these companions are very widely separated ($\sim$12\,700\,au).
By computing the crossing time for the secular perturbation of these external perturbers -- utilizing Eq.~16 from \citet{mustill2009} 
-- we found that they could not have any influence on the disc around HD\,16743 at a time less than the age of 
the system. 


It should be noted that in the previous calculations we focused only on the secular perturbation 
caused by a possible planet or planets formed in situ. The interaction between planets and planetesimals, however, could be 
manifold.  
A migrating massive planet could sweep the
planetesimals from their original orbits by gravitational perturbation, and possibly trap them 
in resonances \citep{wyatt2008}. 
A planet-planet scattering can either lead to a significant 
clearence of leftover planetesimals making the formation of 
a debris disc impossible or -- in case of a mild evolution  -- 
can trigger the debris formation by dynamically exciting the 
planetesimals \citep{marzari2014}.
If the planetesimal
population is massive enough, they also influence the dynamics of the young planet(s), either stabilizing
or destabilizing the planetary system \citet{moore2013}.
A planet formed in the inner protoplanetary disc can scatter 
large embryos in its neighbourhood to the outer disc \citep{goldreich2004}, where it can excite 
the motion of the smaller bodies. These processes could lead the stirring of the outer planetesimals 
  on a shorter timescale than by secular perturbation purely. 
Furthermore, as \citet{mustill2009} noted, their computations are valid for a system where the perturbing planet(s) 
and the planetesimals are far from each other. In case of close planets, the proposed formulae 
overestimate the crossing time. 
Further investigations are needed to explore the possible relevance of
any of these processes for our systems.

\subsubsection{Stirring induced by stellar encounters} 
A close encounter with a nearby star can also stir up the motion of planetesimals in the outer disc.
Investigating the evolution of a planetesimal disc after a moderately close encounter with a passing 
star, \citet{kb2002} concluded that such an event can 
raise minor bodies' eccentricities so that collisions become destructive.
However, the enhanced dust production is not maintained for a long time, since  
collisions of small bodies damp the planetesimal velocities and thereby halt the 
collisional cascade after a time. According to \citet{kb2002}, the dust luminosity 
changed as $L_d / L_{bol} = L_0 / [ \alpha + (\frac{t}{t_d})^\beta],$ where $L_0$ 
is the maximum dust luminosity, $t_d$ is the damping time, while values of $\alpha$ and $\beta$ 
were found to be $\sim$1--2 and $\sim$1, respectively.  
Discs with larger initial masses have shorter damping times and produce more dust.
In a gas-free disc with a surface density corresponding to the MMSN the damping time 
is about 0.1\,Myr at 70\,au and 1.0\,Myr at 140\,au.
If gas is also present, the damping becomes
more efficient due to gas drag, resulting in shorter damping time.
In the following we evaluate this scenario for our cases.

 Close stellar encounters 
are very rare events among field stars, occuring 
once every 10\,Gyrs \citep{wyatt2008}.
Although close encounters are much more probable in a dense cluster \citep{breslau2014}
none of our stars belong to such a stellar cluster now. Nevertheless, we cannot exclude the possibility 
that some of them were born in a dense star forming region.  
However, even in such a case, considering  
the predicted short damping timescales, an encounter-induced dynamical excitation probably 
does not play a crucial role in current stirring of our discs.

By tracing the space motions of more than 21000 stars in the past 1\,Myr, \citet{deltorn2001} investigated     
the possible close stellar encounters for many stars with debris discs. Interestingly, 
the closest encounter was found for HD\,17848, one of our targets. According to their 
calculations, HD\,17848 encountered with another debris-disc bearing star HD\,20010 $\sim$350\,kyr ago 
with a separation of 0.081$^{+0.063}_{-0.049}$\,pc. Based on formulae of \citet{kalas2001}, however, 
such a relatively far encounter may not have significant influence on the planetesimals motion at a 
stellocentric distance of $\langle R_{\rm avg} \rangle$ $\sim$170\,au in HD\,17848.  
Thus, stellar encounter does not look a feasible explanation for our discs.

\subsubsection{Additional considerations}

HD\,9672, HD\,21997 and $\beta$\,Pic, three discs which are unlikely to be self-stirred, 
show another interesting feature: 
they are the only known debris systems where a detectable amount of CO gas was measured 
at molecular rotational lines \citep{zuckerman1995,moor2011b,dent2014}. 
Depending on its quantity, the gas can have influence on the dynamics of dust grains 
\citep{takeuchi2001,besla2007,krivov2009}. 
In the $\beta$\,Pic system both the CO and C gas masses are lower than the measured dust mass 
\citep{dent2014,cataldi2014}.
Since the gas may probably be of 
secondary origin \citep{fernandez2006,dent2014} -- i.e. released from colliding/evaporating 
grains and planetesimals -- the total gas mass may not exceed significantly the mass of these 
constituents and thereby 
the relatively low amount of gas may not have impact on the dust dynamics.
Though the origin of gas in HD\,9672 (49\,Ceti) is still debated, 
recent works proposed that it may also be secondary, implying that its influence on 
dust might not be significant. 
In the disc around the $\sim$30\,Myr old HD\,21997, we detected 
0.04--0.08\,M$_\oplus$ CO gas \citep{kospal2013} 
and argued that the gas may rather be residual primordial implying that a significant amount of H$_2$ gas
may also be present there, thus the gas mass surpasses the dust mass. In such an environment, 
 the gas--dust coupling could be stronger and might induce radial migration of grains 
whose rate depends on the amount of gas \citep{moor2013b}. \citet{kb2008} start 
 their simulations in a disc where gas is also present. The gas density declines exponentially 
 with time, the gas removal timescale was set to 10\,Myr. The fact that HD\,21997
 still harbours large amounts of gas may significantly modify the predictions 
 of the self-stirred model, making the analysis of stirring mechanisms difficult in this system.    
 
Alternatively, in the youngest systems from our sample, 
the outermost portion of the discs can partly be composed of primordial dust grains, because 
 without effective stirring, the low velocity collisions lead to merging, rather than destruction.
 In the self-stirring model, this would mean
  that the largest planetesimals in these 
 outer regions are in a pre-oligarchy phase and still growing, the collisional cascade has not yet
 been initiated, and the 
 bulk of available solids are in moderately large grains that are still observable at 
 far-IR and submillimetre wavelengths. According to \citet{heng2010} and \citet{krivov2013} in a 
 disc where the  
 collisions are not typically destructive, the majority of grains would radiate like blackbodies. 
 The fact that the SED of most targets deviate significantly from a blackbody 
 at long wavelengths ($\beta = 0.8-1.2$) and that the $\Gamma$ factor is $>$1.0, i.e. a significant amount 
 of non-blackbody grains are also present \citep[see also in,][]{pawellek2014},  
makes this scenario less likely. Nevertheless, we cannot exclude the possibility 
that the outermost regions in our targets are partly composed of primordial grains. 
High spatial resolution multiwavelength ALMA observations will answer this question.

Interestingly, HR\,8799 and HD\,95086 -- two discs among the three
that mostly deviate from the predictions of self-stirring
mode -- resemble each other in terms of the structure
of their planetary system as well: they harbour a warm in-
ner dust belt and a very broad colder outer disc and giant
planet(s) between the two dusty regions. Indeed, \citet{su2013}
proposed that a warm and cold debris dust belt with
a large gap between them could be a signature of planets in
the intervening zone. The failure of the self-stirring model
could be another hint for the presence of massive planets
in debris systems. Therefore HD\,9672 and HD\,16743, where
the self-stirring was also found to be unfeasible and that may
harbour multiple dust rings could also be promising candidates for planet searching programmes.

\section{Summary and conclusions} \label{conclusion}

With the aim of investigating the possible stirring mechanisms in debris discs, we
observed 11 targets with the {\sl Herschel Space Observatory} between 70 and 500\,{\micron}. The
discs are among the most massive and extended known debris discs. The analysis of
the excess emission over the photosphere implied that all targets harbour cold
outer dust belts, while seven of them
may also harbour warmer, inner debris dust as well. We found that all outer discs
are spatially extended at 70 and 100\,{\micron}, five of them being resolved for the first
time. 

By fitting a geometrical ring model to the far-infrared images, we determined the
inclination and position angle of the discs and estimated the outer size of
emitting regions, which is typically larger than 190\,au. In all systems, the best
fit to the observed far-IR brightness distribution was achieved by broad dust rings
(dR/R $>$ 1). In the case of HD\,170773 even the inner edge of the disc was well
resolved, revealing a broad outer dust belt between 80 and 270\,au.

In order to learn about the relative importance of the possible mechanisms which
may stir up the planetesimals and trigger debris dust production, we evaluated the
feasibility of the self-stirring scenario for our 11 targets.
To this end, we took into account their
measured disc sizes and the ages of the systems and compared
them to the \citet{kb2008} model predictions.
We concluded that this
explanation might work for seven discs, but is highly unlikely for HD\,9672, HD\,16743,
HD\,21997 and HD\,95086. In the latter young systems the dust producing regions are
located far beyond the maximum stellocentric distances predicted by the
self-stirring model. Using literature data, we claim that self-stirring is also
unfeasible in the well-known young massive debris discs $\beta$\,Pic and HR\,8799.
Taking into account different initial 
planetesimal size distributions 
\citep[e.g. due to rapid planetesimal formation via turbulent 
concentration and gravitational clumping,][]{johansen2014} 
might further refine our conclusions.

The mentioned discs are potential candidates for planetary stirring, since in all
of these systems there exist reasonable planetary configurations in which the
stirring of the outer disc can be occured. $\beta$\,Pic, HR\,8799 and HD\,95086 have
already been known to harbour massive wide-separation planet(s) and in
the latter two cases the known outer planets of the system can dynamically excite
planetesimals in the whole disc. The other three systems, HD\,9672, HD\,16743, and also HD\,21997,
could be prime targets for future planet search programmes. 
Interestingly the discs around
HD\,9672, HD\,21997 and $\beta$\,Pic -- uniquely among known debris systems --  
harbour detectable amount of molecular CO gas as well.

Our study demonstrated that among the largest and most massive debris discs
self-stirring may not be the only active scenario, and potentially planetary
stirring is responsible for destructive collisions and debris dust production in a
number of systems.

\section*{Acknowledgements}
We are grateful to our referee for the useful comments.
This work was supported by the Momentum grant of the MTA CSFK Lend\"ulet Disk Research Group, 
the ESA PECS Contract No. 4000110889/14/NL/NDe
as well as the Hungarian Research Fund OTKA grants K101393 and K104607.
A. M. and Gy.~M. Sz. acknowledges support from the Bolyai Research Fellowship of the Hungarian Academy of Sciences.
This publication makes use of data products from the Wide-field Infrared Survey Explorer, 
which is a joint project of the University of California, Los Angeles, and the 
Jet Propulsion Laboratory/California Institute of Technology, funded by the National 
Aeronautics and Space Administration.
The publication also makes use of data products from the Two Micron All Sky Survey, which
is a joint project of the University of Massachusetts and the Infrared Processing and
Analysis Center/California Institute of Technology, funded by the National Aeronautics
and Space Administration and the National Science Foundation.
Our research has made use of the VizieR catalogue access tool, 
CDS, Strasbourg, France.

\appendix 
\section{HD~182681 -- a candidate member of the $\beta$~Pic moving group?} \label{appendix}
We used the {\sc BANYAN} web tool \citep{malo2013} to perform a first evaluation of 
the membership probability of HD~182681 in the $\beta$~Pic moving group (BPMG). 
Based only on the {\sl Hipparcos} astrometric data (position, proper motion, 
trigonometric parallax) only, {\sc BANYAN} gave a probability of 99.3\% for the membership of the star.
By placing HD~182681 in the colour-magnitude diagram of young moving groups constructed by 
\citet[][Fig.~3]{malo2013}, we found a good match with the locus of BPMG stars, implying that its age could be consistent 
that of the group.
These properties make HD~182681 a potential candidate member of the group.

In order to further investigate its membership status we obtained a high-resolution 
optical spectrum of HD~182681 on 2011 April 16, using the 
Fiber-fed Extended Range Optical Spectrograph \citep[FEROS,][]{kaufer1999} 
on the 2.2\,m MPG/ESO telescope in La Silla, Chile.
The ``object-sky'' mode was used, with one fiber positioned at the target, and the other one on
the sky. The data reduction was
performed with the FEROS Data Reduction System (DRS) tool implemented within {\sc ESO-MIDAS}.
In order to derive the radial velocity of the star, 
we used the IRAF {\sc fxcor} task to cross-correlate
the obtained spectrum with a synthetic template (T$_{\rm eff}$=9750~K, [Fe/H]=0, $\log g=4.0$) from the spectral library 
of \citet{munari2005}. This procedure yielded a heliocentric radial velocity of $-$2$\pm$10~km~s$^{-1}$. The 
large uncertainty is due to the high projected rotational velocity of the star (275~km~s$^{-1}$, Table~\ref{stellarprops}).
Using this new radial velocity measurement, we computed the 
galactic space velocities of the star, obtaining $-$3.7$\pm$9.3, $-$14.2$\pm$1.5, $-$10.2$\pm$3.5\,km\,s$^{-1}$ 
for the U, V, W components, respectively. While the V and W space velocity components match well the 
corresponding characteristic space velocities of BPMG 
\citep[$V_0$ = $-$16.25 and $W_0$ = $-$9.27~km~s$^{-1},$][]{malo2013} the U component (velocity toward the Galactic centre) 
 deviates from that of the group \citep[$U_0$ = $-$10.94~km~s$^{-1}$,][]{malo2013} although it is within the uncertainty.
We obtained a velocity modulus of $d_{UVW} = \sqrt{ (U-U_0)^2 + (V-V_0)^2 + (W-W_0)^2) } = 7.6$\,km~s$^{-1}$.
Analysing the kinematic properties of previously identified members of 
young moving groups 
both \citet{shkolnik2012} and \citet{moor2013mg} requested $d_{UVW} < 5$\,km~s$^{-1}$ as a criterion 
for secure group membership.
Thus, based on the current results HD~182681 cannot be considered as a secure member.



\label{lastpage}

\end{document}